\documentclass[a4paper,fleqn,usenatbib]{mnras}

\usepackage{newtxtext,newtxmath}

\usepackage[T1]{fontenc}
\usepackage{ae,aecompl}

\usepackage{graphicx}	
\usepackage{amsmath}	
\usepackage{amssymb}	
\usepackage{cancel}
\usepackage[utf8]{inputenc}
\usepackage[export]{adjustbox} 
\usepackage{multirow, multicol, makecell}
\usepackage{booktabs}
\usepackage{enumitem}
\usepackage[caption=false]{subfig}
\usepackage{hyperref}
\hypersetup{
    colorlinks=true,
    linkcolor=blue,
    filecolor=magenta,      
    urlcolor=cyan,
}
\usepackage{xcolor}

\newcommand{\psqm}{\,m$^{-2}$}

\newcommand{\ncol}{\,$N_{\text{H}_2}$}
\newcommand{\ncold}{\,$N_{\text{H}_2}^{\text{dust}}$}

\newcommand{\psec}{\,s$^{-1}$}
\newcommand{\kkms}{\,K~km~\psec}
\newcommand{\coab}{\,[CO/H$_2$]}
\defcitealias{champ1}{Paper I}
\defcitealias{champ3}{Paper II}
\defcitealias{champ4}{Paper III}
\defcitealias{champ5}{Paper IV}
\defcitealias{champ2}{BRO13}
\defcitealias{gerner}{GBS14}
\defcitealias{schneider15}{SOC15}

\title[SED Fitting of Massive Molecular Clumps]{On the Diagnostic Power of FIR/Sub-mm SED Fitting in Massive Galactic Molecular Clumps}

\author[R. L. Pitts et al.]{
Rebecca L. Pitts,$^{1}$\thanks{E-mail: rlpitts@ufl.edu}
Peter J. Barnes,$^{1,2}$
Frank Varosi$^{1}$
\\
$^{1}$Dept. of Astronomy, University of Florida, 211 Bryant Space Science Center, Gainesville, FL 32611-2055 USA\\
$^{2}$School of Science and Technology, University of New England, Armidale, NSW 2351, Australia\\
}

\date{Accepted XXX. Received YYY; in original form 15 May 2018}

\pubyear{2018}

\begin{document}
\label{firstpage}
\pagerange{\pageref{firstpage}--\pageref{lastpage}}
\maketitle

\begin{abstract}
We used FIR and sub-millimeter continuum data from \textit{Herschel} and the Atacama Pathfinder EXperiment (APEX) to fit pixel-by-pixel modified Planck SEDs to prestellar and protostellar clumps in the Census of High- and Medium-mass Protostars (CHaMP) ($280^{\circ}<\ell<300^{\circ}$, $-4^{\circ}<b<+2^{\circ}$). We present maps of dust temperature ($T_{\text{d}}$) and H$_2$ column density (\ncol) for molecular clumps in the Carina Nebula complex (Regions 9 through 11), and surrounding RCW 64 (Region 26). We compare the column densities of CO and H$_2$ to chart regional variations in their correspondence, and derive maps of the CO abundance. We find the CO abundance varies by an order of magnitude or more across each region, averaging a few$\times$10$^{-5}$ CO per H$_2$, and that the CO abundance distribution across each clump is correlated in both form and magnitude with environmental conditions, especially $T_{\text{d}}$. This demonstrates that no single CO abundance suffices to convert from $N_{\text{CO}}$ to \ncol, even within a single molecular cloud. We also find that $L/M$ traces $T_{\text{d}}$ almost exclusively, and therefore is not an independent star formation tracer, but minima in $T_{\text{d}}$ almost universally coincide with maxima in \ncol, implying that cooling and density enhancement must be simultaneous steps in prestellar clump evolution. Finally, based on generalized histogram N-PDFs of clump-scale (1-5 pc) and cloud-scale ($\gtrsim10$ pc) samples, we could only obtain dual log-normal and power-law fits to $\sim10\%$ of the clumps. The physical parameters derived from these fits approach theoretical expectations, but have largely unknown uncertainties, so we advise treating the results of N-PDF fitting with caution.
\end{abstract}




\section{Introduction}
\indent Creating an accurate, detailed model of massive star formation remains one of the great challenges of astrophysics \citep[e.g.][]{beuth07,motte}. Massive stars are rare \citep{salpeter,chabr}, and their progenitor clouds evolve on timescales of tens of kyr to a few Myr from the onset of collapse \citep{kahn74,kuip11,schall,yusof} after a period of quiescence whose length is still debated \citep{balls,koda09,champ5}. Being homonuclear, the molecular hydrogen gas in which stars form \citep{ladas} is invisible at typical temperatures in molecular clouds \citep[of order 10 K, e.g.][]{stahlpal,sfnotes}. The nearest massive star nurseries are hundreds to thousands of parsecs distant, and are heavily obscured by intervening dust \citep[e.g.][]{reid09,zinnyork}. Only with the most recent instruments have astronomers created large-area, uniform, unbiased surveys on the scales of prestellar clumps ($\sim$1 pc) and cores ($\sim$0.1 pc) at millimeter and sub-millimeter wavelengths. Among the most noteworthy of these surveys are the \textit{Herschel} infrared Galactic Plane Survey \citep[HiGAL,][]{higal1,higal2,higal3}, the APEX Telescope Large Area Survey of the GALaxy \citep[ATLASGAL,][]{agal1,agal2}, and the FOREST Unbiased Galactic Plane Imaging survey with the Nobeyama 45-m telescope \citep[FUGIN,][]{fugin}. \\
\indent This paper focuses on the Census of High- and Medium-mass Protostars (CHaMP), one of the first large-area (20$^{\circ}$ $\times$ 6$^{\circ}$), unbiased surveys of massive ($\>$100 M$_{\odot}$) molecular clumps to reach sub-parsec-scale resolutions \citep[hereafter Paper I, BRO13, and Paper III, respectively]{champ1,champ2,champ4}, and is likely still the most comprehensive in its multi-wavelength, multi-transition coverage.
As with other radio surveys, CHaMP survey utilizes isotopologues of carbon monoxide, HCO$^+$, HCN, and many other species to estimate the mass budgets of individual molecular cloud clumps, and assess their physical conditions for signs of the earliest stages of star formation.\\
\indent Some of the biggest questions in massive star formation linger over the rates and efficiency of the process; fundamental to answering them is knowing the total mass available and understanding the environmental effects that control which gas goes on to form stars. As the next most abundant molecule in the universe, CO is the most readily detectable line-emitting tracer of H$_2$, with each isotopologue tracing a different column density regime depending on its abundance. Absent multi-isotopologue data, the measured $^{12}$CO line intensity, $W_{\text{CO}}$ (sometimes also denoted $I_{\text{CO}}$), is often assumed to be proportional to the H$_2$ column density, \ncol, by some constant $X_{\text{CO}}$ \citep{dame01}, typically around 2$\times$10$^{24}$ m$^{-2}$(K km s$^{-1}$)$^{-1}$ for the Milky Way \citep[we refer readers to the review by][for more information on its derivation and justification]{bolatto}.\\
\indent This approach has several problems, only gaining widespread appreciation the last few years, that bias both theory and observation toward underestimating the total molecular gas mass. Recent studies by \citet{thrumms}, \citet{wada}, and \citet[][hereafter Paper IV]{champ5} find $^{13}$CO and C$^{18}$O observations indicate that $N_{\text{CO}}$ is higher than expected given $W_{^{12}\text{CO}}$ at both the highest and lowest column densities. \citet{pine08}, \citet{lee14}, and \citet{kongco} find non-linear $X_{\text{CO}}$ dependencies on visual extinction ($A_V$), and---following e.g. \citet{bergin02} and \citealt{ripple}---attribute their results at least in part to CO abundance variations due to local conditions. We focus on better characterizing the variable CO abundance in molecular clumps, and identifying the environmental factors that control it.\\
\indent Traditionally, $X_{\text{CO}}$ encodes a fixed abundance relative to H$_2$, usually about 10$^{-4}$ CO molecules per H$_2$ molecule \citep[e.g.][]{sfnotes}, with a maximum value of about 2$\times$10$^{-4}$ CO per H$_2$ assuming solar metallicity. Yet we know from both observations and laboratory experiments that the amount of CO in the gas phase varies by some three orders of magnitude over the range of gas densities typically observed in massive star forming regions in FIR to mm-wave emission. \citet{caselli99}, \citet{kramer}, \citet{bacmann}, \citet{akhern}, and \citet{fontani} each found that CO gradually freezes out at temperatures below about 20~K by comparing line emission of different isotopologues of CO to dust emission, visual extinction ($A_V$), or other continuum tracers in molecular clouds and cores. \citet{oberg09}, \citet{munoz10}, and \citet{munoz16} showed that CO ice desorbs under UV irradiation well below the lowest temperature at which thermal desorption is seen---15 to 23~K---and that both thermal and UV photodesorption rates at a given temperature are anti-correlated with the temperature of the CO at deposition up to about 30~K. \citet{munoz16} and \citet{cazax17} show the amount of gas-phase CO released via thermal desorption increases by about three orders of magnitude starting between 15 and 23~K out to a peak between 27 and 30~K at typical dense molecular gas densities (10$^9$ to 10$^{10}$ m$^{-3}$). Simulations by \citet{penaloza18} show that strong interstellar radiation fields (ISRFs) or cosmic rays can effect the same range of CO abundances over the same range of densities across sub-parsec-scale clumps. Clearly the CO abundance must be mapped, and its environmental dependencies pinned down, to have any hope of knowing a molecular cloud's mass budget to better than back-of-the-envelope precision.\\
\indent Dust continuum emission provides an alternate estimate of molecular gas column densities, and a means by which to make sorely needed estimates of the CO abundance. Dust is not subject to the same photo-destruction, chemical destruction, freezing out, or depletion onto grains that can bias molecular line tracers at these temperatures (though it has other caveats). It is optically thin at the spatial scales probed by single-dish telescopes, and over nearly the full range of wavelengths needed to model thermal emission of dust $<$100 K. The major drawback is that the typical gas-to-dust mass ratio for the Milky Way is uncertain by about a factor of about two (the uncertainty is itself uncertain), and where measurements are available on the clump scale, there is evidence of wide variation between and within clumps. For instance, while the gas-to-dust ratio is typically assumed to be $\sim$100  \citep[e.g.][]{beckw} or 124 \citep{li2001}, values range from 162 \citep{zubko} in the Solar neighborhood to $\lesssim$10 in the centers of some isolated clouds of order 100 $M_{\odot}$ \citep{reach}. Additionally, dust traces HI as well as H$_2$, so in the absence of 21-cm HI observations or extinction maps, dust continuum emission can only provide an upper limit on the amount of molecular hydrogen. However, below about 100 K, nearly all hydrogen is expected to be in the form of H$_2$, and dust temperature is insensitive to uncertainties in the gas-to-dust ratio as used in spectral energy distribution (SED) fitting. Therefore, dust should predominantly trace molecular gas at the brightness temperatures to which \textit{Herschel} and LABOCA are sensitive. Moreover, \citet{reach2} showed that the cold gas not traced by either CO or 21-cm HI emission (``CO-dark gas'') is most likely to be relatively diffuse H$_2$ where CO is dissociated \citep{wolf}, and that ``cold'' HI with saturated 21-cm emission accounts for little of the gas mass traced by dust.\\
\indent In \S\ref{sec:meth} we describe our data selection and SED-fitting methods; in \S\ref{sec:pms} we discuss the parameter maps resultant from SED-fitting; in \S\ref{sec:disc} we discuss the interpretation of our results with respect to CO observations, $L/M$ ratios, and column density PDFs (N-PDFs); and we conclude with the highlights in \S\ref{sec:conc}.

\section{Data and Methods}
\label{sec:meth}
\subsection{CHaMP Observations}\label{ssec:champ}
Using data from the 4 m NANTEN telescope, Paper I identified 209 clumps in the area of $280^{\circ}<\ell<300^{\circ}$, $-4^{\circ}<b<+2^{\circ}$ in 4$'$ resolution maps of the 3 mm J = 1 $\rightarrow$ 0 transitions of HCO$^+$, $^{12}$CO, $^{13}$CO, and C$^{18}$O. These comprise the Nanten Master Catalog (NMC) started by \citet{yone} in the immediate vicinity of $\eta$ Carinae. Clumps in this catalog are grouped into 27 ``Regions''  (figures 1 and 2 in Paper I). Paper I followed up on the brightest 120 of these with the 22 m Mopra antenna of the Australia Telescope National Facility (ATNF)\protect\footnote{The Mopra telescope is part of the Australia Telescope National Facility which is funded by the Australian Government for operation as a National Facility managed by CSIRO. The University of New South Wales Digital Filter Bank used for the observations with the Mopra Telescope was provided with support from the Australian Research Council.}, mapping the J = 1 $\rightarrow$ 0 transitions of HCO$^+$, N$_2$H$^+$, and other transitions near 90 GHz at 37$''$ resolution, which resolved the 120 clumps into 303 sources. \citet[][hereafter Paper II]{champ3} then fit 2-component SEDs to the median pixel of each clump as seen in dust continuum emission by the InfraRed Astronomical Satellite (IRAS, \citealt{iras}), the Spatial Infrared Imaging Telescope (SPIRIT) III instrument aboard the Midcourse Space EXperiment (MSX) satellite \citep{spirit3,msx}, and other mid- to far-infrared missions not including \textit{Herschel}.\\
\indent Papers III and IV have since used Mopra to map the brightest 141 prestellar and protostellar clumps in the three isotopologues of CO and other species with emission lines near 110 GHz. To those maps in CO we compare parameter maps from pixel-by-pixel SED fitting of the dust continuum emission.

\subsection{Region Selection}\label{ssec:regs}
\begin{figure}
    \hspace{-2mm}
	\includegraphics[width=1.02\columnwidth]{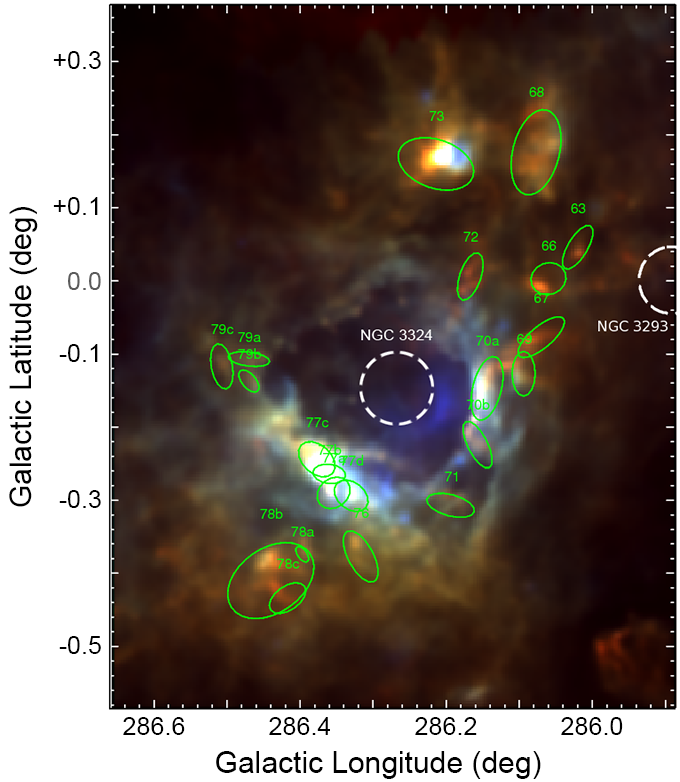}
	\vspace{-4mm}
    \caption{Three-color image of Gum 31, Region 9: \textit{Herschel} PACS 70\micron~(blue), PACS 160\micron~(green), and \textit{Herschel} SPIRE 350\micron~(red), with linear intensity scaling. CHaMP clumps identified in $^{12}$CO are labeled by BYF number, and the green ellipses show their approximate sizes and orientations.}
    \label{fig:gum31}
\end{figure}

\begin{figure}
    \hspace{-2mm}
	\includegraphics[width=1.02\columnwidth]{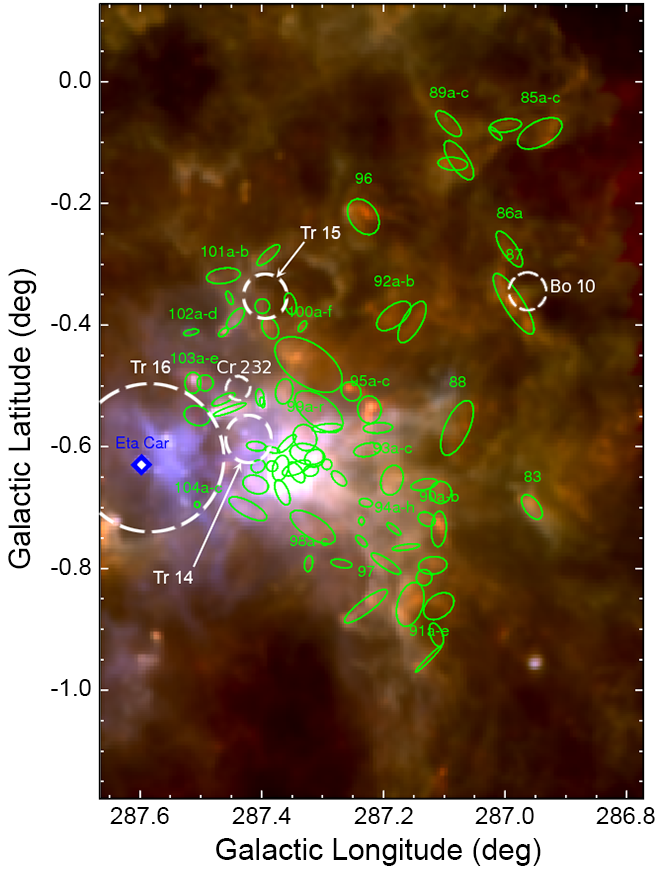}
	\vspace{-4mm}
    \caption{Same as Figure~\ref{fig:gum31}, but for the Region 10 and with square root intensity scaling. Due to crowding issues, clumps are labelled by group rather than individually. Some star clusters are marked for reference.}
    \label{fig:r10rgb}
\end{figure}

\begin{figure}
    \hspace{-2mm}
	\includegraphics[width=1.02\columnwidth]{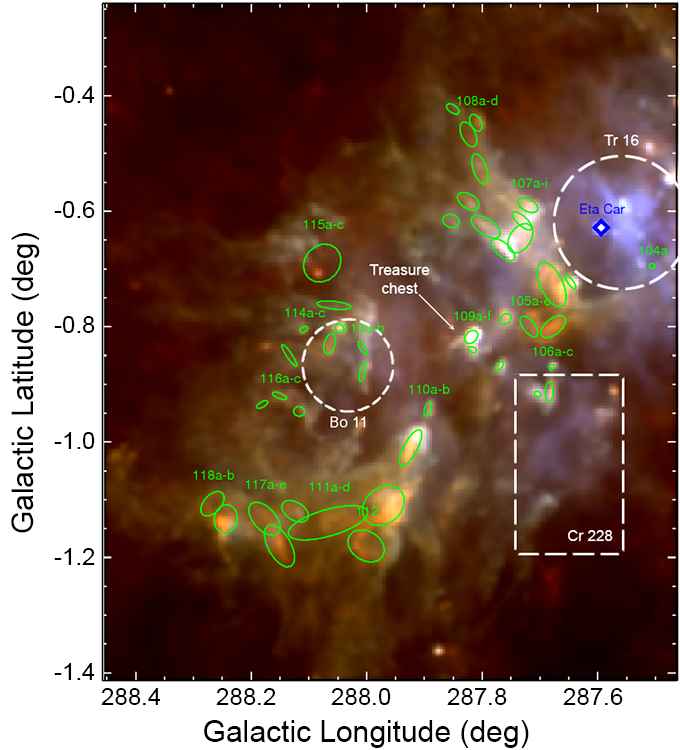}
	\vspace{-4mm}
    \caption{Same as Figure~\ref{fig:gum31}, but for the Region 11 and with square root intensity scaling. As with Figure~\ref{fig:r10rgb}, clumps are named by group and star cluster locations are included for reference.}
    \label{fig:r11rgb}
\end{figure}

\begin{figure}
    \hspace{-2mm}
	\includegraphics[width=1.02\columnwidth]{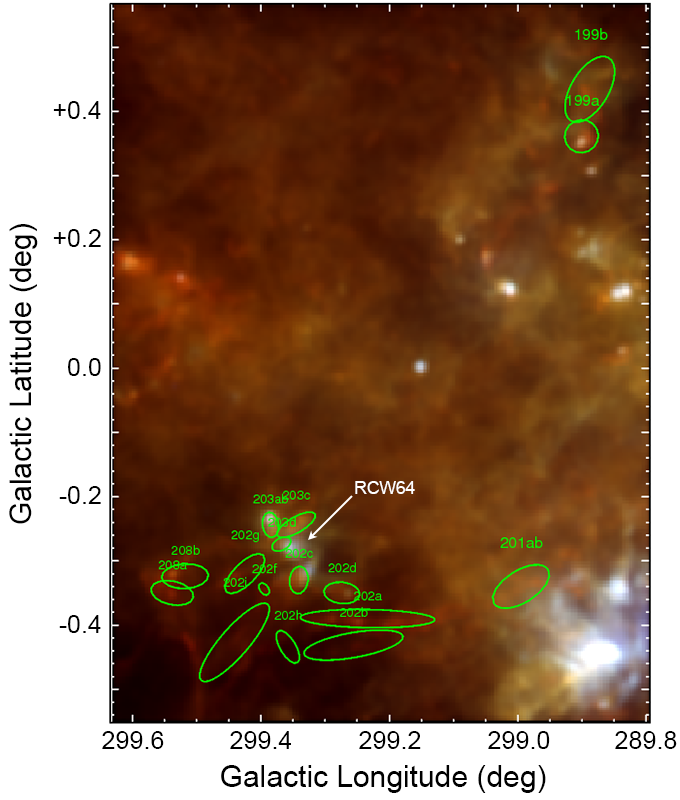}
	\vspace{-4mm}
    \caption{Same as Figure~\ref{fig:gum31}, but for the Region 26 and with square root intensity scaling.}
    \label{fig:r26rgb}
\end{figure}

For this paper we chose four test regions to benchmark our SED-fitting methods and obtain first comparisons with our molecular line data. These were Regions 9 through 11, together enclosing the entire Carina Nebula Complex (CNC) including Gum 31, and Region 26, a 0.8$^{\circ}\times$1.1$^{\circ}$ box enclosing foreground objects in the easternmost third of the more distant Dragonfish Nebula. Figures~\ref{fig:gum31} through \ref{fig:r26rgb} are RGB images of each region with clumps in the BYF catalog mapped in CO shown as labelled ellipses. For clump locations and 2D Gaussian decomposition parameters based on the $^{12}$CO data cubes, we refer readers to Table 1 of \citetalias{champ4}.\\
\indent Gum 31 \citep{gum31} is a diffuse HII region in the northwest corner of the CNC, widely thought to be associated with the open cluster NGC 3324 \citep{dias}. Distance estimates for the various clumps along the edges of Gum 31 are about 2.5$\pm$0.3 kpc with velocities relative to the local standard of rest in the range $-24\leq\,v_{\text{LSR}}\leq-21$~km~s$^{-1}$ \citep{yone}. This puts it at about the same distance as the rest of the Carina Nebula, and within the Carina Nebula's radial velocity range of $-26\leq\,v_{\text{LSR}}\leq-18$~km~s$^{-1}$ \citep{preib}. Gum 31's angular diameter of about 30$'$ translates to a physical size of slightly over 10 pc. The clumps around its periphery and to the galactic north are grouped into Region 9 of the CHaMP survey (see Figure~\ref{fig:gum31}). NGC 3293 also appears along the western periphery of Region 9, and its distance is consistent with the CNC's to within their mutual margins of error \citep{ngc3293}, but there is limited evidence of influence on the nearest CHaMP clumps (see \S\ref{sec:pms}).\\ 
\indent Regions 10 and 11 cover the rest of the CNC east of Gum 31. This part of the CNC is associated with a wealth of young stellar clusters, including: Trumpler 14, 15, and 16 (Tr14, Tr15, and Tr16), Collinder 228 and 232 (Cr228 and Cr232), Bochum 10 and 11 (Bo10 and Bo11), and the embedded Treasure Chest cluster \citep{mv75,smith00,smith05}. We refer readers to the review by \citet{smooks08} and the VISTA survey publications by \citet{vista1} and \citet{vista2} for more information on the stellar content of the CNC. Region 10 covers the so-called Northern Cloud, molecular gas between Gum 31 and Tr16, surrounding Tr14, Tr15, Cr232, and Bo10 (see Figure~\ref{fig:r10rgb}). The clumps in Region 11 comprise the Southern Cloud and the South Pillars \citep{smith00}, irradiated by Tr16 to the northwest, Cr228 to the southwest, and Bo11 to the southeast (see Figure~\ref{fig:r11rgb}), with the Treasure Chest cluster embedded in BYF 109 (the Treasure Chest). The clumps in Region 10 cover a wide range of $v_{\text{LSR}}$, from $-$6.3 to $-$31.5~km~\psec, but kinematic distances from Paper I are all well within a standard deviation of 2.5~kpc away. The same is true of the distances to clumps in Region 11, where $-11.6\leq\,v_{\text{LSR}}\,\leq-30$~km~\psec. The Keyhole Nebula near $\eta$ Carinae ($\eta$ Car) was marginally detected in C$^{18}$O by \citet{yone}, but was not sufficiently bright to be chosen for follow-up in the CHaMP survey.\\ 
\indent Region 26 (Figure~\ref{fig:r26rgb}) focuses on the RCW 64 complex \citep{rcw}, the surrounding nebulae Bran 386a-g \citep{bran}, and the 40 pc-long IRDC arcing due westward from these features, as well as a small pair of molecular clumps to the upper right at a similar $v_{\text{LSR}}$. The $v_{\text{LSR}}$ of these objects range from $-$27.1 to $-$44.0~km~\psec, which places them unambiguously on the near side of the galaxy in the Carina-Sagittarius Arm \citepalias[][and references therein]{champ1}. Some of the unlabeled bright features in Region 26 did appear in the first CHaMP survey publication, but were not followed up with Mopra observations in the three CO isotopologues, since they did not make the cut in HCO$^+$ intensity.\\
\indent In the background of Region 26 is the Dragonfish Nebula, recently found to be the most luminous star-forming complex in the Milky Way \citep{ramen1,ramen2,murray}. Its radial velocity range, $+20\lesssim\,v_{\text{LSR}}\,\lesssim+40$ \citep{russeil,fuente}, places it on the far side of the galaxy about 12.4$\pm$1.7 kpc away. The oblate complex consists of two nested cavities with semi-major axes of 150 pc and 430 pc in the plane of the sky, with 5 compact HII regions studded along the periphery of the inner cavity \citep{murray,fuente}. Two of these HII regions, [WMG70] 298.9$-$00.4 and 298.8$-$00.3 \citep{wmg70}, appear prominently in the lower right corner of figure (\ref{fig:r26rgb}). The entirety of Region 26 lies between the rims of the inner and outer cavities, and given that background contamination can significantly inflate estimates of \ncol\,\citep[see, e.g.][]{schneider15}, background subtraction became a necessity.

\subsection{Dust Continuum Images}\label{ssec:contdat}
To the CO maps made with Mopra as part of CHaMP, we compare the dust emission at the same coordinates using public image data from the Infra-Red Science Archive (IRSA) in the 5.8 to 870 $\mu$m range. The instruments that took the original data include the Large APEX BOlometer CAmera (LABOCA, \citealt{laboca}), the Spectral and Photometric Imaging REceiver (SPIRE, \citealt{spire}) and the Photodetector Array Camera and Spectrometer (PACS, \citealt{pacs}) on the \textit{Herschel} satellite \citep{hso}. For Region 9 only, we also used the Wide-field Infrared Survey Explorer (WISE, \citealt{wise}) to fit the warm dust surrounding NGC 3324, to isolate it from the cold thermal dust emission that was our focus. Data from the Multiband Imaging Photometer for Spitzer (MIPS, \citealt{mips}) were processed but ultimately not included in the final SED fits.
LABOCA images were originally published as part of ATLASGAL \citep{agal1}. \textit{Herschel} PACS and SPIRE data are mostly science and calibration images from the HiGAL, a 5-band survey of the region $-$60$^{\circ}\lesssim\ell\lesssim\,$60$^{\circ}$, $-$1$^{\circ}\lesssim\,b\,\lesssim\,$1$^{\circ}$ which has since expanded to include much of the CHaMP field \citep{higal1,higal2,higal3}. Table~\ref{tab:specs} shows a summary table detailing the specifications of all filters we used. The PACS 100 $\mu$m filter is omitted because no images in this filter were found to overlap regions with CO coverage in CHaMP.\\
\begin{table*}
 \centering
 \caption{Filter Specifications}
 \label{tab:specs}
  \begin{tabular}{  l |  l |  l |  l |  l |  l |  l  }
Telescope	&	Instrument	&	Filter name	&	$\lambda$ ($\mu$m)$^{\star}$	&	FWHM ($''$)	&	$\geq$50\% peak transmission ($\mu$m) & Source\\ \hline
APEX	&	LABOCA	&	(345 GHz)	&	870	&	19.2 $\pm$ 0.7$^{\ddagger}$	&	806 - 958	& 1\\
\textit{Herschel}	&	SPIRE	&	PLW	&	500	&	35.2$^{\ast}$	&	420 - 630	& 2\\
\textit{Herschel}	&	SPIRE	&	PMW	&	350	&	23.9$^{\ast}$	&	290 - 410	& 2\\
\textit{Herschel}	&	SPIRE	&	PSW	&	250	&	17.6$^{\ast}$	&	210 - 290	& 2\\
\textit{Herschel}	&	PACS	&	R	&	160	&	11.60 $\times$ 15.42	&	125 - 210	& 3\\
\textit{Herschel}	&	PACS	&	B	&	70	&	5.83 $\times$ 12.12	&	60 - 85	& 3\\
WISE	&	WISE	&	W4	&	22.1	&	12	&	...$^{\dagger}$	& 4\\
WISE	&	WISE	&	W3	&	11.56	&	6.5	&	...$^{\dagger}$	& 4\\
\multicolumn{6}{ l }{ }\\
\multicolumn{6}{ l }{\textbf{Notes:}} \\
\multicolumn{6}{ l }{1. \citet{laboca}; $^{\ddagger}$Average and standard deviation over all 248 bolometers}\\
\multicolumn{6}{ l }{2. \citet{spire,prespire}; $^{\ast}$geometric mean. Beams have ellipticities of $\lesssim$10\%.}\\ 
\multicolumn{6}{ l }{3. \citet{pacs}}\\ 
\multicolumn{6}{ l }{4. \citet{wise,wisecal}; $^{\dagger}$50\% peak transmission is a poor descriptor of WISE filter band edges}\\
\multicolumn{6}{ l }{$^{\star}$Characteristic wavelengths are isophotal for WISE filters and central for all others}\\
  \end{tabular}
\end{table*}
\indent We used ATNF's Miriad \citep{miriad} data reduction tasks to create mosaics of all images in a given Region and filter, convolve the images to the 37$''$ resolution of the Mopra maps, and regrid them with 12$''$ pixels using the CHaMP survey's $^{12}$CO maps as templates.\\
\indent To LABOCA images, limited by atmospheric water vapor emission and already background-subtracted, we applied a simple smooth-and-mask \citep[SAM,][]{thrumms} algorithm with a low threshold to exclude noise from fitting, while preserving as much as possible the filamentary emission verified as real by comparison to \textit{Herschel} SPIRE images. The mask was made by convolving the 37$''$ images to 74$''$, computing the standard deviation of the smoothed image ($\sigma_s$), and flagging as good only those pixels in the 37$''$ image with flux densities at least 2$\sigma_s$ above the mean of the smoothed image. The threshold was determined by trial and error. The threshold was set low because, while the flux density distribution in each image is a Gaussian centered on zero after background subtraction, the negative pixels tend to cluster around the perimeters of bright sources, which can suppress diffuse emission in a convolved image. However, the LABOCA data have such large uncertainties everywhere but at the centers of the molecular cloud clumps that, where these negative pixels might have had an impact, the LABOCA data become unimportant to the fit. Therefore, the most uncertain data essential to the single-temperature fit were the SPIRE data, with an absolute calibration error of about 15\% \citep{spire}.\\
\indent For some regions where 24 \micron~data from the 278 deg$^2$ MIPSGAL survey \citep{mipsgal} are available, their inclusion helped to tie down the high-frequency end of the SED in some locations, but in others caused discontinuities in the parameter maps, depending on the number of free parameters in fitting. However, for all but a handful of pixels in Region 26, removing MIPSGAL data improved the SED fits. We also tried using WISE data from the All-Sky release \citep{wisesup} at 22 \micron~for a similar purpose in Region 9, where MIPSGAL data were not available through IRSA. For single-temperature SED fits, including WISE data caused widespread discontinuities across the parameter maps of temperature and column density (see \S\ref{sec:pms}). However, in two-temperature fits, including WISE data at 12 and 22~\micron\,helped isolate the components of the gas that is presumably warm enough to contain atomic hydrogen, and improved the fit at 70~\micron.

\subsection{SED-Fitting with \texttt{Mosaic-Math}}\label{ssec:code}
\begin{figure}
    \centering
    \subfloat[BYF 77c]{\includegraphics[width=\columnwidth,trim={2mm 7mm 0 0},clip]{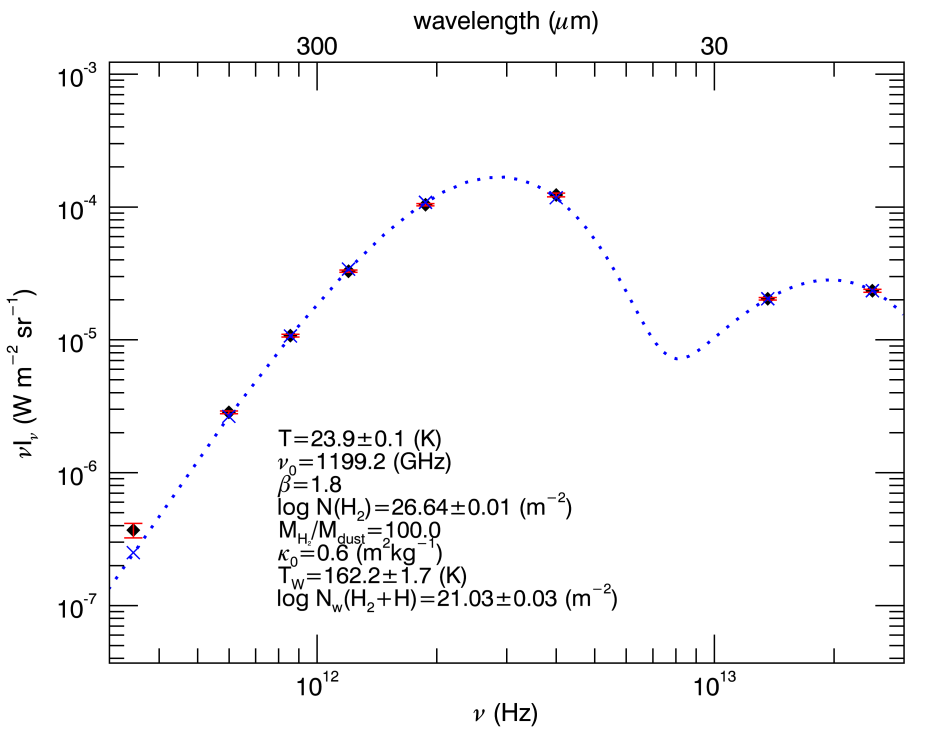}}\\ \vspace{-2mm}
    \subfloat[BYF 99a]{\includegraphics[width=\columnwidth,trim={2mm 6mm 0 0},clip]{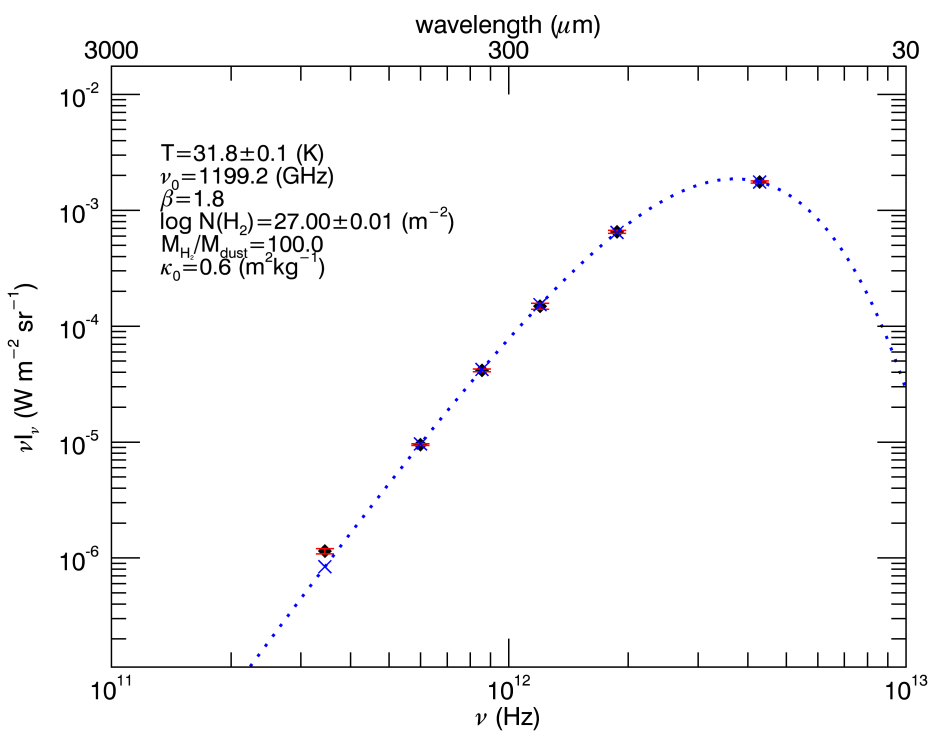}}\\ \vspace{-2mm}
    \subfloat[BYF 202a]{\includegraphics[width=\columnwidth,trim={2mm 7mm 0 0},clip]{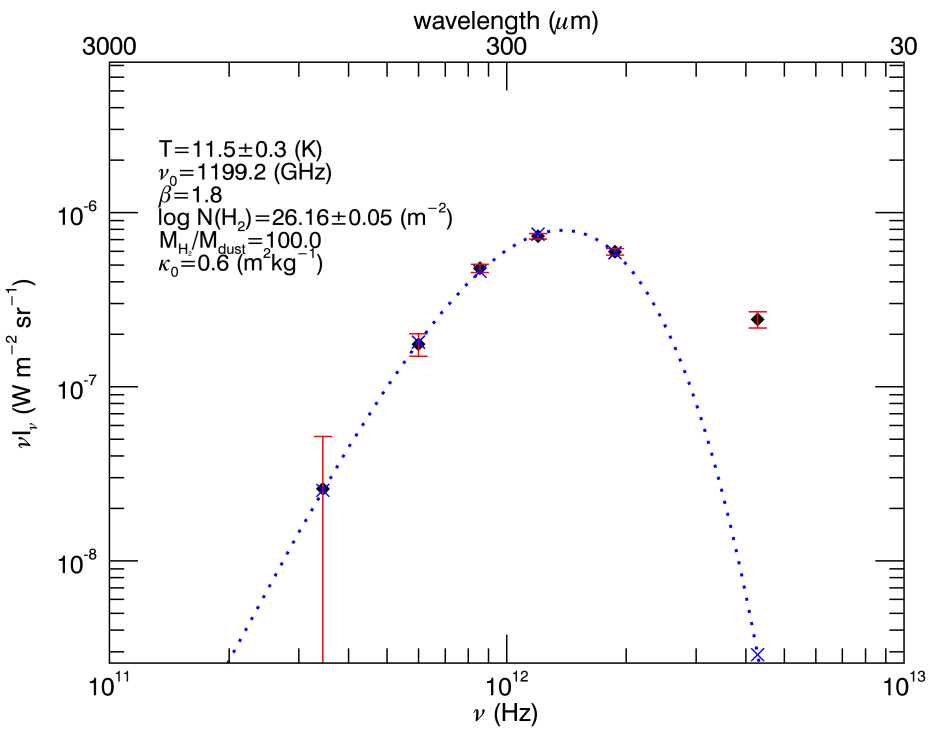}}
    \caption{Sample SED fits for an individual pixel in BYF 77c (top), 99a (middle) and 202a (bottom). A two-component fit was used for Region 9, so most clumps in Region 9 (BYF 63-79) have this form of SED. Error bars are random errors computed by MPFIT using naive uncertainty maps in intensity (more details in Appendix~\ref{sec:apu}).}
    \label{fig:seds}
\end{figure}

We designed our SED-fitting routine around \texttt{MOSAIC-Math} \citep{mosmath}, an extension of the IDL-based \texttt{MOSAIC} software package released by \citet{mosaic}, as a more generalized alternative to the native SED-fitting routine optimized for internally-heated sources imaged between 1 and 20 \micron. Following the prescription of \citet{hild}, our routine uses \texttt{MPFIT} \citep{mrkwdt} to fit a single-temperature modified black-body of the form
\begin{equation}
    I_{\nu}\approx B_{\nu}(T_{\text{d}})[1-e^{-\tau}]
	\label{eq:mbb}
\end{equation}
 where 
\begin{equation}
    \tau=\left(\frac{\nu}{\nu_0}\right)^{\beta}\frac{N_{\text{H}_2}\mu m_{\text{H}}\kappa_0}{\gamma},
	\label{eq:tau}
\end{equation}
$B_{\nu}(T_{\text{d}})$ is the Planck function at dust temperature $T_{\text{d}}$, $\beta$ is the dust emissivity index, $\nu_0$ is the frequency at which the dust opacity coefficient $\kappa_0$ (also called the mass absorption coefficient) is measured, \ncol\,is the fitted column density of (mostly molecular) hydrogen, $\mu$ is the mean molecular weight per hydrogen molecule \citep[$\sim$2.8, see][Appendix A for a detailed derivation]{kauff}, $m_\text{H}$ is the mass of a hydrogen atom, and $\gamma$ is used to symbolize the gas-to-dust mass ratio. Only the dust temperature and hydrogen column density were allowed to vary for this study. The code optimizes log$_{10}$[\ncol] because \texttt{MPFIT} varied \ncol\,too slowly to find the optimal value within the maximum allowed number of iterations. Our code offers several options for additional temperature components, but for this paper we only used a second component in emission with variable $T_{\text{d}}$ and \ncol. As $\beta$ is often noted to be degenerate with $T_{\text{d}}$, so were $\kappa_0$ and $\gamma$ found to be degenerate with the \ncol. Assuming $\gamma$=100, taking $\nu_0$=250\micron, and using the typical dust opacity value in the Galactic Plane at that wavelength from \citet{planck21}, $\tau/N_{\text{H}}$=9.2$\pm$0.5$\cdot$10$^{-30}$ m$^2$, we solved 
\begin{equation}
    \left(\frac{\tau}{N_{\text{H}}}\right)=\left(\frac{\nu}{\nu_0}\right)^{\beta}\frac{m_{\text{H}}\kappa_0}{\gamma},
	\label{eq:tau2}
\end{equation}
for $\kappa_0$ \citep{bianchi13}. We used the result, $\kappa_0$=0.6$\pm$0.4 m$^2$g$^{-1}$, as a constant of the SED fit. Nearly all of the formal uncertainty in $\kappa_0$ is propagated from the the factor of $\sim$2 uncertainty in the gas-to-dust mass ratio. Uncertainties on reported $T_{\text{d}}$ and \ncol\,values are computed by \texttt{MPFIT} based on input of naive uncertainty maps included as extensions of the science images\protect\footnote{\texttt{Miriad} is incompatible with multi-extension FITS files and \texttt{Mosaic-Math} offers partial support for them, so the uncertainty maps were extracted and processed separately. Python code that makes single-extension working copies of fits images from a variety of mid- and far-IR missions converted to the user's choice of intensity unit, among other functions, is available as-is on \href{https://github.com/rlpitts/Hello-World}GitHub.}. Sample SED fits are shown in Figure~\ref{fig:seds}.\\
\indent The major uncertainties for dust-derived gas masses are the gas-to-dust ratio, the size distribution of the grains, and their microscopic properties, and we can only currently quantify the first. Given the approximate factor of 2 in uncertainty, and the apparent tendency of the gas-to-dust ratio to decline toward the centers of cold clumps, using a ratio of 100 seemed more appropriate than the more precise value of 124 from \citet{li2001} that contemporary SED-fitting studies often assume. Since continuum emission data are absent between about 30 and 60 \micron, the wavelength range where the effects of grain size and composition could potentially be distinguished from the effects of a changing gas-to-dust ratio (we refer the reader to the demonstrative models of \citealt{galliano} and \citealt{dustem}), these factors will have to be addressed in future works.\\
\indent Our SED fitting pipeline incorporates a color-correction routine that fits an SED to uncorrected data, uses the resulting fit parameters to interpolate on published grids of $T_{d}$ and dust $\beta$ \citep{pacsman,spireman,wisesup}, divides the intensities at each wavelength by the appropriate color-correction factor, and then repeats the SED-fitting with the corrected intensities. If using a two-temperature SED model, the routine automatically calculates the transition point between the two components to determine which temperature to use in interpolation. It also estimates an uncertainty to propagate through the color-corrected intensity by computing color correction factors for temperatures and $\beta$ one $\sigma_T$ above and below the fitted parameters and taking the largest deviations. All data featured on Regions 9, 10, and 11 are derived from color-corrected fits. We also examined the color-corrected fits to Region 26, but found that where the fits were successful, the fitted $T_{\text{d}}$ and \ncol\,changed by less than 0.1 K and 0.01 dex, respectively, both less than the margins of error throughout the image. Since the color-corrected SED-fitting also frequently failed perilously close to important areas, we decided to continue analysis on the uncorrected data.\\
\indent The form of equation~(\ref{eq:mbb}) assumes foreground and background contributions are negligible compared to the source of interest, but allows for self-absorption. The line-of-sight to the Carina-Sagittarius Arm, in which all CHaMP sources located, does not pass through any other arms, so we felt justified assuming any foreground contamination would be roughly constant across the field. We neglected zodiacal light because the survey regions are all more than 35$^{\circ}$ from the ecliptic plane \citep[see dust distributions by inclination in][]{hahnzodi}, and because, based on typical SED fits and the raw flux density maps, zodiacal light accounts for a few percent of the flux density at 70 \micron, and much less at all other wavelengths \citep{leinert,hsomanual}. There is severe background contamination in Region 26, however, and Regions 9 through 11 lie near the tangent point where local diffuse emission is expected to be relatively thick, so we tried several methods of background subtraction for each.\\
\indent The first method we tried was to make a look-up table of patches of sky in each region devoid of obvious clumps or filamentary emission sources, compute the average flux density at each wavelength weighted by the inverse of the uncertainty at each pixel, subtract the averages (propagating the uncertainties), and then fit SEDs to the difference. That approach sufficed for Regions 9, 10, and 11. The resulting column density and temperature parameter maps also showed that \citet{schneider15}'s approach, calculating the average H$_2$ column density of the background patch and subtracting that from the rest of the column density map, is not equivalent. Compared to \ncol\,calculated from SED-fitting after background subtraction, subtracting a background \ncol\,after SED-fitting leads to values of \ncol\,ranging from a factor of a few lower in the centers of clumps and a factor of a few higher in the most diffuse regions (see Appendix~\ref{sec:apbg}). The two approaches tend to agree only in narrow strips far from the sources of interest. Additionally, when the background is subtracted before SED-fitting, the fitted values of $T_{\text{d}}$ tend to increase slightly.\\
\indent In Region 26, a plurality of the clumps sit in a local minimum in the background structure, evident from the gaps between filaments, so subtracting a flat background also erased most of the CHaMP clumps. Using as a mask the brightest 90\% of the pixels in the CO maps and manually adding polygons around attached filaments that the Mopra maps cut off, we constructed a cube of background emission maps at each wavelength by blocking out the CHaMP clumps and bi-linearly interpolating over the masked areas. We subtracted this background image cube from the original image cube and fit SEDs pixel-by-pixel to the difference. Because the differences are so close to zero along the edges of the on-source pixels, the SED fits can become highly uncertain with unstable temperatures. Therefore, throughout most of our analyses, we began by masking pixels where the uncertainty returned by \texttt{MPFIT} was greater than a factor of two (0.3 in the log) for H$_2$ column density, and greater than 5 K in dust temperature. These limits turned out to be too liberal as uncertainties are propagated from naive error maps that themselves tend to understate the total uncertainties, which left many clumps encircled by rings of unrealistically high CO abundances (see \ref{ssec:cocomp}). Realizing that a 15\% uncertainty in the SPIRE data should translate to about the same uncertainty in \ncol\,in an optically thin medium, we also masked data where the background-subtracted \ncol\,was less than 15\% of the fitted \ncol\,before background subtraction. Parameter maps for Region 26 where SED-fitting was performed without background subtraction are available in Appendix~\ref{sec:apu}.\\
\indent We found all regions to be optically thin everywhere in all passbands except 70~\micron, in which the highest optical depth observed was about 0.4. Since allowing for optical depths approaching 1 did not noticeably affect computing time, we saw no compelling reason to use the optically thin approximation for either component.

\section{Parameter Maps}
\label{sec:pms}
\begin{figure*}
  \includegraphics[width=\textwidth]{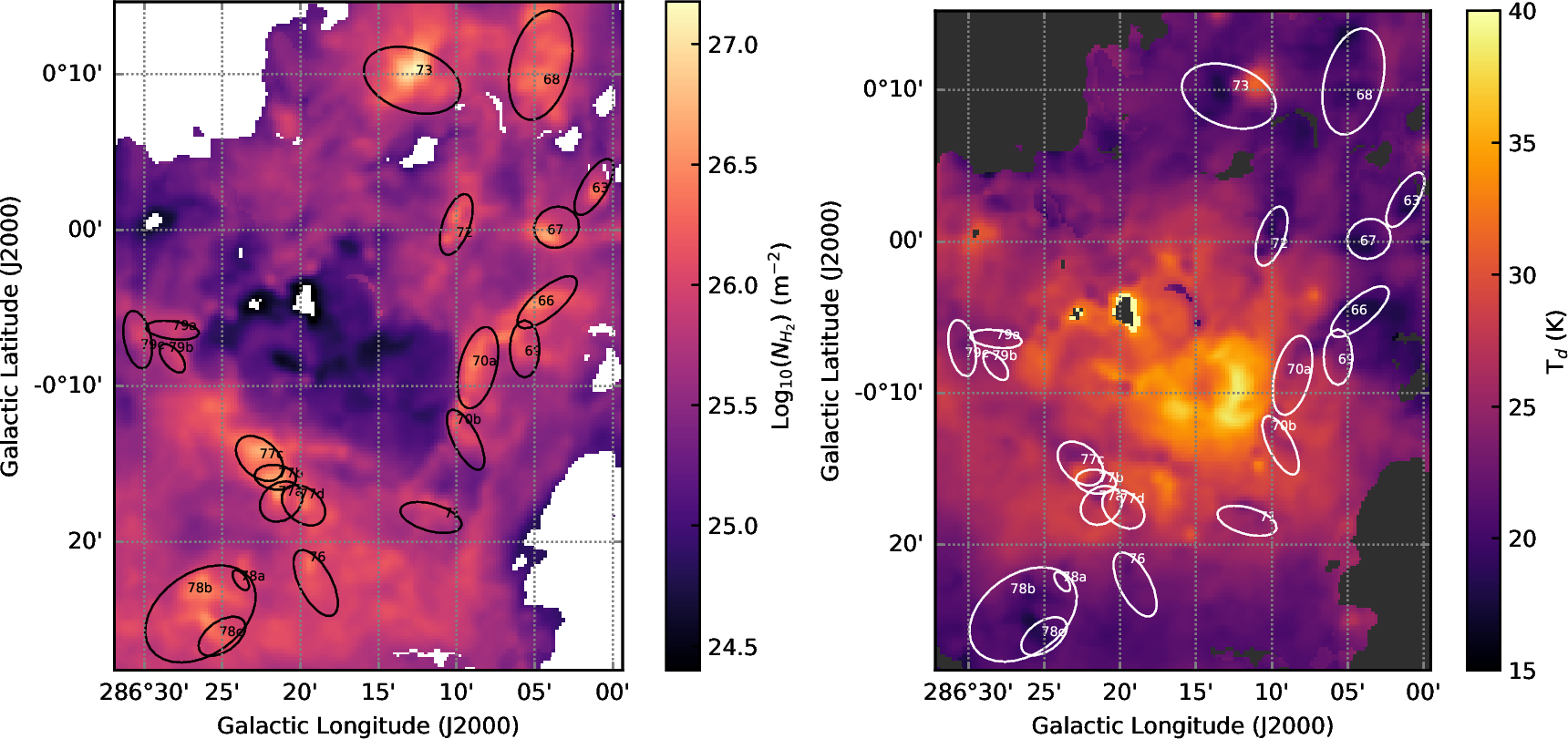}
  \vspace{-4mm}
    \caption{Parameter maps for Region 9 (Gum 31). \textbf{Left:} log H$_2$ column density (log m$^{-2}$) derived from dust emission.\textbf{ Right:} dust temperature (K).}
  \label{fig:pmaps9}
\end{figure*}
\begin{figure}\hspace{-2mm}
	\includegraphics[width=\columnwidth]{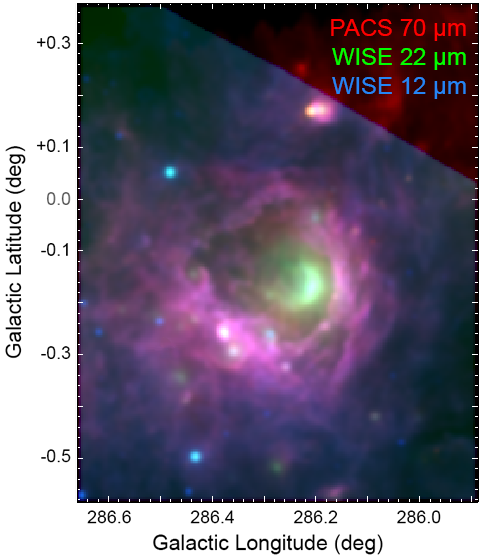}
	\vspace{-1mm}
    \caption{Three-color image of Region 9 in WISE bands 3 (12 \micron, blue) and 4 (22 \micron, green), and PACS-B (70 \micron, red) to show the bright arc of dust around NGC 3324, likely an amorphous silicate feature.}
    \vspace{-1mm}
    \label{fig:wise9}
\end{figure}

\indent Figure~\ref{fig:pmaps9} shows parameter maps for Region 9: H$_2$ column density on the left, and dust temperature on the right. The ionization front of Gum 31 is difficult to see in the temperature map, but is apparent as a series of clumpy arcs around a cavity in the H$_2$ column density map. As expected, the clumps embedded in the compressed gas layer just outside of the ionization front are warmer on average than clumps like BYF 68 and 78a-c still far from the PDR. The global elevation of temperature due to the expanding PDR extends beyond the shell of compressed gas visible in the column density map, but stops just shy of BYF 66 and 76, leaving BYF 63, 68, and 78 untouched. BYF 63 and 66 do not show any evidence of heating from NGC 3293 to their west. The warmest clumps on average, BYF 77b-c, are associated with the UCHII IRAS 10361-5830 \citep{vazz}, seen as a couple cool spots bracketing a local hot ($\sim$30~K) spot along the (galactic) southeast edge of ionization front in the dust temperature map, and as a bright orange spot the lower left of Figure~\ref{fig:wise9}. BYF 77c and b appear to be smaller analogs of BYF 73, while BYF 77a hugs the northeast side of a fainter but much bluer source visible in Figure~\ref{fig:wise9} just southwest of IRAS 10361-5830. BYF 73 varies in dust temperature from 17.8$\pm$0.2~K to $\sim$31~K because it is wrapped around the east side of a compact (0.9 pc) HII region \citep{byf73}. This 2-3$\times$10$^{4}$ $L_{\odot}$ clump is noted for potentially having the highest mass infall rate ever observed, $\sim$3.4$\times$10$^{-2}$ $M_{\odot}$ $yr^{-1}$ \citep{byf73}. We derive a peak H$_2$ column density of 1.5$\pm$0.1$\times$10$^{27}$ m$^{-2}$ for BYF 73, within a margin of error of the 1.4$\times$10$^{27}$ m$^{-2}$ that \citet{ogum31} derived by fitting \textit{Herschel} data with the models of \citet{robit}.\\
\indent On dust temperature and H$_2$ column density maps, our derivations more or less reproduce the results of \citet{ogum31} at Mopra's resolution. The range of values and the peak values in the $H_2$ column density map for Region 9 are also very similar to those of the Galactic Plane HII regions RCW 120 \citep{rcw120} in the tail of Scorpius, and RCW 79 \citep{rcw79} near the southern edge of Centaurus. For all three HII regions, \ncold\,ranges from about 10$^{25}$ to slightly over 10$^{27}$ \psqm, with the highest values just outside the warm PDR boundary where the gas is compressed by the expanding bubble of ionizing radiation. Gum 31 has a similar temperature range and distribution to RCW 79; both are about 10 K warmer along the PDR boundary than RCW 120. Peaks in the map of H$_2$ column density in Region 9 are associated with local reductions in temperature, so the two are somewhat anti-correlated.
\indent The PDR boundary is outshone in the dust temperature map by a crescent of sparse but warm dust near the middle of the PDR. The temperature map in \citet{ogum31} reassuringly shows the same feature at the same temperature. Indeed, Figure~\ref{fig:wise9} shows an arc of emission brightest at about 22 \micron, which does not correspond to any known PAH features, but is similar to the peak wavelength of the dust SED of the Orion "Bright Bar" \citep{cesar}, attributed to hot ($\sim$130 K) amorphous silicates. The same arc is still visible, albeit faintly, at 70\micron~(blue) in Figure~\ref{fig:gum31}. Our program cannot yet fit stochastic dust emission or PAH emission, so column density and temperature data in this feature are not well-defined. However, for the scope of this paper, we only needed the second component to isolate thermal dust emission from other types of emission at shorter wavelengths. Fuller functionality to separate different thermal components will be included in a future study.\\
\begin{figure*}
  \includegraphics[width=\textwidth]{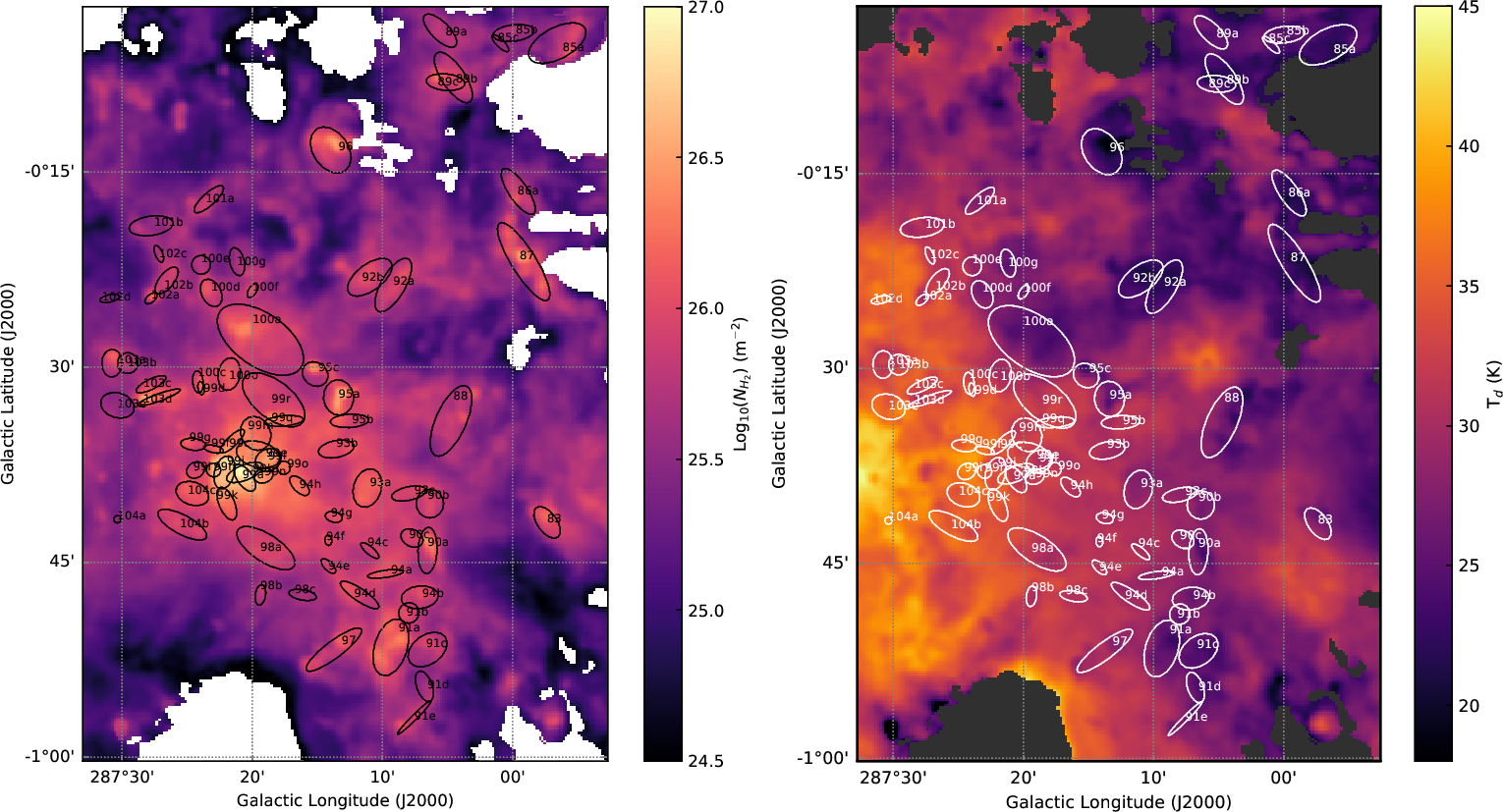}
  \vspace{-4mm}
    \caption{Parameter maps for Region 10 (CNC between Gum 31 and Trumpler 16). \textbf{Left:} log H$_2$ column density (log m$^{-2}$) derived from dust emission.\textbf{ Right:} dust temperature (K).}
  \label{fig:pmaps10}
\end{figure*}
\indent In Region 10, a spatially large over-density about 15$'$ due galactic west of $\eta$~Car, identified as the BYF 99 group, dominates the H$_2$ column density map, shown in Figure~\ref{fig:pmaps10}. The high-column clump in this group has a peak H$_2$ column density of about 1.02$\pm$0.05$\times$10$^{27}$ \psqm, comparable to BYF 73 in Region 9, and not unexpected for the PDR boundary of Tr14 \citep{brooks03}. Most of the other clumps in the region have peak column densities about half a decade lower, comparable in density to most of the clumps along the PDR boundary of Gum 31. BYF 99f-l, 102d, 103c, and 104a-c are some of the warmest clumps in any of the regions studied, all between 33 and 40 K, and appear to lie inside a bubble of 40-45 K diffuse gas about 15$'$ in radius ($\eta$ Car is about 3$'$ off the left edge of the frame in Figure~\ref{fig:pmaps10}, due east of the BYF 99 group). Between them and the cooler (20-25K) chain of clumps BYF 88, 90a-c, and 91a-e, many of the clumps are nearly indistinguishable from the background in the temperature map. Indeed, BYF 88, 90a and b, and 91a-c all have $v_{\text{LSR}}$ between about $-$18 and $-$20~km~s$^{-1}$, so they are likely part of the same filament. The temperature map makes it look like the chain continues up to BYF 86a and 87, and BYF 86a is within the same $v_{\text{LSR}}$ range as the others. However, the morphology of the column density map suggests BYF 86 and 87 are more closely connected to BYF 92a and b, which also lie in the same $v_{\text{LSR}}$ range. Despite harboring more stellar clusters than Region 11, or perhaps because there are so many blowing stellar winds in different directions, Region 10 has only two cometary or pillar-like formations containing only three clumps between them: BYF 102a and b, and 103d. The evacuated area north and west of the BYF 100-102 groups, and those clumps' low column densities, independently suggest this part of the CNC may be more eroded by past star formation. Indeed these clumps surround Tr15, which, at an estimated age of 5-10 Myr, is substantially older than Tr14 or Tr16, and may have given rise to one or more supernovae \citep{wang}.\\ 
\begin{figure*}
  \includegraphics[width=\textwidth]{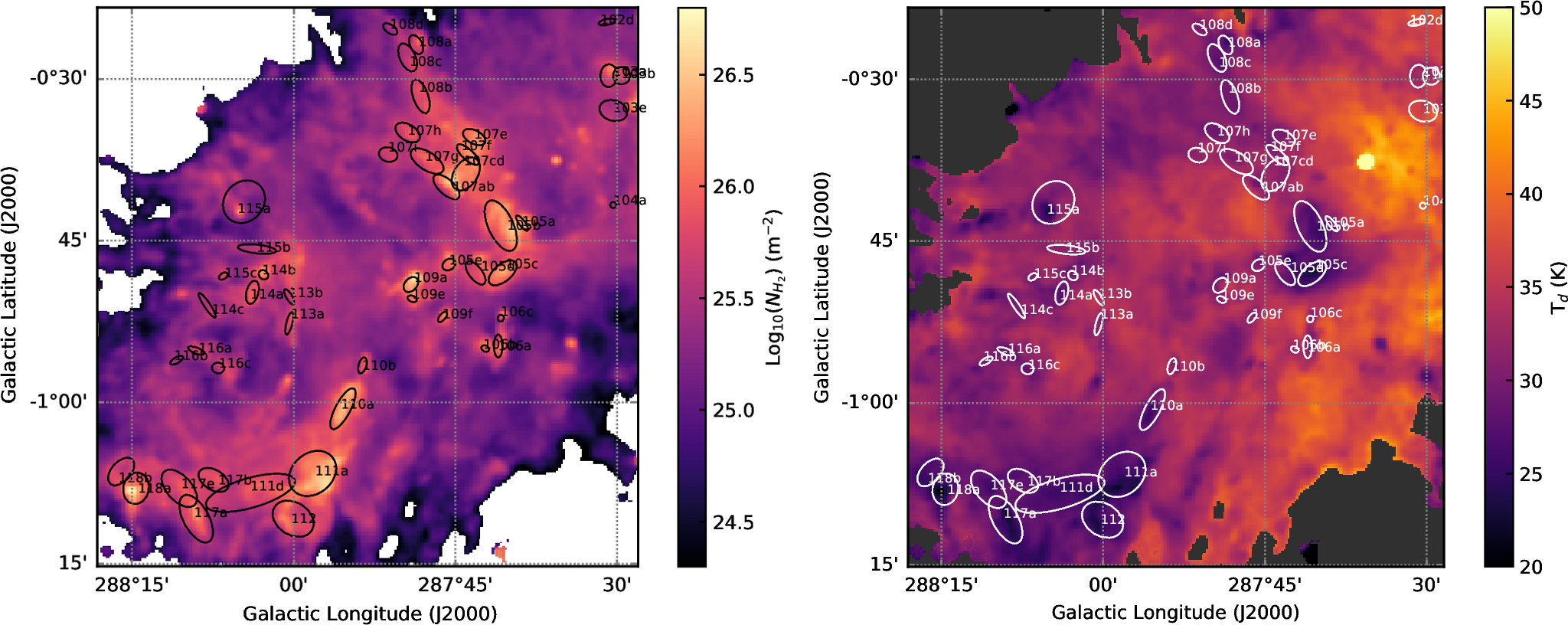}
  \vspace{-4mm}
    \caption{Parameter maps for Region 11 (Carina Nebula complex east of Trumpler 16). \textbf{Left:} log molecular hydrogen column density (log m$^{-2}$) derived from dust emission.\textbf{ Right:} dust temperature (K).}
  \label{fig:pmaps11}
\end{figure*}
\indent Region 11, shown in Figure~\ref{fig:pmaps11}, shows a number of pronounced pillars with cool centers and high-column-density tips. The South Pillar---containing BYF 110a and b, 111a and d, 112, 117a-e, and 118a and b---is the most prominent, but has several companions, including one that was not mapped in CO despite having a higher column density tip than some of the clumps in the BYF 108 group. The range of temperatures across the clumps and diffuse gas is smaller than for Region 10: no clump is warmer than 35 K or cooler than 20 K. The range of column densities is similar to Region 10, but the highest-column clump in Region 11, BYF 118a at the base of the South Pillar, has a peak column density of 6.3$\pm$0.3$\times$10$^{26}$ \psqm, about half that of BYF 99 at its peak. The Southern Cloud, containing BYF 105a-e, 107a-i, and 108a-d, is strikingly different in morphology from the South Pillar, lacking the highly directional ``fingers'' pointing toward a source of ionizing radiation. However, since both clouds have similar median and average temperatures ($\sim$30 K) and H$_2$ column densities ($\sim$4$\times$10$^{26}$ \psqm), but different $v_{\text{LSR}}$, the Southern Cloud's apparent lack of pillar-like features is probably a geometric effect. The Treasure Chest, BYF 109a, has the second-highest H$_2$ column density in the region, peaking at 5.8$\pm$0.3$\times$10$^{26}$ \psqm\,in the (galactic) northernmost end of the C-shaped over-density surrounding the embedded cluster and falling by about a factor of 5 toward the cluster's center. \citet{smith05} give the Treasure Chest's highest observed $A_v$ as 50, corresponding to a H$_2$ column density of 5.5$\times$10$^{26}$ \psqm, so our results agree to within one standard deviation. BYF 109a's temperature rises from a minimum of 28.7$\pm$0.3~K in the C-shaped envelope to 35.6$\pm$0.4~K at the location of the embedded cluster. Its $T_{\text{d}}$, \ncol, and morphology are notably similar to BYF 77a-c (IRAS 10361-5830) in Region 9 despite containing many times more stars and protostars. 
BYF 115a and nearby clumps with little CO data are also associated with a handful of young stellar objects (YSOs) \citep{povich,gacz,vista2}, but their small sizes and evacuated environment may be due in part to the A0Iab supergiant HD 93737 slightly over 5 arcmin away away. \citet{humph72} puts the distance to HD 93737 at 2.9 kpc, but that is within the margin of error for distance measures to the CNC. In any case, the direction that the pillar containing BYF 115a is pointing suggests Tr16 is not the only source of the stellar winds that sculpted it, and the only other nearby cluster containing OB stars is Bo11 \citep{smith06}. Bo11 is the most obvious potential driver of the morphological evolution of BYF 113-116, as all but 115a are found in one of two arcs around an apparent cavity in the \ncol\,map where Bo11 sits. Of BYF 113-116, only 115a is noteworthy for its low $T_{\text{d}}$ (22.5$\pm$0.7~K at the \ncol\,peak) considering its environment, and none have particularly high column densities at the pixel resolution of our images, though the handful of YSOs in BYF 115a could individually have higher H$_2$ column densities in their disks.\\
\indent Compared to the results of \citet{preib}, we find temperatures across our parameter maps up to 5 K higher across both Regions 10 and 11, with lower discrepancies in the coolest clumps. Our map of \ncol\,agrees with theirs to within about 30\%, with a tendency to be slightly lower where discrepant. The temperature discrepancies may be due to their use of a different dust $\beta$, because color corrections were applied later in \citet{rocca}, and/or because the areas we used to determine a background level were closer to the sources and therefore our background subtraction was more aggressive. Compared to \citet{rocca}, our temperature maps of Regions 10 and 11 are in better agreement, but our column densities are larger by about a factor of six. \citet{rocca} were even more conservative about background subtraction than \citet{preib}, so it's rather surprising that their H$_2$ column densities should be so much lower than not only ours, but their group's previous work. They tried other background subtraction methods besides the one that we used, but concluded all approaches produced similar results to within their adopted margin of uncertainty. The rest of our fitting and color-correction procedures were either the same or mathematically equivalent to those in \citet{rocca} and \citet{preib}.
\begin{figure*}
  \includegraphics[width=\textwidth]{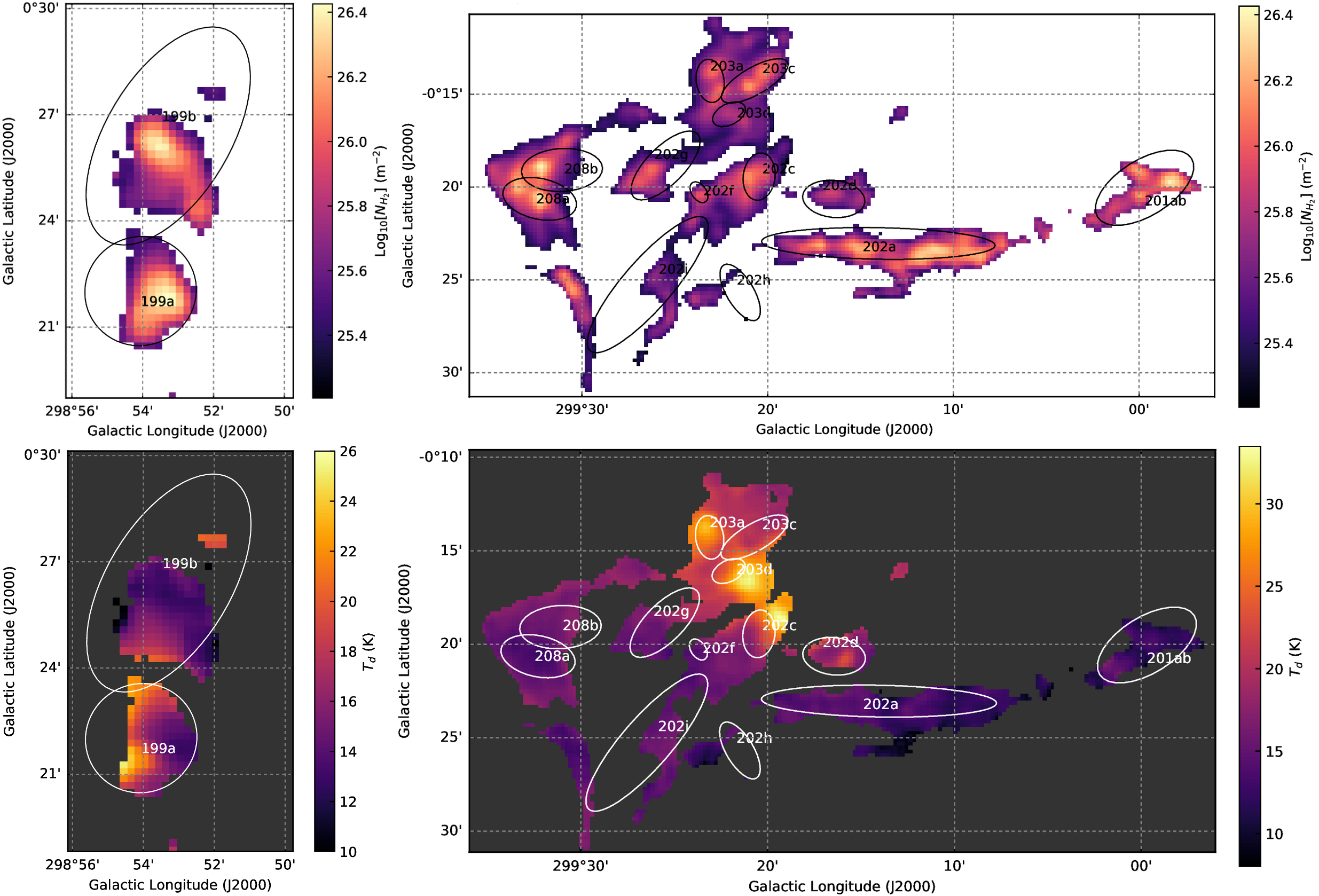}
  \vspace{-4mm}
    \caption{Parameter maps for Region 26a (BYF 199a and b, left) and 26b (BYF 201-208, right). \textbf{Top:} log molecular hydrogen column density (log m$^{-2}$) derived from dust emission. \textbf{Bottom} dust temperature (K)}
  \label{fig:pmaps26}
\end{figure*}
\indent Because of the large separation between BYF 199a-b and the rest of the clumps in Region 26, we divide it into Region 26A (BYF 199) and Region 26B (BYF 201-208). Figure~\ref{fig:pmaps26} shows the H$_2$ column density map (top) and dust temperature map (bottom) of Region 26A in the left-hand column and 26B in the right-hand column.\\
\indent In Region 26, as with Regions 9 through 11, most of the column density maxima correspond closely to temperature minima, except where the clumps border on one of the several reflection nebulae, infrared bubbles, or compact HII regions in the area. Most important are the three that appear in a line of bright spots between 25 and 30 K in the temperature map, Figure~\ref{fig:pmaps26}, associated with the brightest clumps in CO: BYF 203ab, 203d, and 202c. The brightest of these, in the center, is RCW 64, or Bran 386E. To the upper left is Bran 386G \citep[which may be the same as MWP1G299389-002316---][]{bubbly}, and to the lower right is Bran 386C. These clumps are warmer than most---around 26 K for BYF 203ab and 203d, and 22 K for 202c---but not exceptionally high or low in density. The rest of the clumps are substantially cooler, with average temperatures between 12 and 16 K, and appear as dark filaments in Figure~\ref{fig:pmaps26}.\\
\indent Typical column densities seen in Region 26 after background subtraction are nearly an order of magnitude smaller than in Region 9. Rather than Galactic Plane HII regions, clumps in Region 26 are more consistent with lower-mass star-forming complexes, like the Aquila Rift complex \citep{aquila} or the Taurus star-forming region \citep{kirk}, or filaments and IRDCs near high-mass star-forming regions \citep[see, e.g.][]{schneider13,fontani,planck22}. This is to be expected since most of the clumps in Region 26 lie in one of at least two IRDCs in 26B, one containing BYF 201 and most of the 202 group, and the other containing BYF 208a-b. The range of column densities and temperatures in Region 26 are also more consistent  with the Cold Clump catalog of Planck Objects \citep[C3PO][]{planck23} and its follow-up with \textit{Herschel} \citep{feher}, although it's worth noting also that the earlier C3PO results may have suffered from beam dilution.

\section{Discussion}\label{sec:disc}
\subsection{Comparison with CO}\label{ssec:cocomp}
\begin{figure*}
	\includegraphics[width=\textwidth]{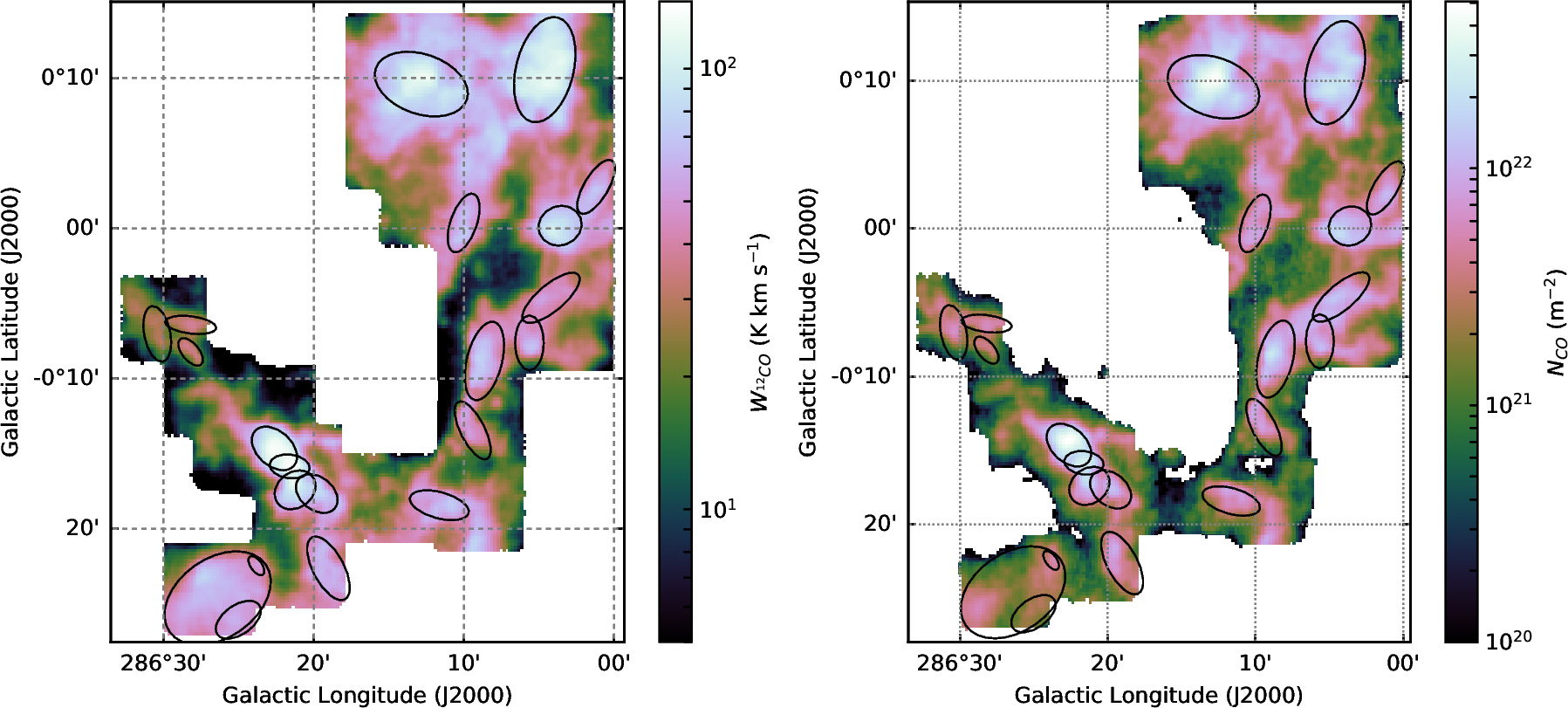} \\
	\includegraphics[width=0.9\textwidth]{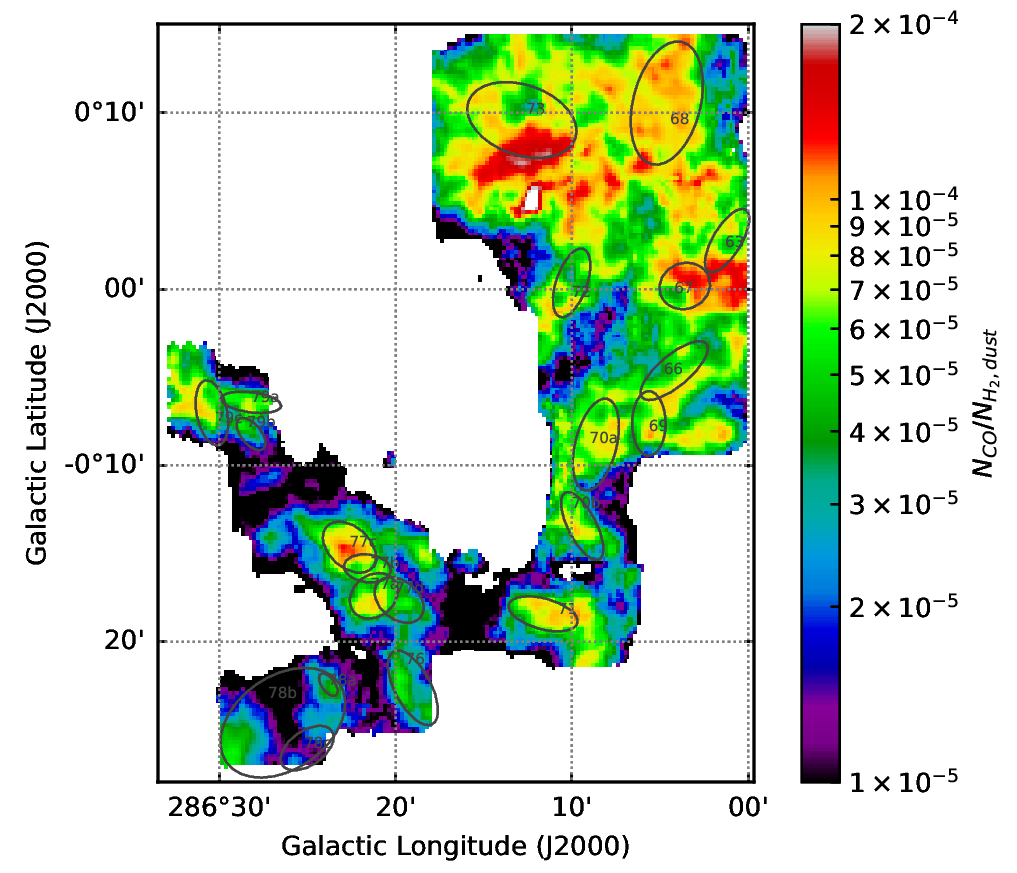}
    \caption{\textbf{Top left:} Map of integrated $^{12}$CO line intensity in \kkms~ over Region 9. \textbf{Top right:} Map of CO column density, calculated according to \citetalias{champ5}. \textbf{Bottom:} Ratio of $N_{\text{CO}}$ to \ncold.}
  \label{fig:r9co}
\end{figure*}
\begin{figure*}
	\includegraphics[width=\textwidth]{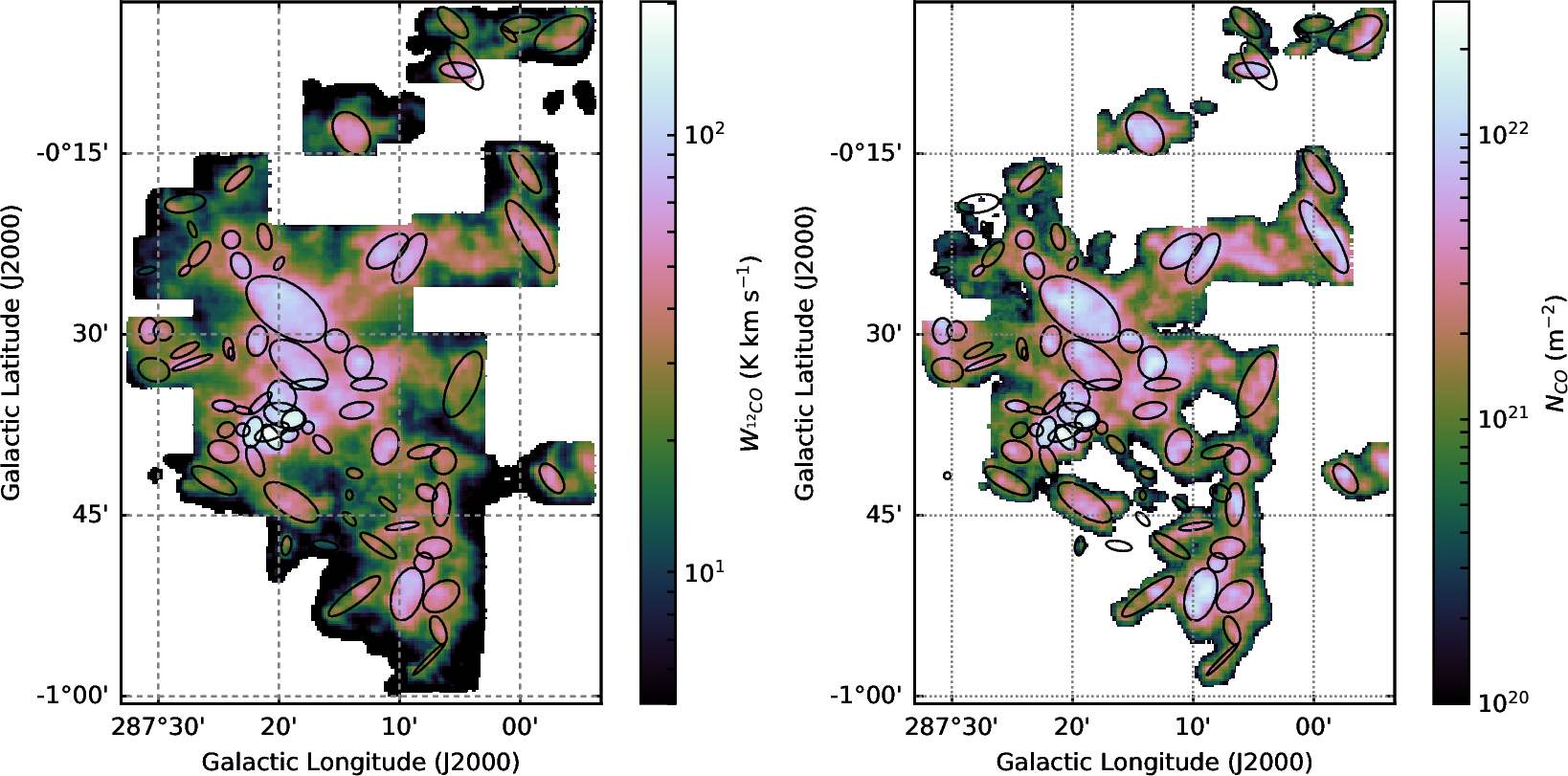} \\
	\includegraphics[width=0.8\textwidth]{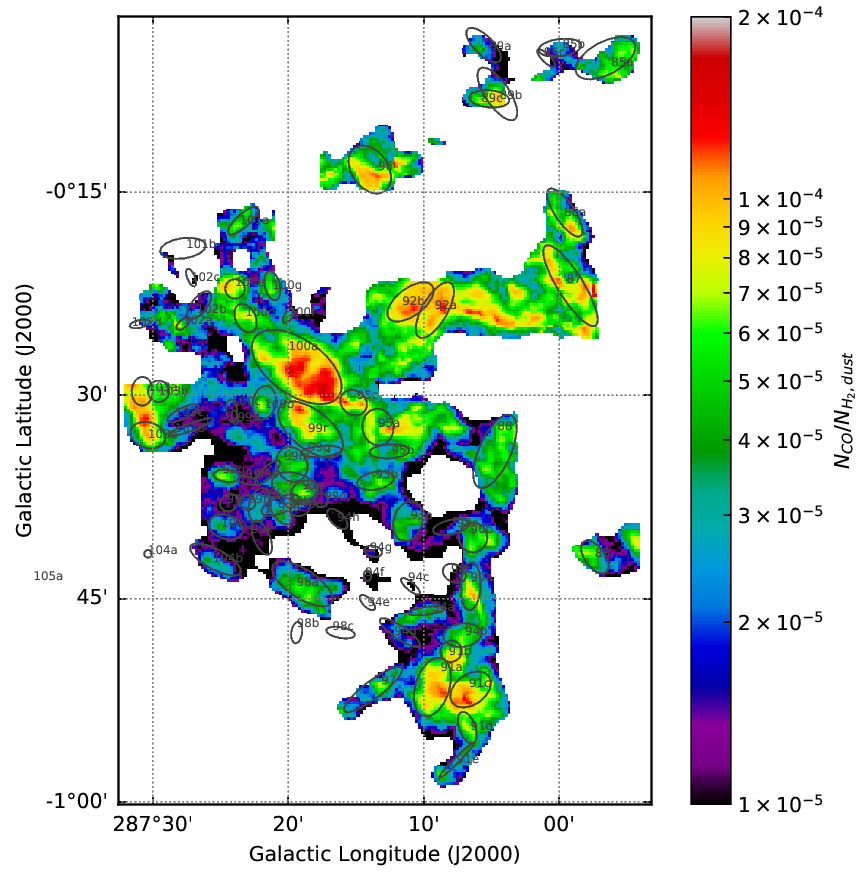}
    \caption{Same as Figure~\ref{fig:r9co} but for Region 10 and over a different velocity integral (refer to \citetalias{champ4}).}
  \label{fig:r10co}
\end{figure*}
\begin{figure*}
	\includegraphics[width=\textwidth]{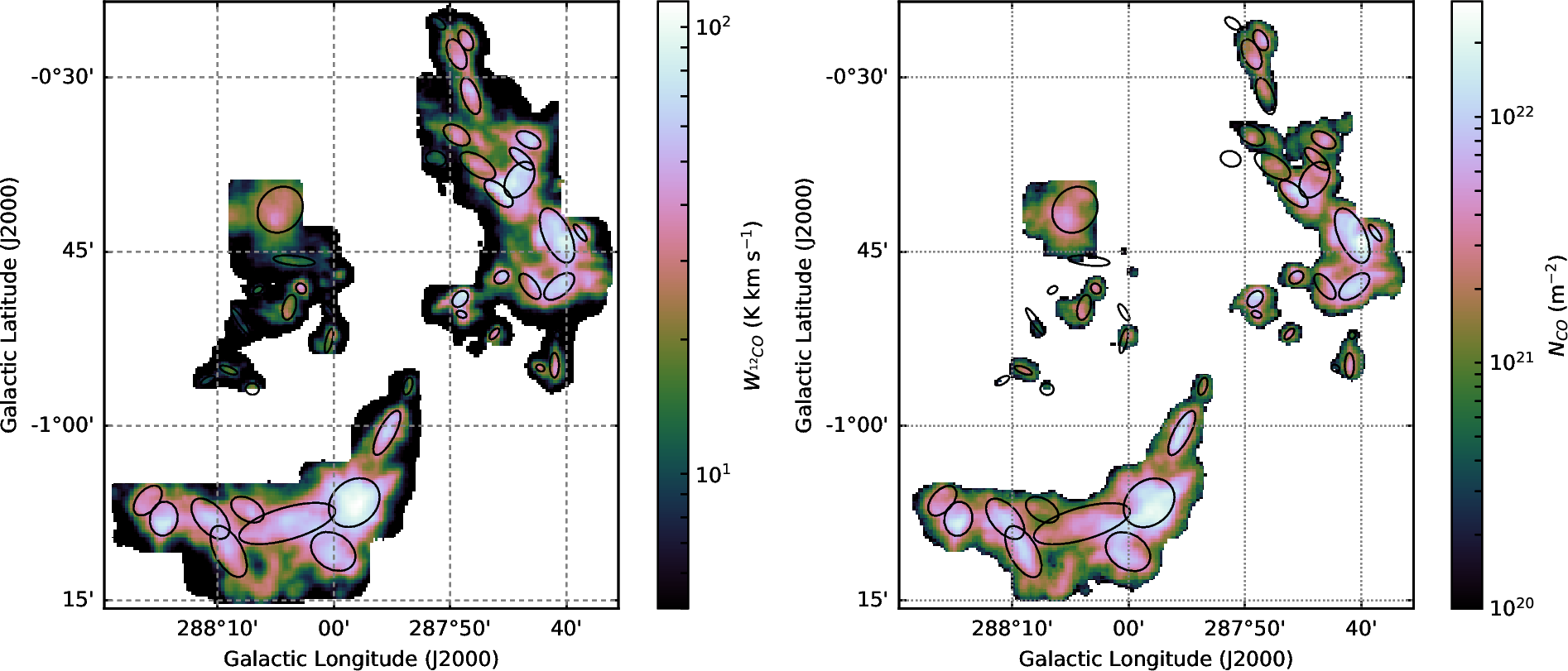} \\
	\includegraphics[width=\textwidth]{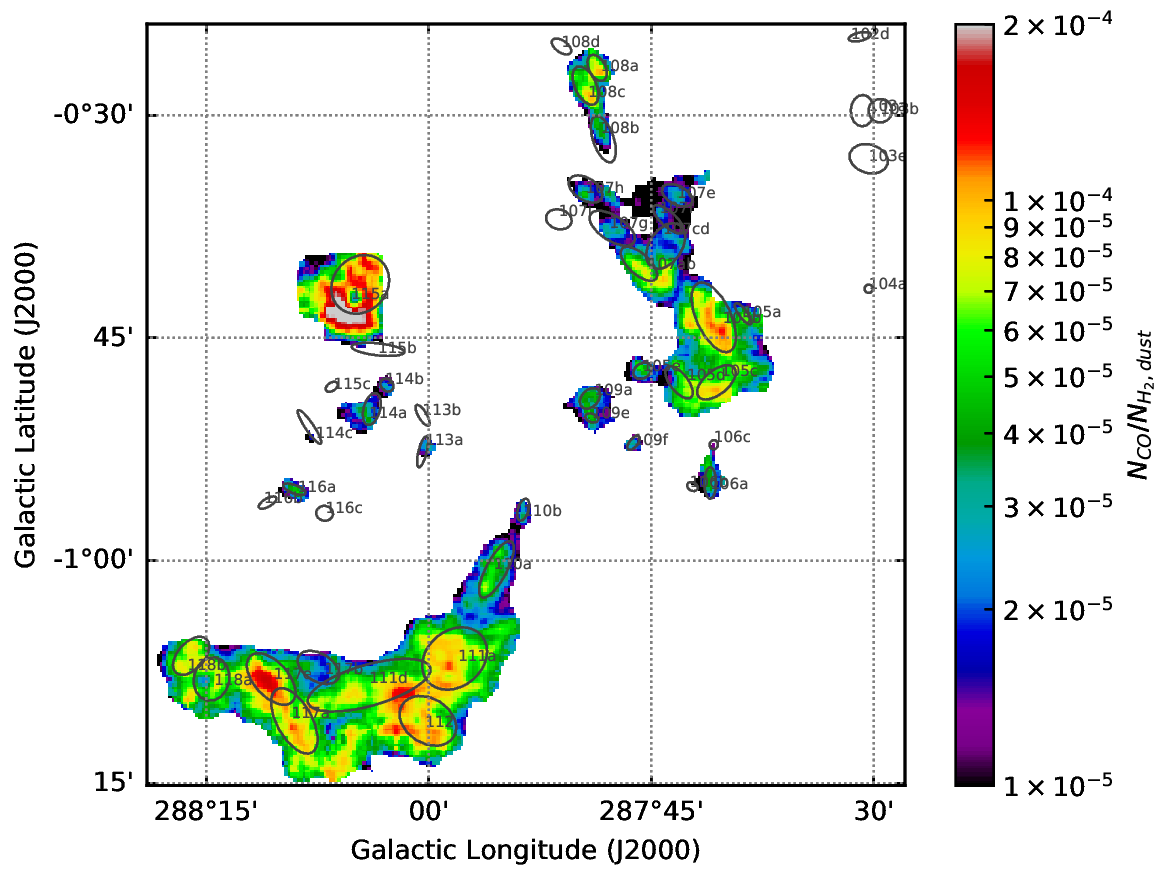}
    \caption{Same as Figure~\ref{fig:r9co} but for Region 11.}
  \label{fig:r11co}
\end{figure*}
\begin{figure*}
 	\includegraphics[width=\textwidth]{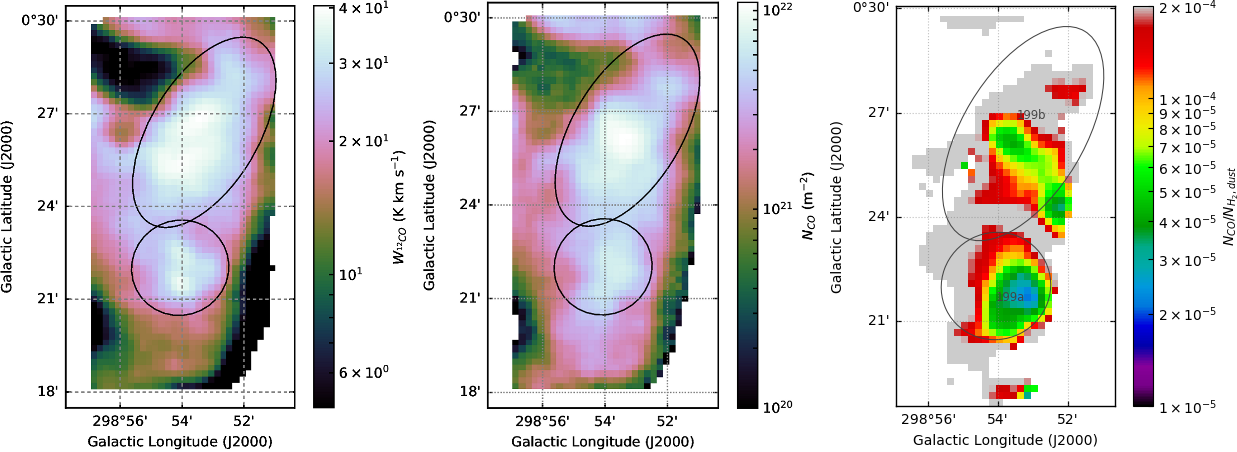} \vspace{-4mm}
    \caption{\textbf{Left:} Map of integrated $^{12}$CO line intensity in \kkms~ over Region 26a (BYF 199), where $v_{\text{LSR}}$ $\approx$ -28 km \psec. \textbf{Middle:} Map of CO column density, calculated according to \citetalias{champ5}. \textbf{Right:} Ratio of $N_{\text{CO}}$ to \ncold.}
  \label{fig:r26aco}
\end{figure*}
\begin{figure*}
	\includegraphics[width=.9\textwidth]{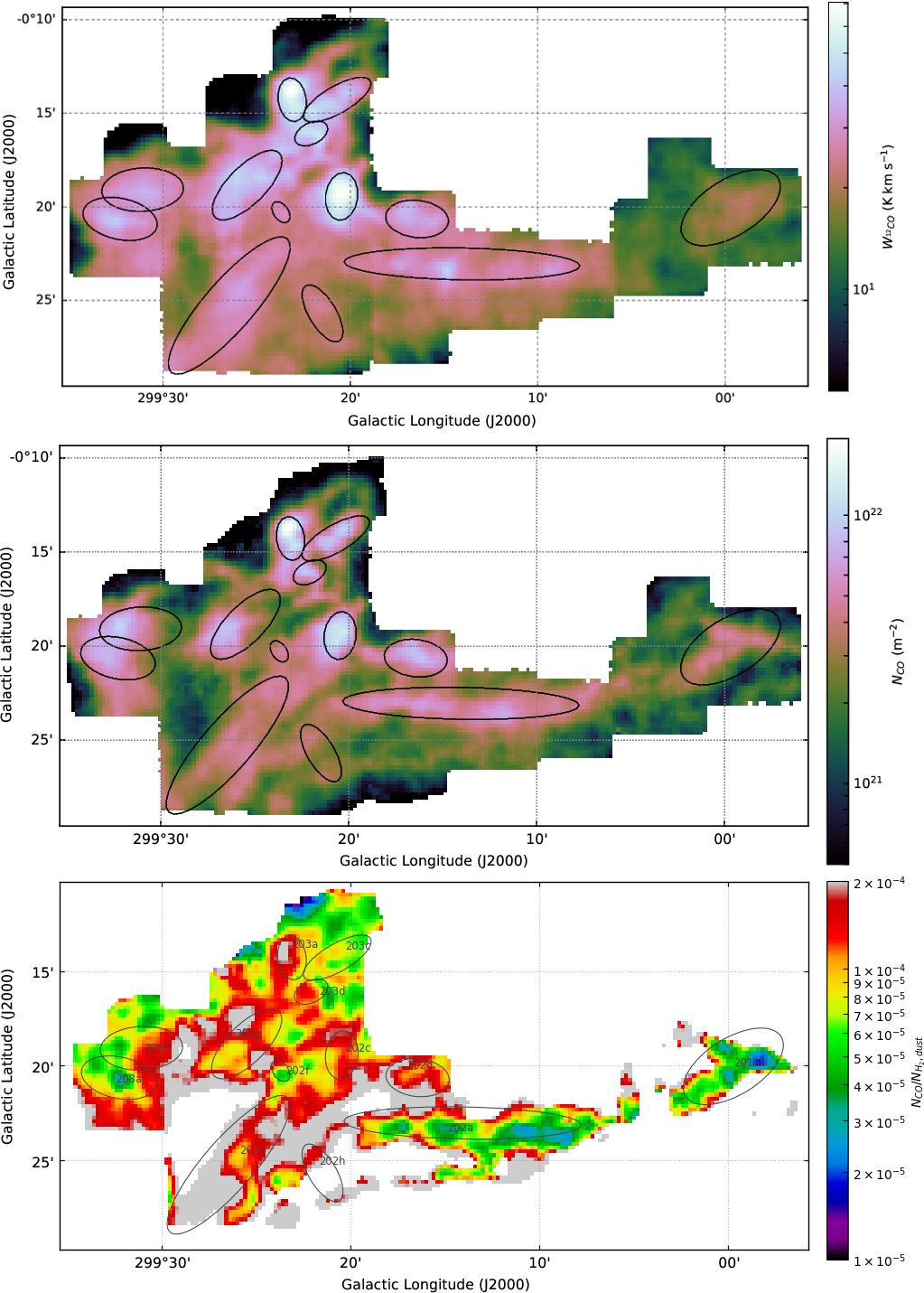}
    \caption{\textbf{Top:} Map of integrated $^{12}$CO line intensity in \kkms~ over Region 26b (BYF 201-208). This group of clumps has $v_{\text{LSR}}$ $\approx$ -38 km \psec. \textbf{Middle:} Map of CO column density, calculated according to \citetalias{champ5}. \textbf{Bottom:} Ratio of $N_{\text{CO}}$ to \ncold.}
  \label{fig:r26bco}
\end{figure*}
\indent The brightest clumps in the $W_{^{12}\text{CO}}$ maps (Figures~\ref{fig:r9co}, \ref{fig:r10co}, \ref{fig:r11co}, \ref{fig:r26aco}, and \ref{fig:r26bco}) are not necessarily the highest column density clumps in the \ncold~ maps. For instance, in Region 9, BYF 73 has the highest \ncold\,of any clump in all regions in this study, whereas BYF 68 appears large in size but otherwise unremarkable in either temperature or column density. In the $W_{^{12}\text{CO}}$ map, BYF 68 and BYF 73 are similar in size and peak intensity ($\sim$110 \kkms; see Papers III and IV for velocity integration ranges). Both are outshone in CO emission by BYF 77b, which peaks at $\sim$140 \kkms. In Region 10, the BYF 99 group is brightest clump in $^{12}$CO emission anywhere in any of the four regions, but its CO column density and its peak \ncold\,are still lower than in BYF 73. In Region 11, BYF 105b and 111a are most prominent in the $W_{^{12}\text{CO}}$ map. In the map of \ncold, however, BYF 118a has the highest \ncold, followed by BYF 109a and then 111a and 105a. BYF 115a, while fainter in $W_{^{12}\text{CO}}$ than 108a-c or 117a-c, has a higher \ncold\,than any of the BYF 108 and 117 groups, and appears in the CO abundance map as a sharply-defined hole of low abundance through a relatively high-CO medium. The $W_{^{12}\text{CO}}$ map in Region 26 is dominated by BYF203ab and BYF 202c, but while they are the hottest and third-hottest clumps in the dust temperature map, respectively, their peak densities are some of the lowest in the region, and their average densities are within the error margins of the background column densities in those areas.\\
\indent Most of the clumps in Regions 9 and 11 have a one-to-one correspondence between the $N_{\text{CO}}$ map and the \ncold\,map, where $N_{\text{CO}}$ exist, in terms of the number of peaks in the ellipse defined by Gaussian 3D decomposition of the CO emission. The only clump in the $N_{\text{CO}}$ map of Region 9 without a counterpart in \ncold\,is BYF78a; Regions 10 and 11 have a handful of CO clumps between them with no clear \ncold\,counterpart. However, the shapes and orientations of some clumps in both the parameter maps and far-infrared images bear little resemblance to their CO counterparts. In Region 9, BYF73 in particular seems to be X-shaped, and while the far-IR emission and \ncold\,maps seem to pick up the north-to-south axis more strongly, in CO it appears the emission is stronger along the east-northeast to west-southwest axis. BYF78c appears to be same length in dust emission as its CO counterpart, but rotated 90$^{\circ}$. Region 10 has BYF 91a, ovoid in CO emission/column density but shaped like a less-than sign in \ncold. In Region 11, many of the most prominent clumps have peak \ncold\,values offset by more than an arcminute from the centroids of their CO emission, but not in consistent directions that would suggest poor coordinate calibration. For BYF 105-109, the dust-derived column density peaks tend to be systematically slightly closer to $\eta$ Car and the Tr16 cluster than the CO emission centroids, which would make sense if dust were more robust to photodestruction than CO as expected. Further away, in BYF 111-118, such offsets are not seen, at least not consistently.\\
\indent It's harder to tell if the clumps in $N_{\text{CO}}$ and \ncold\,have a one-to-one correspondence in Region 10 because crowding is such a severe problem. BYF 85c, 98c, 102d, 103e, and 104a are isolated enough to see that they lack counterparts in the \ncold\,map. BYF 83, 85a-b, 86a, 87, 88, 89a-c, 92a-b, 95a and c, 96, and 100 all have easily identifiable counterparts. BYF 88 appears to have two peaks in the map of \ncold. Among the remaining clumps, there are at least as many if not more single \ncold\,peaks spanning multiple clumps in the $N_{\text{CO}}$ maps as there are clumps that resolve into multiple components in the \ncold\,map. We know from the original $W_{^{12}\text{CO}}$ maps that several clumps overlap in the line of sight, and most of these still have separate column density measurements in \citetalias{champ4}. However, these measurements have limited use for determining how much of the continuum emission is attributable to each clump, as the lower panels of Figures~\ref{fig:r9co} through \ref{fig:r26bco} show.\\
\indent All of the CHaMP clumps in the $N_{\text{CO}}$ map of Region 26 have local column density peaks in the same locations in the \ncold\,map, and many resolve into multiple clumps despite the parameter maps being convolved to the same resolution as the CO maps. Most of the clumps have several peaks in the ellipses defined by Gaussian 3D decomposition of the $N_{\text{CO}}$ map. BYF201ab has at least two, possibly three density peaks, and at the other end of the filament, BYF 202a has at least four maxima. BYF202c and BYF202f seem to be part of the same filament with three to four density peaks between them. BYF202d, 203a, 203c, and 208a each have at least two density peaks. BYF208b seems to have an extension to its immediate north that shows up in both the \ncold\,map and the $N_{\text{CO}}$ map but was not labelled separately or included as part of BYF208b.

\subsubsection{Temperature Dependence of CO Abundance}\label{sssec:morpho}
\begin{figure}
	\includegraphics[max width=\columnwidth, max height=8in]{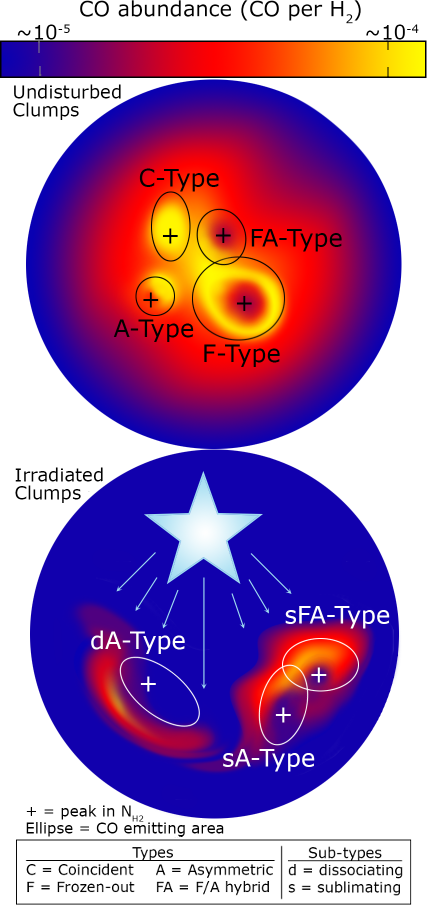} \vspace{-3mm}
    \caption{Artistic renditions of the different clump Morphotypes with a semi-realistic color scale for reference. No actual data were used in the production of this image, except as qualitative reference. Peculiar (P) type not shown. Made with Photoshop CS6.}
    \label{fig:ctypeart}
\end{figure}
\begin{figure}
    \hspace{-0.2cm}
	\includegraphics[width=1.02\columnwidth]{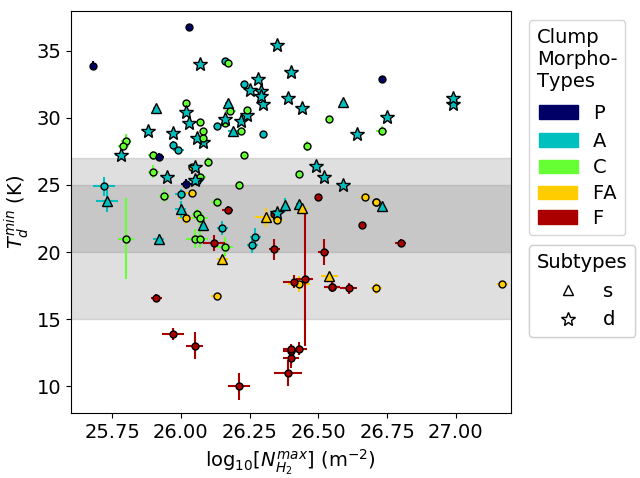}
    \caption{Clump-by-clump plot of minimum $T_{\text{d}}$ versus maximum log$_{10}$[\ncold], color-coded by clump Type and shape-coded by sub-type. Blue points are F-Type, magenta points are FA-Type, green points are Type-C, yellow points are Type-P. Circular points have no associated sub-type, star-shaped markers indicate the ``dissociating'' sub-type, and upward-pointing triangles indicate the ``sublimating'' sub-type. Shaded regions highlight the expected (light) and observed (dark) temperature range where the gas-phase CO abundance is maximized depending on local conditions.}
    \label{fig:alltypes}
\end{figure}
\begin{figure*}
	\includegraphics[max width=\textwidth, max height=8in]{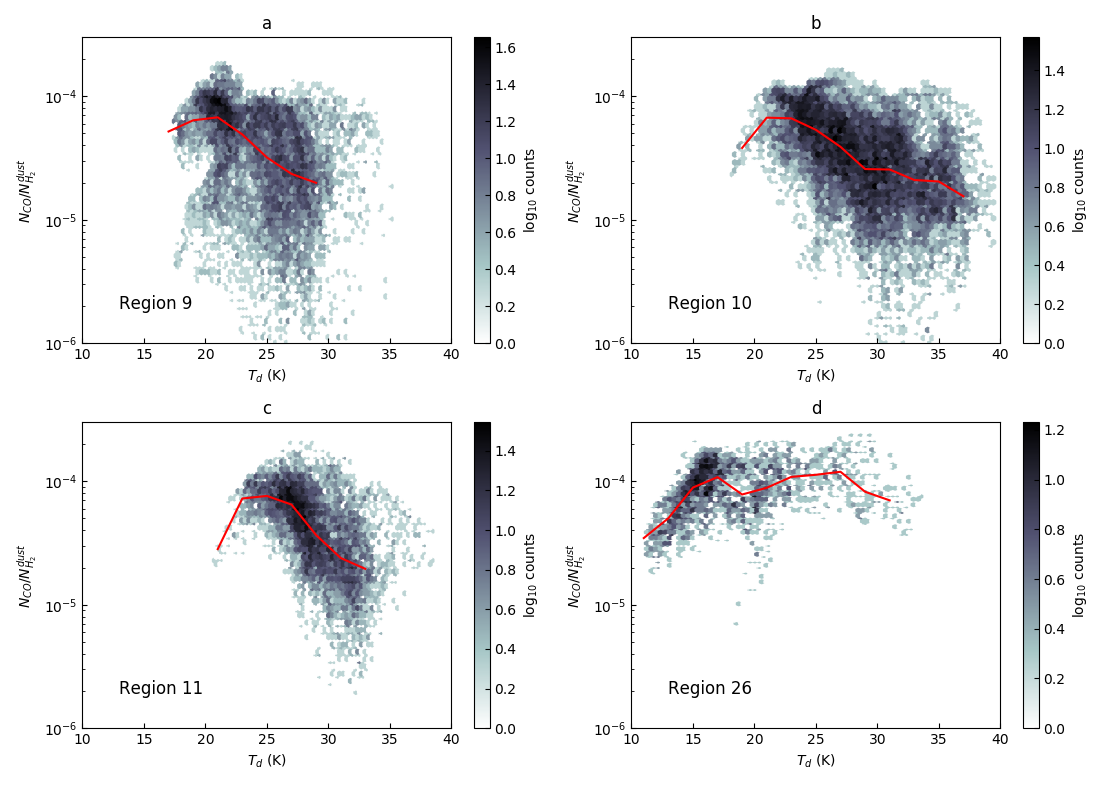} \vspace{-7mm}
    \caption{Pixel density map of \coab\ versus $T_{\text{d}}$ for all four regions. Unexpectedly high CO abundances in Region 26 are likely due in part to over-subtraction in edge pixels. Red lines are rolling medians binned in 2~K-wide bins}
    \label{fig:allnratvt}
\end{figure*}

\indent In the bottom rows of Figures~\ref{fig:r9co}, \ref{fig:r10co}, \ref{fig:r11co}, and \ref{fig:r26bco}, and the rightmost column of Figure~\ref{fig:r26aco}, we take the ratio of the CO column density, $N_{\text{CO}}$, to \ncold, assuming a gas-to-dust ratio of 100. Again for reference, the typical ISM abundance of CO is 1.1$\times$10$^{-4}$ per H$_2$ molecule \citep{sfnotes}. We interpret the maps as showing real variations in the \emph{gas-phase} CO abundance ($N_{\text{CO}}$/\ncold, denoted \coab\ hereafter) in the plane of the sky. The CHaMP clumps in our \coab\ maps (with \ncold\,maps for comparison) all fit one of a handful of morphological profiles illustrated in Figure~\ref{fig:ctypeart} and described as follows:
\begin{itemize}
    \item \textbf{Frozen-out (F-Type)}: clumps where, at the location of the local \ncold\,maximum, \coab\ declines to a few$\times$10$^{-5}$ molecules of CO per H$_2$. In most cases, multiple clumps share a larger envelope or filament of gas with \coab$\sim10^{-4}$ across most of its width, falling away along the outer edges where the $N_{\text{CO}}$ maps are cropped sufficiently far from the emission sources. The higher-\coab\ envelope need not have the same CO abundance on all sides, so there is necessarily some overlap with the A-Type (we denote these as FA-Type).
    \item \textbf{Asymmetric (A-Type)}: clumps where \coab\ and \ncold\,have offset peaks, or where \ncold\,has a local maximum but \coab\ across the same clump has a different, neither directly nor inversely correlated, distribution. The majority of these clumps have FUV sources nearby in the form of one or more massive young star clusters, although the CO abundance can peak on either side of the local \ncold\,maximum with respect to the FUV source. If we assume that the position of maximal CO abundance relative to a nearby FUV source (if there is one) and the nearest H$_2$ column density maximum says something about the degree to which the FUV radiation has penetrated the cloud, we can further subdivide this category as follows:
    \begin{enumerate}
        \item \textbf{Sublimating Asymmetric (sA-Type)}: clumps where \coab\ reaches a peak, typically $\gtrsim10^{-4}$ CO per H$_2$, \textit{between} the location of the local maximum in \ncold\,and a known FUV source. \coab\ at the \ncold\,peak may be a factor of a few lower and may either rise to a second, much lower maximum (sFA-Type), or continue to fall with increasing distance from the FUV source.
        \item \textbf{Dissociating Asymmetric (dA-Type)}: clumps where \coab\ rises \textit{on the opposite side} of the local \ncold\,maximum from a known FUV source. \coab\ is often exceptionally low ($\lesssim10^{-5}$ CO per H$_2$) at the \ncold\,peak and often falls short of the canonical ISM abundance even at its highest.
    \end{enumerate}
    \item \textbf{Coincident (C-Type)}: clumps where \coab\ and \ncold\,have coincident peaks, or where both quantities have such flat distributions across the clump's apparent area that no asymmetries or local extrema can be identified. Many of these clumps have semi-minor axes of $\lesssim$1', so the appearance of this type could be partly a resolution effect.
    \item \textbf{Peculiar (P-Type)}: clumps that appear as local maxima in CO abundance maps, but are either indistinguishable from the background in maps of \ncold, or appear as local deficits. Some appear more closely associated with other named clumps nearby, so these may be artifacts of how the clumps were separated out in earlier studies that did not have the benefit of \ncold\,maps.
\end{itemize}
\indent Classifying many of the clumps required examination at higher contrasts and magnifications than were suitable for publication. Zip files containing the original maps of \ncold\,and \coab\ in FITS format are available upon request. Table~\ref{tab:typestats} lists how many clumps were classified as each Type, and the group averages and dispersions of their minimum $T_{\text{d}}$. The best statistic to characterize a clump's CO abundance has not yet been established, and no consistent Region-independent correlation was found between clump Type and maximum H$_2$ column density (see Figure~\ref{fig:alltypes}).\\
\begin{table}
    \vspace{-1mm}
    \centering
    \caption{Sample statistics on clump Types, subtypes, and their inverse-variance-weighted average minimum dust temperatures, $\langle\!T_{\text{d}}^{\text{min}}\rangle$. The dispersion around this variance-weighted mean is $\langle\sigma_T\rangle$.}
    \begin{tabular}{l|lll} \hline
        Type (\emph{Subtype}) & Count & $\langle\,T_{\text{d}}^{\text{min}}\rangle$ (K) & $\langle\sigma_T\rangle$ (K)\\ \hline
        P & 5 & 33 & 6\\
        A & 55 & 29 & 4\\        
         \hspace{3mm}\textit{sA} & \hspace{3mm}\textit{12} & \hspace{3mm}\textit{27} & \hspace{3mm}\textit{4} \\
         \hspace{3mm}\textit{dA} & \hspace{3mm}\textit{30} & \hspace{3mm}\textit{30} & \hspace{3mm}\textit{3} \\
        C & 30 & 28 & 5 \\
        FA  & 15 & 22 & 3\\
         \hspace{2mm}\textit{sFA} & \hspace{3mm}\textit{5} & \hspace{3mm}\textit{20} & \hspace{3mm}\textit{3} \\
        F  & 21 & 18 & 5 \\\hline
        \textbf{Total} & 126 & 27 & 7\\\hline
    \end{tabular}
    \label{tab:typestats}
\end{table}
\indent If we plot each clump's minimum $T_{\text{d}}$ against its maximum H$_2$ column density, color-code them by Type, and shape-code them by sub-type (Figure~\ref{fig:alltypes}), they show clear differentiation in temperature. Only for $20\lesssim\,T_{\text{d}}\lesssim25$~K can all five morphotypes can be found. There are no purely F-Type clumps, and only one FA-Type, above 25~K. There are no C- or pure A-Types below 20~K, and there are no P-Type clumps and only one dA-Type clump below 25~K. Column density appears not to be a factor, assuming a constant gas-to-dust ratio. While the statistics of each individual Type and Sub-Type are somewhat sparse, these preliminary results suggest there may be something special about the $20\lesssim\,T_{\text{d}}\lesssim25$~K range. This becomes clearer when CO abundance is plotted against $T_{\text{d}}$ on a pixel-by-pixel basis (Figure~\ref{fig:allnratvt}). Laboratory studies show that if there is more than one monolayer of CO on a grain or other substrate, thermal desorption of CO must begin between 15 and 23 K, and increases with temperature up to $\sim$30~K \citep{munoz16,cazax17,noble12}. We also know from \citet{munoz10} and \citet{cazax17} that freeze-out starts at about 27~K, though preexisting CO ice may persist to higher temperatures. And we know UV photodesorption occurs for all $T\lesssim80$~K with a rate dependent on the temperature at which the CO ice formed, but independent of the temperature at desorption. There must be a temperature above which the ISRF and cosmic rays dissociate gas-phase CO faster than thermal desorption or UV photodesorption can replenish it. Therefore, there must be a $T_{\text{d}}$ where \coab\ is maximized between 15 and 30~K depending on the balance of desorption and dissociation, exactly as Figure~\ref{fig:allnratvt} suggests.
\subsubsection{F- and FA-Type Clumps}\label{sssec:ftypes}
\indent The F-Type clumps are the coldest on average, and largely confined to Region 26. Ten of the 21 clumps in Region 9 are of F- or FA-Type, as are all but two clumps in Region 26, but Regions 10 and 11 have only about 10 Type F/FA clumps between them. F-Type clumps are not restricted to the coolest, most sheltered parts of the nebulae, but most are located at least a couple of parsecs away from the nearest PDR front (BYF 70a is one notable, and relatively warm, exception). All but one F-Type clump in all four Regions have $T_{\text{d}}\lesssim25$~K at their centers, but the clumps with $T_{\text{d}}\lesssim20$ K at their highest-column-density points have uniformly lower \coab\ ($\sim$2-4$\times$10$^{-5}$ CO per H$_2$) than clumps that are warmer ($\sim$5-9$\times$10$^{-5}$ CO per H$_2$), and are less likely to be FA-Types. Even BYF 73, a large prototypical FA-Type on the side of an HII region, has a minimum $T_{\text{d}}$ of 17.8$\pm$0.5~K at its \ncold\,peak and a CO abundance of 2.6$\pm$0.1$\times$10$^{-5}$ CO per H$_2$. The FA-Types as a group seem to be slightly warmer than the pure F-Types, with all FA-Types having minimum temperatures of $>$17 K, but the sample sizes are small and their same standard deviations overlap. Notable examples of FA-Type clumps in the CNC include: BYF 67, 73, 95c, 96, 100a, and 115a. BYF 67 is especially interesting because it shows no evidence of being asymmetrically heated by the closest stellar cluster, NGC~3293, but its \coab\ is nevertheless enhanced facing that direction. Some prominent F-Type clumps in the CNC are: BYF 72, 78b\protect\footnote{Pixels toward the galactic southeast in Region 9, in particular around BYF78a-c, may be somewhat under-subtracted due to the assumption of a flat background}, 87, 91a, and 118a.\\
\indent The F-Type clumps of Region 26 in particular, as parts of IRDCs, are consistent with the simulations of \citet[][hereafter GBS14]{gerner}. Their simulations suggest that when a molecular clump is in its IRDC phase of life, the typical \coab\ is a few$\times$10$^{-5}$ at the center, rising to something close to the ISM value of \coab\ at a distance of about half a parsec. Where BYF 199a, 208a and b, and the IRDC containing BYF 201ab and 202a have local minima in $T_{\text{d}}$ and maxima in \ncold, \coab\ can dip to $\sim$2$\times$10$^{-5}$ CO per H$_2$. These clumps also all have a minimum $T_{\text{d}}\lesssim15$~K, and thermal CO desorption only starts to become active for temperatures above $\sim$15~K in a cloud with a typical \ncol\,between 10$^{26}$ and 10$^{27}$ \psqm\,\citep{munoz10}. If the gas-to-dust ratio is approximately constant as assumed, then the best explanation for the apparent CO depletion in these clumps is that the CO has frozen out onto dust grains to varying degrees. We favor this explanation because it agrees with the works of \citet{caselli99}, \citet{kramer}, \citet{bacmann}, \citet{akhern}, and \citet{fontani}. However, since those studies relied on some of the same assumptions we used, there are alternative explanations. The gas-to-dust ratio, $\gamma$, could be declining toward the center of the clumps, as inferred in \citet{reach} and \citet{ripple}. This could inflate the derived H$_2$ column density by up to an order of magnitude. The dust emissivity, $\kappa_0$, could also be changing as the grains accrete or lose CO ice \citep{ossen, bergin}. For example, over the range of densities observed in nearby molecular clouds in Taurus, \citet{stepnik} showed that $\kappa_0$ increased by a factor of about 3 toward the cloud centers, holding $T_{\text{d}}$ constant. Underestimating the dust $\kappa_0$ would also inflate \ncold, but the relationship between $\kappa_0$ and $\gamma$ makes these two explanations degenerate.

\subsubsection{A-Type Clumps}\label{sssec:atypes}
\indent We observe more A-Type clumps than any other type, due in large part to the fact that the majority of the clumps in this study are in Regions 10 and 11, where FUV fluxes are strong and highly directional. In future works, we might expect to see fewer A-Type clumps in the remaining CHaMP Regions, which contain fewer nearby OB star clusters. Most of the warmest clumps in Regions 10 and 11---BYF 93a-c, 98a, 99a-q, 100c and f, 102a-d, 103c and d, 104 b and c, 106-109---have CO abundances of a few $\times$10$^{-5}$ per H$_2$ or lower at their \ncold\,peaks, with little or no nearby gas of ISM-like CO abundance. What little CO is present tends to be opposite the \ncold\,peak from the nearest FUV source, if a peak can be found in either the map of \coab\ or the map of \ncold, making these clumps dA-Type. Only one clump with a minimum temperature below 25~K was identified as dA-Type. There are no dA-Type clumps in Regions 9 or 26. Some of the most striking examples are in BYF 99, where \coab\ can be $\lesssim2\times$10$^{-5}$ in the clumps closest to Tr14. Most of BYF 99 is in the part of the expanding PDR front around Tr14 situated between Tr14 and us \citep{brooks03}, of which very little is left. Given Tr14's age of 0.3-0.5 Myr \citep{sana10}, BYF 99 has had 0.3 to 0.5 Myr to have its CO dissociated and/or ionized; indeed, \citet{brooks03} find that some 36\% of the total carbon in Tr14 PDR front is in the form of C$^0$ and C$^+$. The dA-Types in the Southern Cloud---BYF 105 a, 107ab-f, and 108b---are 5 to 10 times lower in column density than those in the Northern Cloud, but have a similar range of \coab with a higher minimum. Perhaps as the Southern Cloud has faced irradiation from Tr16 for longer than the Northern Cloud has been irradiated by Tr14, more of its original dust content has been destroyed along with its CO content, creating a floor in \coab.\\
\indent The behavior of dA-Type clumps seem to support the simulations of \citet{penaloza18} (hereafter PCGK) that demonstrate the importance of external radiation sources (the local ISRF and cosmic rays) in determining the as density needed for CO to self-shield. Over the range of typical gas number densities calculated for the CHaMP clumps in \citetalias{champ1}, 10$^8\lesssim\,n\lesssim10^{10}$ m$^{-3}$, PCGK show the fraction of the carbon in CO should drop about three orders of magnitude, which covers the full range of \coab\ we see, for ISRF fields comparable in strength to the parts of the CNC outside the Tr14 and Tr16 PDRs. Higher ISRF values could dissociate CO at even higher densities. Nearly all of the North and South Clouds have FUV fields above 100$G_0$ in Habing (1968) units (1~$G_0$ is approximately the ISRF strength in the solar neighborhood), and within about a parsec of $\eta$ Car and Tr14, the ISRF can top 10$^4$~$G_0$ \citep{brooks03,rocca}. The fraction of hydrogen in the form of H$_2$ in the low-density envelopes of these clouds also falls with increasing $G_0$, which limits how much \coab\ can drop. However, thermal dust emission technically traces both atomic and molecular hydrogen, and the best HI maps available for the CNC area have resolutions of 2$'$ or lower \citep{sgps1}. PCGK's model predicts that, compared to a 1$G_0$ ISRF, a 100$G_0$ ISRF will reduce the fraction of hydrogen in H$_2$ by about 10\% at $n\sim10^{10}$ m$^{-3}$, 30\% at $n\sim10^9$ m$^{-3}$, and 60\% at $n\sim10^8$ m$^{-3}$. At that density, only about half the total hydrogen content would be in H$_2$ even in a 1$G_0$ ISRF. Our assumption that thermal dust emission traces hydrogen mostly in the form of H$_2$ may lead to underestimating the CO abundance in regions with strong ISRFs, but since HI is several orders of magnitude less dense than molecular gas, its contribution integrated along the line of sight should be small. In any case, our CO abundances for all clumps everywhere are all within or above the range of values predicted by PCGK. They also find that CO fails to trace any gas warmer than about 40~K. In both Regions 10 and 11, while there was no one temperature at which \coab\ was maximized, \coab$\lesssim10^{-5}$ almost everywhere that $T_{\text{d}}>$35~K.\\
\indent Fifteen of the A-Type clumps showed CO abundances increasing on the side of a clump facing an FUV source, identifying them as the sublimating (s) subtype. These tended to have slightly lower minimum temperatures and higher peak CO abundances than the dA-Type clumps, as expected, although the overlap is significant. BYF 95 and c are the most easily-identifiable examples. No sA-Type clumps had minimum $T_{\text{d}}<$20~K, but two of the four sFA-Types had 15$<T_{\text{d}}<$20~K, BYF 66 and 73. These clumps are relatively large in both mass and extent, such that sub-structure is easier to see and more cool CO is insulated in the centers, and they inhabit relatively undisturbed diffuse gas that the Gum 31 PDR front has not reached yet.\\
\indent Several clumps classified as sA-Types contain known protostars, protoclusters, masers, or other star-forming activity \citep[see][etc.]{rath,povich,vista1} that one would expect should cause the CO abundance to rise toward the \ncold\,peak. By definition this is the case for C-Type clumps, not A-Types. \citetalias{gerner} predict that \coab\ should monotonically increase to about 10$^{-4}$ by the end of the high-mass protostellar object (HMPO) phase---wherein a clump contains at least one mid-IR emission source of $\gtrsim$8$M_{\odot}$ \citep{srid}---and keep climbing to a few$\times$10$^{-4}$ by the time a clump becomes an ultra-compact HII (UCHII) region. This seems consistent with what we see in the three UCHII regions in Region 26, and the relatively high-\coab\ sA-Type clumps associated with them, but we know of only two other UCHII regions in our data neither of which agree with the prediction. There are few UCHII regions in the CNC because the strong radiation from Tr14 and Tr16 peels away the envelopes of embedded clusters very quickly \citep{brooks03}. Furthermore, whether or not we can justly make any further comparisons to \citetalias{gerner} is unclear given that the CNC is, despite its ongoing production of stars, an evolved region whose physical conditions are overwhelmingly driven by just a few tens of OB and WR stars. What Figures~\ref{fig:r9co} through \ref{fig:r26bco} suggest is that prestellar clumps in as energetic an environment as the CNC will sample the entire range of possible CO abundances predicted by \citetalias{gerner} regardless of any embedded stellar content. The minimum temperature in a clump generally coincides with the highest-\ncold, but one can easily see from the parameter maps in \S\ref{sec:pms} that the range of $T_{\text{d}}$ across the clump depends on the temperature of the ambient diffuse gas.

\subsubsection{C- and P-Type Clumps}\label{sssec:cptypes}
\indent C-Type clumps generally occupy the same space in the $T_{\text{d}}$-\ncold\,plane as A-Types appear, tending only to be slightly lower in maximum H$_2$ column density. The existence of an apparent lower temperature cut-off that rises with column density suggests that this type, like the P-type, may be an artifact of the limits of human discrimination. Unlike the P-Types, where a clump in CO emission fails to map to a similar structure in dust emission, C-Type clumps are probably real but misidentified. For an A-Type clump directly in front of or behind an ionizing radiation source, it may not be possible to distinguish any segregation of dust from CO. If an observer happens to see a molecular cloud irradiated from the side while CO desorption and dissociation happen to balance along the observer's line of sight to the peak in the dust column density, they would reasonably identify it as a C-Type when historical or far future observations would have revealed the cloud penetrated by radiation to a different depth. Hypothetically, a very small or poorly-resolved F-type could appear as a C-Type, but the average and dispersion of temperature for the C-Types suggests no F-Types have been so mistakenly classified.

\subsection{Luminosity-to-Mass Ratios}\label{ssec:lumar}
\begin{figure}
    \centering
    \includegraphics[width=\columnwidth,trim={0 0.3cm 0 0},clip]{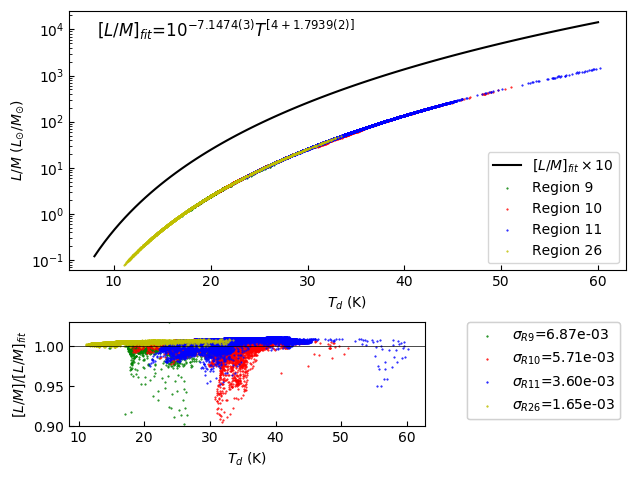} \vspace{-4mm}
    \caption{\textbf{Top panel:} Pixel-by-pixel plot of $L/M$ ($L_{\odot}/M_{\odot}$) versus $T_{\text{d}}$ across all four Regions, fitted with the equation shown (upper left), \textbf{displaced upward by a factor of 10} so as not to cover the data. The fit recovers our assumed constant $\beta=1.8$ to within 1\%. \textbf{Bottom panel}: same as above, but with the data divided by the fit and showing the standard deviation of each Region. Deviations toward $L/M$ values below the fit come from sky background-dominated areas where background subtraction is systematically more aggressive for redder filters. This makes $\beta$ steeper than assumed and reduces the integrated luminosity.}
    \label{fig:lmvt}
\end{figure}

\begin{figure*}
    \includegraphics[max width=\textwidth, max height=8in]{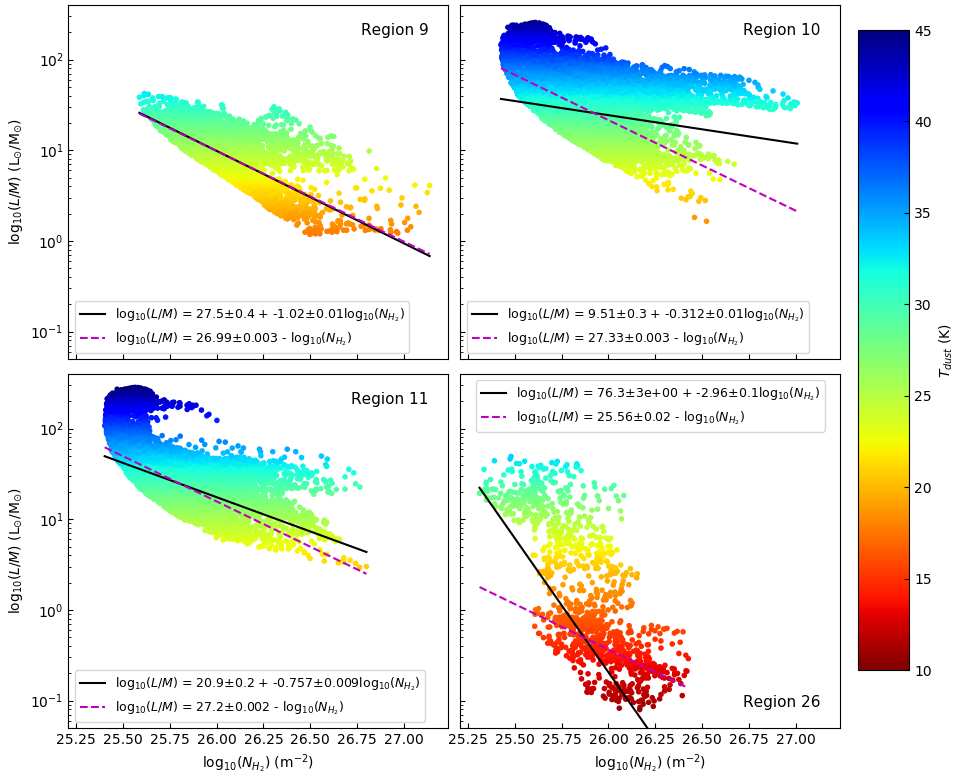} \vspace{-4mm}
    \caption{Scatter plots for each Region, as labelled, of log$_{10}$(\ncol) (\psqm) versus $L/M$ ($L_{\odot}/M_{\odot}$), color-coded by $T_{\text{d}}$, with curve fits weighted by the inverse of the uncertainty in each point. Top row: Regions 9 (left) and 10 (right); bottom row: Regions 11 (left) and 26 (right). We suspect the apparent cutoff in the data in the lower left corner of each panel is due to the sensitivity limits of \textit{Herschel}. The extremely high $L/M$ branch in the plot for Region 11 was identified as belonging to $\eta$ Car, most likely the Homunculus Nebula. These are likely lower limits as $\eta$ Car is saturated at some wavelengths.}
    \label{fig:nvlm}
\end{figure*}

\begin{figure*}
    \includegraphics[max width=\textwidth, max height=8in]{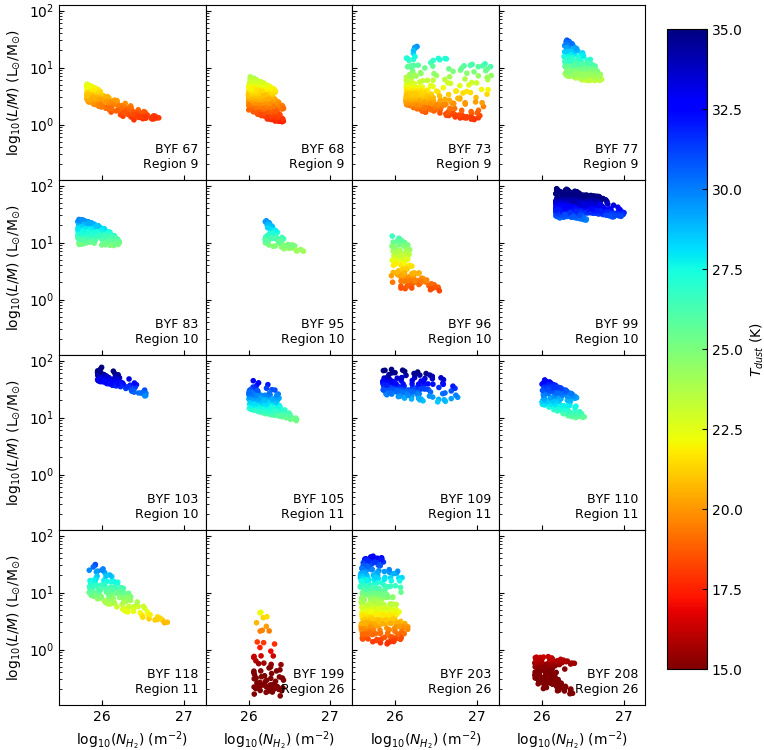}\vspace{-4mm}
    \caption{Scatter plots for some sample clumps, as labelled,  of log$_{10}$(\ncol) (\psqm) versus $L/M$ ($L_{\odot}/M_{\odot}$), color-coded by $T_{\text{d}}$, as in Figure~\ref{fig:nvlm}.}
    \label{fig:exnvlm}
\end{figure*}

\indent It has been popular in recent studies of clump-scale star-forming regions (e.g. \citetalias{champ3}, \citet{rcw120,rcw79,agal3}, etc.) to use $L/M$ as an indicator of evolution. $L/M$ is expected to be associated with prestellar and protostellar evolution, increasing as a molecular clump transitions from quiescent to star-forming. As protostars form and begin to drive outflows, most of the gas in the clump is gradually dispersed, revealing more and more of the embedded young stellar objects (YSOs). The work of \citet{molinariLM} appears to provide observational support by showing that for prestellar/protostellar clumps identified in 70~\micron~emission, CH$_3$C$_2$H detections\protect\footnote{CH$_3$C$_2$H is an optically-thin dense ($n\sim10^{11}$ m$^{-3}$) gas tracer expected to probe only the parts of massive molecular clumps where YSOs should already be forming \citep{molinariLM}.} are strongly correlated with $L/M$ such that no sources are detected for $L/M\lesssim$1~$L_{\odot}/M_{\odot}$, and all targets are detected for $L/M\gtrsim$10~$L_{\odot}/M_{\odot}$. We looked to see if $T_{\text{d}}$ was was as well-correlated with $L/M$ as the CH$_3$C$_2$H excitation temperature. Unlike \citet{molinariLM}, however, our plots of $L/M$ versus $T_{\text{d}}$ (Figure~\ref{fig:lmvt}) had virtually no scatter and never departed from an exponential curve.\protect\footnote{Figures~\ref{fig:r9lm} through~\ref{fig:r26lm} in the Appendix also show maps of $L/M$ for Regions 9, 10, 11, and 26 to demonstrate that the log-scaled distribution of $L/M$ almost perfectly mirrors each regions' corresponding temperature map.}\\
\indent Many studies that start with least-squares SED-fitting to far-IR/submm data, like ours, assume constant $\beta$, $\kappa_0$, and $\gamma$ because they are all mutually interdependent and computationally expensive to determine without fixing the more commonly desired parameters, $T_{\text{d}}$ and \ncol. However, in the optically thin limit (where $1-e^{-\tau_{\nu}}\approx\tau_{\nu}$), one can mathematically prove that $L/M$ becomes approximately independent of \ncol. $L/M$ can be calculated via
\begin{equation}
    \begin{split}
    \frac{L/M}{L_{\odot}/M_{\odot}} &\approx \frac{4\pi}{\cancel{N_{\text{H}_2}\mu m_{\text{H}}}}\int_{0}^{\infty}\bigg(\frac{\cancel{N_{\text{H}_2}\mu m_{\text{H}}}\kappa_0}{\gamma}\bigg)\bigg(\frac{\nu}{\nu_0}\bigg)^{\beta}B_{\nu}(T)d\nu\\
     &=\frac{4\pi\kappa_0h}{c^2\gamma\nu_0^{\beta}}\bigg(\frac{kT}{h}\bigg)^{\beta+4}\Gamma(\beta+4)\zeta(\beta+4)
    \end{split}\label{eq:loverm}
\end{equation}
where $\Gamma$ is the integral form of the gamma function, and $\zeta$ is the Riemann Zeta function. If $\beta$, $\gamma$, $\kappa_0$, and $\nu_0$ are held constant, $L/M\propto\,T_{\text{d}}^{4+\beta}$ and nothing else. Figure~\ref{fig:lmvt} demonstrates the algebraic identity of $T_{\text{d}}$ and $L/M$---the same power law fits the data from all four Regions simultaneously (the fitted curve is shifted upward by a factor of 10 for visibility). Depending on one's choice of free SED-fitting parameters and the optical depth regime, comparing $L$ to $M$ or mapping $L/M$ may be merely an abstruse representation of the temperature distribution. More generally, $L/M$ \emph{is not an independent measure of star formation activity} when its calculation involves SED-fitting.
\indent At large optical depths (generally not expected for far-IR/submm dust emission), the power law modification of the Planck function disappears entirely such that $L/M\propto\,T_{\text{d}}^4N_{\text{H}_2}^{-1}$. One might expect, then, that a plot of $L/M$ versus \ncol\,would show no correlation at low column densities, and gradually converge toward a log-linear trend in $N_{\text{H}_2}^{-1}$ with increasing optical depth---that is, $L/M\propto\,N_{\text{H}_2}^{\varphi}$ with -1$<\varphi<$0 for all realistic optical depths. Figure~\ref{fig:nvlm} shows that, using weighted least-squares to fit the data, lines with slopes within the predicted range of $\varphi$ \emph{can} be fit some of the time, but the orders of magnitude of scatter and complex morphology raise doubts about their utility. Region 26 is especially inconsistent, and the reason why is demonstrated in Figure~\ref{fig:exnvlm} where $L/M$ is plotted pixel-by-pixel against \ncol\,and color-coded by $T_{\text{d}}$ for a selection of clumps whose boundaries are set by the last closed contour in \ncol. Every clump is different.\\
\indent Most clumps in Figure~\ref{fig:exnvlm} show $L/M$ ($T_{\text{d}}$) declining, with a variety of slopes, as \ncol increases. This broadly inverse relation between temperature and column density is the opposite of the trend one typically observes for \ncol\,and $T_{\text{ex}}$ for CO \citep[e.g.][]{kongco}, and generally inconsistent with the idea that $L/M$ is tracing pre-stellar evolution. Even clumps that are known to contain or border on active star-forming regions---like BYF 73, 77, 109, and 203---sometimes still show higher $L/M$ or temperatures at lower column densities. One can select for sources known to be internally heated by YSOs and trivially find that the peak $T_{\text{d}}$ and \ncol\,are positively correlated, but it seems dangerous to extrapolate those findings to clouds where YSOs are not known to be present, or are known not to be present. 
More generally, Figures~\ref{fig:nvlm} and \ref{fig:exnvlm} suggest cloud evolution in the $T_{\text{d}}$-\ncol\,plane is more complex than any one power law can capture, which is as unsurprising as it is frustrating because, if
\begin{equation}
    L/M\propto
    \begin{cases}
        T_{\text{d}}^{4+\beta}\kappa_0\gamma^{-1} & \text{for $\tau\ll1$} \\
        T_{\text{d}}^4N_{\text{H}_2}^{-1} & \text{for $\tau\gg1$}
  \end{cases} \label{eq:lmprop}
\end{equation}
is continuous across the transition from the small to large optical-depth regimes, then
\begin{equation}
    N_{\text{H}_2}\propto\,T^{-\beta}\gamma\kappa_0^{-1}. \label{eq:lmvt}
\end{equation} Given the results of such studies as \citealt{reach,ossen,bergin}, and others, and the availability of the integral of the Bose-Einstein distribution on Wolfram Alpha\protect\footnote{Wolfram Alpha LLC. 2009. Wolfram|Alpha. http://mathworld.wolfram.com/Bose-EinsteinDistribution.html (access December 8, 2017).}, none of this is new information. Knowing that $\gamma$ can vary by a factor of two, and that $\beta$ can vary by about 0.5, either could account for the entire range of observed scatter on its own, so both are variables that deserve more attention. On the other hand, equation~\ref{eq:lmvt} suggests more productive avenues for numerical analysis.

\subsection{Column Density PDFs}\label{ssec:npdfs}
\indent The H$_2$ column density probability distribution function (N-PDF) of a molecular cloud is widely thought to be diagnostic of cloud kinematics \citep{semadeni,nordlund,burklaz}, star formation rate (SFR) \citep{krum05,hennchar08,hennchar11}, and on larger scales, the stellar initial mass function (IMF) \citep{padoan02,nordlund03,hennchar13}. Simulations by \citet{semadeni01,padoan97,feder08}, and others assert that N-PDFs of molecular gas containing no star formation (as well as neutral atomic and ionized gas) are well-fit by a pure log-normal distribution,\\
\begin{equation}
    p(\eta)d\eta = \frac{1}{\sqrt{2\pi\sigma^2}}\text{exp}\bigg(-\frac{(\eta-\Bar{\eta})^2}{2\sigma^2}\bigg)d\eta
    \label{eq:npdflog}
\end{equation}
where we take $\eta$ to be ln($N$), and the logarithmic dispersion $\sigma^2$ is given by 
\begin{equation}
    \sigma^2=ln(1+b^2\mathcal{M}^2).\label{eq:dispers}
\end{equation}
Kinematically, the most important parameter to the fit is $\sigma$ because of its theoretical relation to the sonic Mach number, $\mathcal{M}$ \citep{passot}, and a turbulent forcing parameter, $0.3\lesssim\,b\lesssim1$ depending on the curl and divergence of the magnetic field, assuming the gas is nearly isothermal \citep{feder10,burklaz}. In theory, $\sigma^2=2|\Bar{\eta}|$ if $\Bar{\eta}$ is defined as either $\text{ln}(N/\Bar{N})$ or $\text{ln}(N/\widetilde{N})$, where $\Bar{N}$ and $\widetilde{N}$ are the mean and median of $N$, respectively. We chose not to scale ln($N$) by a mean or median in our definition of $\eta$, as both the mean and median become difficult to establish when crowding, map edges, and/or background subtraction make the boundaries of a given cloud or clump ill-defined.\\
\indent At the onset of cloud collapse, N-PDFs of molecular gas are expected to develop a power-law tail with a functional form
\begin{equation}
    p(\eta) = p_c\; \text{exp}\left[s\left(\frac{\eta}{\eta_c}\right)\right] = p_c\left(\frac{N}{N_c}\right)^s
    \label{eq:npdfplaw}
\end{equation}
where $s$ is the slope of the power-law tail, $\eta_c$ is the power law's lower-$\eta$ cutoff, and $p_c=p(\eta_c)$. Many observational studies relying on these models appeared to show that real molecular clouds have (sometimes multiple) log-normal N-PDFs toward lower column densities \citep[e.g.]{schneider13,rcw79,tremblin}, and many seem to transition to a power law (or two) at higher column densities \citep[e.g.]{kain09,aquila,balls,schneider15}. \\
\indent A few recent papers raise some questions about the utility of N-PDFs as applied to real molecular clouds, where finite map size, noise, and line-of-sight contamination can strongly effect the measured N-PDF distribution, especially at low column densities. \citet{balls} showed that the slope of the power-law part of the N-PDF depends on the cloud's depth along the line of sight. \citet{chenh17} find that the power-law tail of a cloud is a summation of the power laws of individual clumps, all shaped differently from each other and from the distribution as a whole. \citet{alves14} observed that at least in nearby molecular clouds like Lupus and Taurus, for \ncol$>$3$\times$10$^{25}$~\psqm, the N-PDFs are about equally well-fit by the wing of a broad log-normal, or a power-law. \citet{lombardi} and \citet{alves} find that the log-normal part of the N-PDF systematically vanishes when one defines the area of the clump for which to plot the N-PDF by closed contours (i.e., not broken by the boundaries of the map) in column density. The \citet{lombardi} and \citet{alves} studies may not be as troubling, because at some appropriate signal-to-noise (S/N) limit in any practical observation, closed contours in column density may visually select for the denser condensations that will be part of a gravitationally-dominated clump with a power-law PDF.\\
\begin{figure*}
    \begin{tabular}{cc}
    \includegraphics[width=\columnwidth]{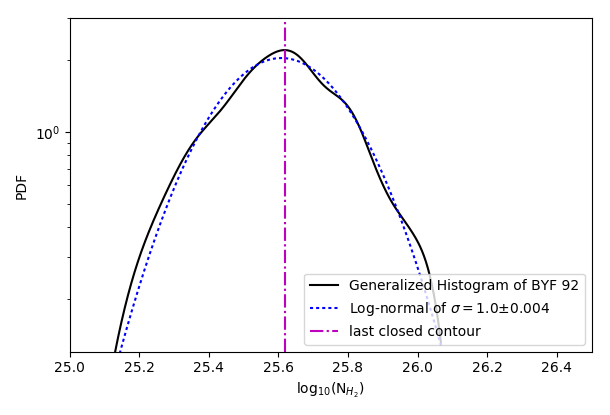} & 
	\includegraphics[width=\columnwidth]{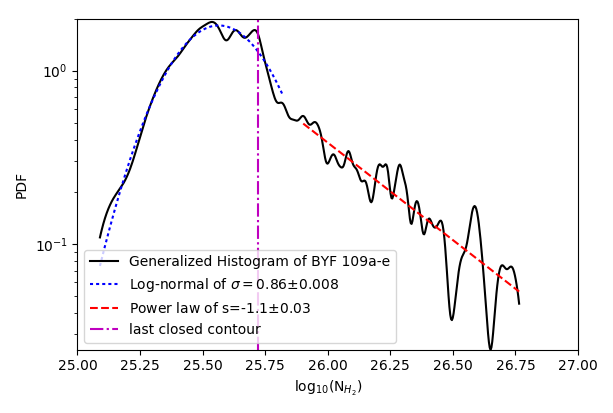}\\ \vspace{-4mm}
	\includegraphics[width=\columnwidth]{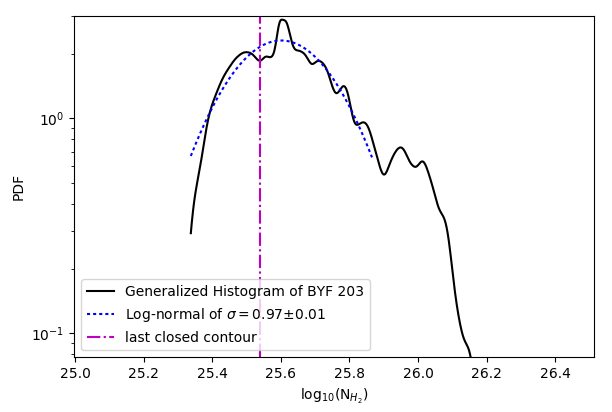} & 
	\includegraphics[width=\columnwidth]{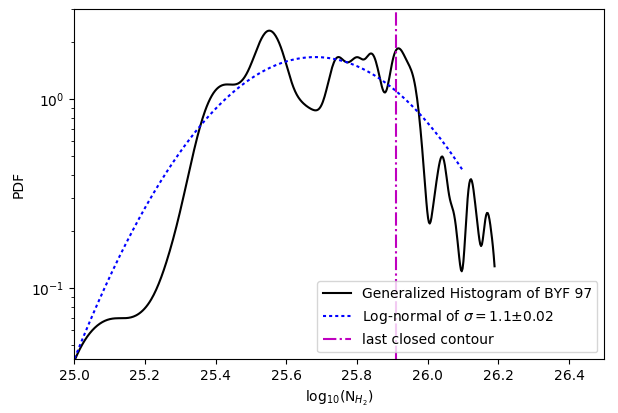}
	\end{tabular}
    \caption{Sample generalized histogram N-PDFs showing a high-quality pure-power-law fit (top left, BYF\,92), a good-quality log-normal plus power-law fit (top right, BYF\,109), a typical fit where something of a tail was observed but only the log-normal could be fit (bottom left, BYF\,203), and a poor-quality fit where optimization returns parameters and error estimates but the histogram and fit bear little resemblance to each other (bottom right, BYF\,97). Note the poorly-constrained transition from log-normal to power law for the N-PDF of BYF\,109.}
    \label{fig:cnpdfs}
\end{figure*}
\indent In our view, the important point is that the sampling of the log-normal part of the PDF, and to some extent the power-law part as well, is strongly subject to observational limitations. These limitations mean that one cannot assume either part of the N-PDF is well-sampled, or that the parameters of $p(\eta)$ derived from fitting the N-PDF are physically meaningful. Specifically, as \citet{lombardi} and \citet{schneider15} point out, the shape and width of the log-normal component depends on both the aggressiveness of one's background subtraction procedure and one's definition of a cloud's boundary. These factors in turn depend on map size, resolution, detector sensitivity, and the severity of crowding. The prestellar clumps in our sample are seldom more than a few beam widths in any dimension, and so close together that blending makes separation of the power-law and log-normal components (when both are identifiable, which is often not the case) difficult and the parameters derived from the resulting fits less reliable than their calculated uncertainties reflect.\\
\indent We determined that the shape of the mask used to isolate a clump and draw an N-PDF affects the width of the log-normal much less than the total sample area, so we isolated large clumps and groups of small clumps according to the catalog numbers of Paper I where possible, using iso-\ncol\ contours as a guide to help separate clumps with connecting filaments. The resulting N-PDFs were startlingly heterogeneous. At the scales of large clumps and clump groups ($\sim$1-5 pc), we observed every possible combination of zero to two log-normal components and zero to two power-law components, and even a couple N-PDFs that seemed to have log-normal distributions with power-law tails on the low-\ncol\ side. Figure~\ref{fig:cnpdfs} shows examples of generalized\footnote{We used generalized histograms because well-known numerical artifacts can be introduced by the choice of binning (i.e. sampling). To make a generalized histogram, each data point is modeled as a unit-height Gaussian with a mean equal to the datum and a standard deviation equal to the uncertainty in that datum. The model for each point is evaluated over the entire range of the data, sampled at a sufficiently large number of intervals ($n\gtrsim100$). Then the model Gaussians are summed and divided by the total number of points. The histogram and the parameters of any function fitted to it are thus relieved of any dependence on bin width, and the histogram is automatically normalized as a PDF.} histogram N-PDFs with good, typical, and poor $p(\eta)$ fits respectively.\\
\indent About $10\%$ were well-fit with a log-normal and a power-law, albeit with a poorly-defined transition between the two components. This is a rather evocative percentage: if 5-10$\%$ of clumps are actively star-forming, as \citetalias{champ5} suggests, we expect a similar fraction of N-PDFs to have strong power-law tails indicating gravitational dominance across a large fraction of the dense gas. Another $\sim10\%$ of the clumps' N-PDFs were exceptionally well-fit with a pure log-normal. The clumps with pure log-normal distributions tended to have low to middling temperatures and densities, and virtually all log-normal fits had $0.7\le\sigma\le1.1$ in units of log$_{10}$(\ncol). About half of all clumps had N-PDFs with a somewhat well-fit log-normal component and significant excess on one or both sides, usually at the high-column-density end, that could not be fit. Most of the remaining $\sim30\%$ of N-PDFs had two or more poorly defined peaks, flat tops, or other peculiarities such that curve-fitting failed or returned obviously untrustworthy parameters. Not all clumps with known embedded protostars had recognizable power-law tails in their N-PDFs, and not all clumps with a power-law component in their N-PDFs are known to contain protostars (though none are known for certain \emph{not} to contain protostars). It's hard to say what constitutes an appropriate sample to generate an N-PDF, but the sampled area of the clumps in Figure~\ref{fig:cnpdfs} are $\sim10^2$ beams each, and that appears to be only sufficient in ideal cases where the N-PDF is overwhelmingly log-normal in shape. Fitting a power-law tail unsurprisingly seems to require much more data. It's also important to remember there are arguments to be made for partitioning the clumps differently than we have done, and doing so can result in vastly different N-PDFs.\\ 
\begin{figure*}
	\includegraphics[max width=\textwidth]{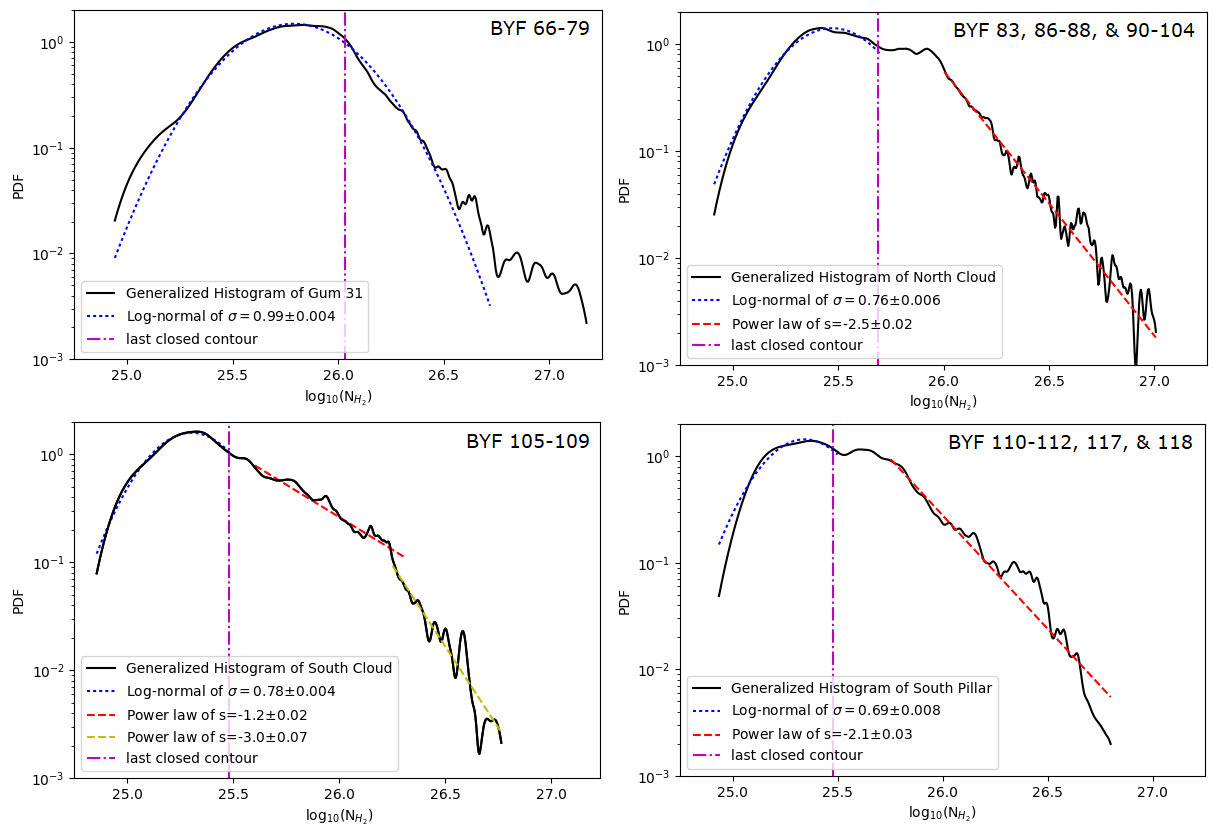} \vspace{-4mm}
    \caption{N-PDFs (\psqm) of the large-scale ($\sim10$ pc) features of the CNC: Gum 31 (upper left), the North Cloud (upper right), the South Cloud (lower left), and South Pillar (lower right). Each feature was selected using a polygonal mask around the lowest closed iso-\ncol\ contour containing the entire feature but not connected to any other large-scale features.}
    \label{fig:rnpdfs}
\end{figure*}
\indent Like \citet{alves}, we find that the log-normal part of clump-scale N-PDFs, where identifiable, peaks where \ncol\,is at or below that of the last closed contour. If the sample sizes are expanded to encompass the largest ($\gtrsim10$ pc) features of the CNC---e.g. the North Cloud (BYF 83, 86-88, and 90-104), South Cloud (BYF~105-109), or Southern Pillar (BYF~110-112, 117, and 118) as shown in Figure~\ref{fig:rnpdfs}---we do sometimes observe a turnover slightly above the column density of the last closed contour. However, there is not enough data on either side of the turnover to fit a convincing log-normal component without the data outside the last closed contour.\\
\indent Usually the transition point between log-normal and power-law components is so close to the peak of the log-normal or so poorly defined that, if the data are masked below the lowest closed contour, what was originally interpreted as the wing of a log-normal plus a power-law could be just as well-fit with one or occasionally two power-laws. Our methodology explicitly attempts to remove at least some of the diffuse component without erasing filaments well-associated with our selection of clumps, so to the extent that background subtraction limits the sampling of diffuse gas, the derived log-normal widths should be taken with a grain of salt even if the fit and the available data are well-matched. The nominal uncertainties in the slopes of the power-law components were only marginally larger than those of the log-normal widths, but the constants of the fit to equation~\ref{eq:npdfplaw}, $N_c$ and $p_c$, were essentially unconstrained. Attempts to fit the log-normal and power-law components simultaneously always resulted in complete nullification of the power law. The best guess for a power-law cut-off had to be input manually, well-separated from the log-normal so that neither component would be distorted near the transition.\\
\indent If we take at face value the log-normal fits to the N-PDFs of $\sim10$ pc-scale features in the CNC, shown in Figure~\ref{fig:rnpdfs}, and solve equation~\ref{eq:dispers} for the Mach number, their sonic Mach numbers range from 0.78 to 3.9 depending on the value of forcing parameter $b$. In the pure solenoidal forcing case ($b\sim1/3$), $\mathcal{M}$ takes approximate values of 3.9, 2.7, 2.8, and 2.3 for Gum 31, the North Cloud, the South Cloud, and the South Pillar, respectively, which would make them all slightly supersonic. In the case of pure compressive forcing ($b\sim1$, i.e. the previous results divided by three), the four clouds would be transonic. Neither of these sets of results makes much sense in the context of studies like \citet{schneider13} and \citet{chenh17} where the estimated Mach numbers of the active star forming regions are 5-10 (hypersonic turbulence). Also, based on \citet{schneider15}, residual background contamination should have the effect of heightening and narrowing the log-normal, but while our background subtraction method was more aggressive than theirs, it has the tendency to create a wider range of column densities than their method (see Appendix~\ref{sec:apbg}). On the other hand, \citet{feder10} find that sonic Mach number tends to decrease with increasing density; since we systematically select for denser gas on average than \citet{schneider13} and \citet{chenh17}, our smaller Mach numbers may not be so unreasonable after all.\\
\indent Since the N-PDFs of the North Cloud and South Pillar have troughs (at log\ncol$\sim25.7$ and 25.5 respectively) near the peaks of their apparent log-normal components, it may be that the log-normal fits needed to span the trough and encompass the secondary maximum. Still, we think it more pertinent to ask if these sonic Mach numbers are meaningful. Clearly the discrepancies are in how the background subtraction is done and how each clump/cloud is isolated to do the N-PDF analysis, but there is no independent way to find the optimal balance between including the diffuse envelope of a source and excluding unrelated material in the line of sight. If observational data can (or worse, must) be tuned to match theory, the data lack predictive power. More broadly, what does the sonic Mach number, or slope, or any other N-PDF fitting parameter of a cloud tell us when the cloud's N-PDF is the sum of the NPDFs of every clump within it, each with different parameters from each other, and from the cloud as a whole? We cannot comment on the soundness of the theory behind N-PDFs, but the results when put into practice are hard to trust: the samples are biased from the outset by observational limitations, are further skewed by any standard method of correcting for line-of-sight contamination, and suffer from several unquantifiable sources of uncertainty. The qualitative character of the N-PDF may still be useful, though, in the sense that the rate of occurrence of believable power-law tails roughly matches our expectations if 90-95$\%$ of clumps are quiescent \citepalias{champ5}.

\section{Conclusions}
\label{sec:conc}
In this paper, we present parameter maps, derived from SED fits to \textit{Herschel} and LABOCA dust continuum emission maps, of the far-IR/submm counterparts of molecular clumps in four of the 35 regions in the CHaMP survey. We compare \ncol\,derived from dust emission to the same derived from CO emission using the region-specific power laws derived from simultaneous radiative transfer on three isotopologues of CO in the manner of \citep{thrumms} and \citetalias{champ5}. We present maps of CO abundance in each Region derived by dividing $N_{\text{CO}}$ from the 3 isotopologues of CO by \ncold, and we present a morphological classification scheme to sort clumps by appearance in maps of CO abundance. We also map $L/M$ and compare it to \ncold, and we examine N-PDFs for both regions, sampled both from the entire Mopra coverage area and from areas bounded by the lowest closed contour in \ncold. Our main results are as follows:
\begin{enumerate}[label=(\arabic*)]
    \item H$_2$ column-density maxima almost always coincide with dust temperature minima, implying that density enhancement and cooling must be contemporaneous phases in the evolution of prestellar clumps.
    \item Gas-phase CO abundances can vary by one to three orders of magnitude across each Region, and by up to an order of magnitude across some large molecular gas clumps. There is no single CO to H$_2$ abundance ratio that is appropriate for all molecular gas densities, temperatures, or locations within the Milky Way, even in the same galactic arm.
    \item The CO abundance distribution over a molecular clump depends in both form and value on environmental factors (like the presence of nearby OB stars). The coldest clumps tend to be relatively CO-poor (2-4$\times$10$^{-5}$ CO per H$_2$) at their \ncold\,peaks, surrounded by envelopes with CO abundances $\gtrsim10^{-4}$ CO per H$_2$. We interpret this as CO depletion due to freezing out, although alternative explanations are possible, such as reduction in the gas-to-dust ratio. Warmer clumps tend to have \coab\ and \ncold\,maxima either in the same place or offset from each other. For many clumps, knowing the locations of nearby OB star clusters, one can infer from maps of \coab\ and \ncold\,the depth into the clump to which CO-desorbing or -dissociating radiation has penetrated.
    \item There appears to be a Region-specific temperature or range of temperatures for which the CO abundance is maximized due to the competition between CO desorption from ice and dissociation as the rates of both increase with temperature. This maximum must occur between 15 and 30~K; for Regions 9 through 11 and 26, the turnover point seems to be somewhere between 18 and 25~K. Region 26 was the only Region to have molecular clumps cold enough to show thermal the desorption rate increasing with temperature. 
    \item $L/M$ transitions slowly from being independent of \ncol\,for $\tau_{\nu}<<$1 to a function of \ncol$^{-1}$ for $\tau_{\nu}>>$1; when only temperature and \ncol\,are allowed to vary, $L/M$ becomes merely a complicated restatement of the temperature, and is therefore not an independent tracer of star formation. Variation in the $T_{\text{d}}$-\ncol\,plane is more complex than any one power law can capture, which presents both a crisis and an opportunity for modelling star formation because \ncol$\propto\,T^{-\beta}\gamma\kappa_0^{-1}$.
    \item Generalized histogram N-PDFs over the large-scale ($\gtrsim10$ pc) features of the CNC (Regions 9 through 11) generally take the form of log-normal distributions with power-law tails and poorly-defined transitions between the two functions. At the clump-scale (1-5 pc), a majority of the clumps have log-normal N-PDFs with some excess at large \ncol, but only in $\sim10\%$ of cases can the excess be fit with a power-law. The samples used to produce them necessarily suffer from several selection biases, and unquantifiable uncertainties due to crowding and the poor definition of the boundary of a molecular cloud/clump. Setting aside the merits of the theoretical background, N-PDF fitting may be qualitatively helpful, but we question the reliability of parameters derived from the fits, especially in crowded fields.
\end{enumerate}
\indent These results show that parsec-scale prestellar clumps are more than just brightness enhancements in FIR and mm-wave maps: they are physically, chemically, and thermodynamically distinct from the ambient diffuse ISM. Properly interpreted, the structures of prestellar clumps in density, temperature, and CO abundance should yield a wealth of new information about their star-formation history and ongoing evolution.

\section*{Acknowledgements}
 The CHaMP project is funded by the National Aeronautics and Space Administration through grant ADAP-NNX15AF64G, which we gratefully acknowledge. We wish to thank Mark Wieringa, Eric Pantin, and Ana Duarte-Cabral for their advice on data processing with IDL and Miriad. We also thank the anonymous referee for their prompt and helpful comments. This research makes use of the NASA/IPAC Infrared Science Archive, which is operated by the Jet Propulsion Laboratory, California Institute of Technology, under contract with the National Aeronautics and Space Administration. This research also used the Spanish Virtual Observatory (http://svo.cab.inta-csic.es) supported by the Spanish MICINN / MINECO through grants AyA2008-02156, AyA2011-24052.




\bibliographystyle{mnras}
\bibliography{RLPppr1bib} 


\clearpage
\appendix
\section{Uncertainty Maps}\label{sec:apu}
\begin{figure*}
	\includegraphics[width=\textwidth]{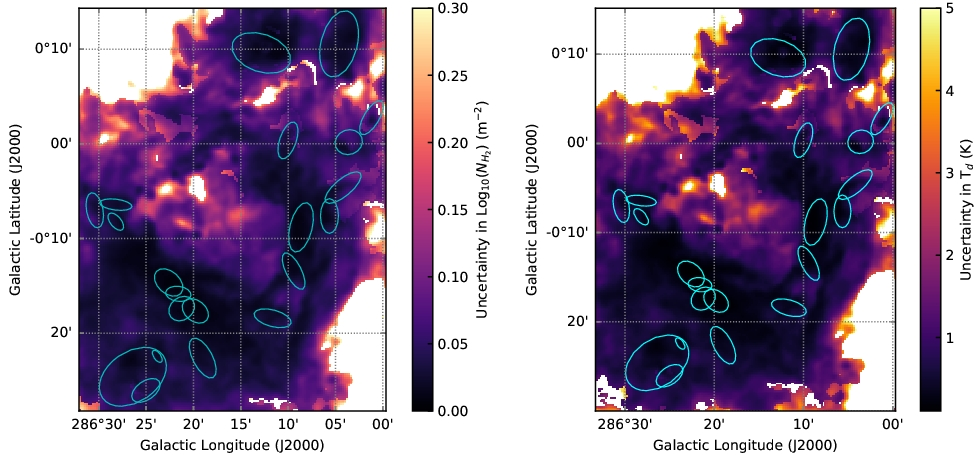}\vspace{-4mm}
    \caption{Uncertainty maps for Region 9 corresponding to Figure~\ref{fig:pmaps9}. \textbf{Left:} log$_{10}$[\ncol] (log \psqm). \textbf{Right:} $T_{\text{d}}$ (K).}
    \label{fig:errs9}
\end{figure*}

\begin{figure*}
	\includegraphics[width=\textwidth]{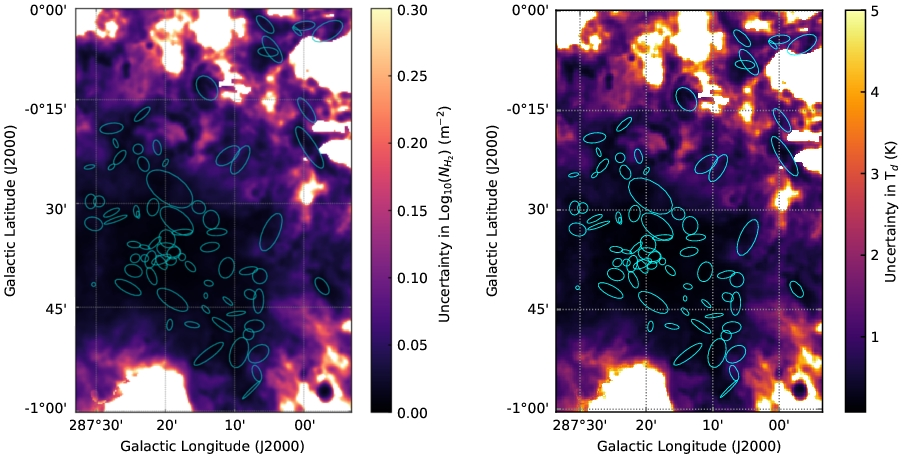}
	\vspace{-4mm}
    \caption{Same as Figure~\ref{fig:errs9} but for Region 10 and Figure~\ref{fig:pmaps10}.}
    \label{fig:errs10}
\end{figure*}

\begin{figure*}
	\includegraphics[width=\textwidth]{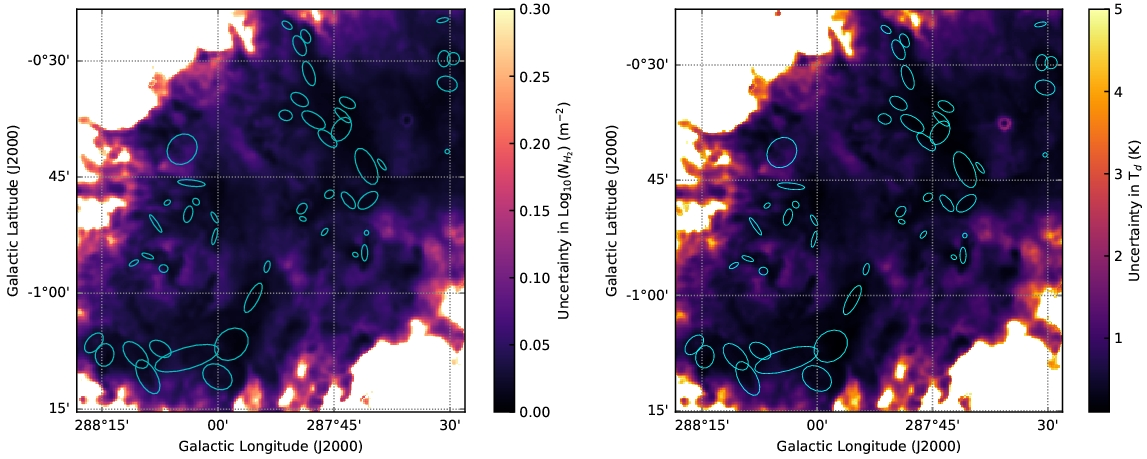}
	\vspace{-4mm}
    \caption{Same as Figure~\ref{fig:errs9} but for Region 11 and Figure~\ref{fig:pmaps11}.}
    \label{fig:errs11}
\end{figure*}

\begin{figure*}
	\includegraphics[width=\textwidth]{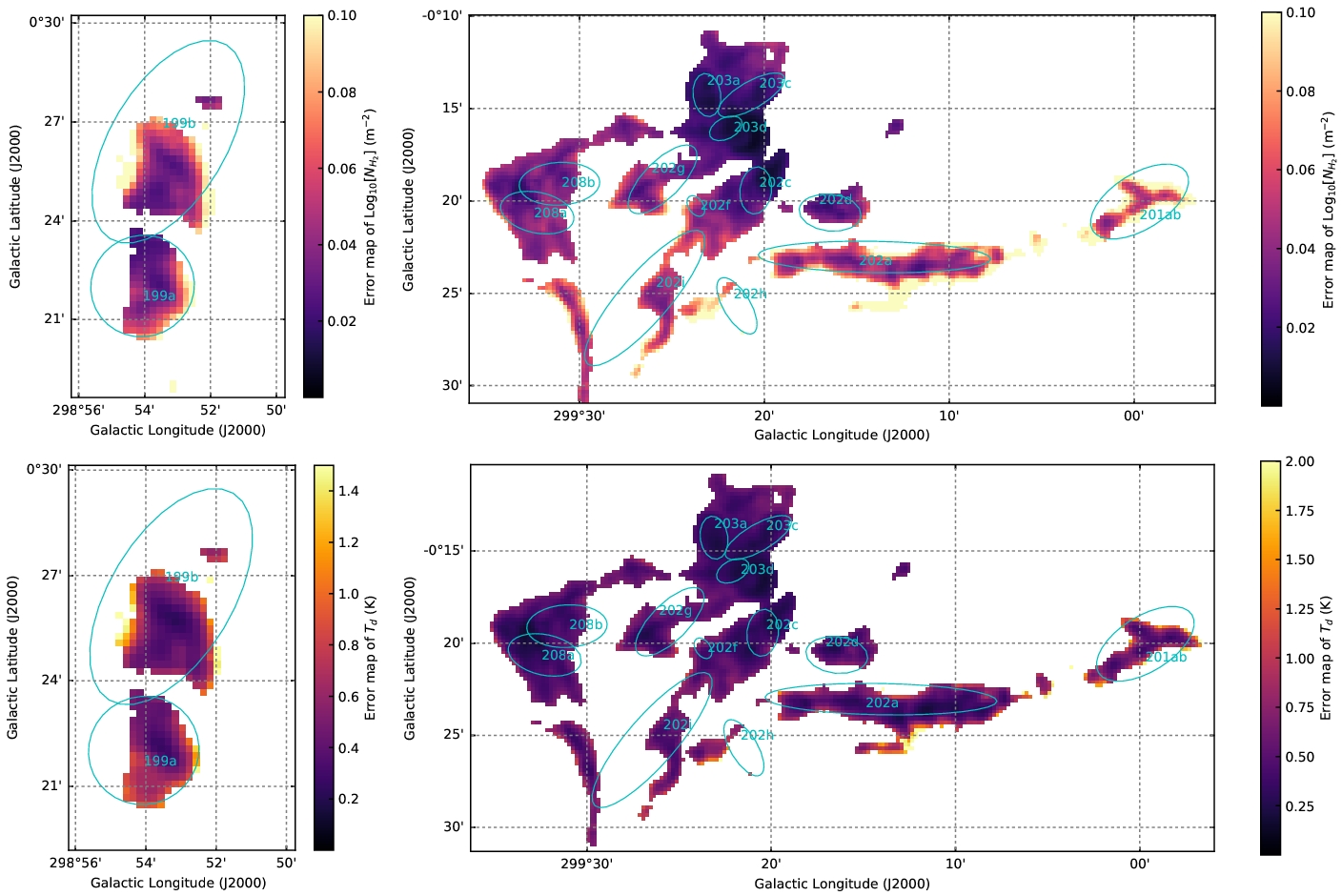} \vspace{-4mm}
    \caption{Uncertainty maps of log$_{10}$[\ncol] (top row) and $T_{\text{d}}$ (bottom row) for Region 26a (left column) and 26b (right column)}
    \label{fig:errs26}
\end{figure*}

\begin{figure}
	\includegraphics[width=\columnwidth]{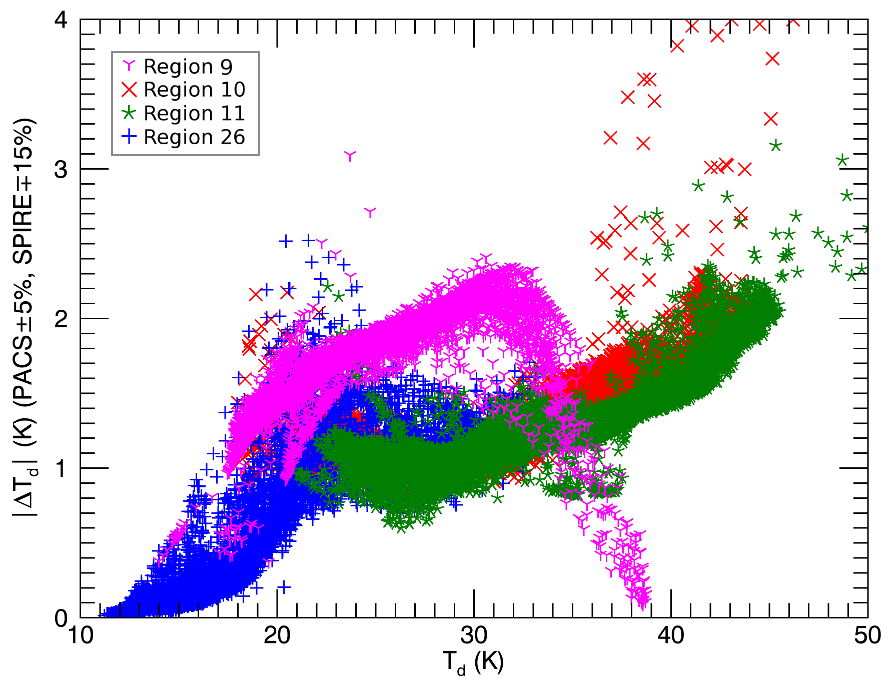}
	\vspace{-5mm}
    \caption{Calculated change in fitted $T_{\text{d}}$, denoted $\Delta T_\text{d}$, after raising/lowering the $I_{nu}$ in the \textit{Herschel} bands by their absolute calibration uncertainties and refitting each Region. Data from Region 9 are marked by magenta Y's, Region 10 by red X's, Region 11 by green stars, and Region 26 by blue crosses.}
    \label{fig:tdelt}
\end{figure}

\indent All archival continuum emission data used in this paper came with naive uncertainty maps, which we processed the same way as the science images. \texttt{MPFIT} automatically calculates uncertainties in fitted parameters based on input intensity uncertainties. Figures~\ref{fig:errs9} through \ref{fig:errs26} show the maps of uncertainty calculated this way for each parameter, corresponding to the images in \S\ref{sec:pms}. When subtracting a flat background, we propagated the standard deviation of the background value ($\sigma_{\text{BG}}$) to the rest of the error cube via ($\sigma^2$+$\sigma_{\text{BG}}^2$)$^{0.5}$ assuming the two uncertainties are uncorrelated (this assumption should fail near the selected background region, but this leads to overstating the uncertainty rather than understating it). Our color-correction module approximates its contribution to the total uncertainty by calculating a grid of color correction factors where every free parameter is varied by plus or minus one standard deviation (calculated by \texttt{MPFIT} in the initial SED-fit), taking the maximum possible deviation from the applied color correction as the uncertainty in $k_{\text{cor}}$, and updating the intensity uncertainty at that pixel by standard error propagation rules. Uncertainties contributed by both background subtraction and color correction are deliberately liberal estimates---the naive error maps they compound upon do not include systematic errors, and our code cannot yet include uncertainties in the central wavelength of each filter.\\
\indent As discussed previously in \S\ref{ssec:contdat} and \S\ref{ssec:code}, the most uncertain data important to the SED fit have an absolute calibration uncertainty of about 15\%. That translates to an uncertainty in log$_{10}$[\ncol] of about $\pm$0.07 (more precisely, +0.06, $-$0.08). To estimate how much $T_{\text{d}}$ can vary, we ran the SED fitting code on copies of the data cubes where PACS intensity data were raised/lowered by their absolute flux calibration error, 5\% \citep{pacsman}, while the SPIRE data were lowered/raised by their 15\% uncertainty. The uncertainties for each instrument are correlated across filters \citep{hsomanual,spire}, and the PACS and SPIRE filters tend to trace opposing sides of the SED. Thus the worst case scenario is that the intensities measured by PACS and SPIRE are each off by their respective calibration errors in opposite directions. If the PACS intensity measurements are higher than expected while the SPIRE intensities are lower, the fitted $T_{\text{d}}$ will be elevated; if the reverse is true, $T_{\text{d}}$ will be reduced. Unsurprisingly, we find that the uncertainty in $T_{\text{d}}$ ($\Delta\!T_{\text{d}}$) estimated this way is mostly an increasing function of the original fitted $T_{\text{d}}$, as shown in Figure~\ref{fig:tdelt}.\\
\indent The distributions of $\Delta\!T_{\text{d}}$ for Regions 10, 11, and 26 all link up with each other; only Region 9's $\Delta\!T_{\text{d}}$ distribution has a substantially different shape, and a higher $\Delta\!T_{\text{d}}$ overall due to the larger number of free parameters. All four Regions' distributions have a ``knee'' at $T_{\text{d}}\sim$20~K where $\Delta\!T_{\text{d}}$ where the slope flattens. For Regions 10, 11, and 26, $\Delta\!T_{\text{d}}$ has a local maximum and relatively large scatter at $T_{\text{d}}\sim$20~K, and declines in magnitude to a local minimum at 25$\lesssim\,T_{\text{d}}\lesssim$30~K before resuming its upward trajectory. The central wavelengths of the PACS filters (70 and 160\micron) mean that those filters straddle the peaks of any SED with a fitted $T_{\text{d}}$ between and 18 and 40~K. Thus, $T_{\text{d}}$ is most stable to correlated variations in $I_{\nu}$ in those bands when the SED's peak wavelength is about halfway between them. In Region 9, $\Delta\!T_{\text{d}}$ continues to rise with increasing $T_{\text{d}}$ (the cold temperature component) to a maximum of $\Delta\!T_{\text{d}}\sim$2~K at $T_{\text{d}}\sim$33~K, and thereafter declines with $T_{\text{d}}$. Blending between SED components began to occur at these temperatures, so part of the uncertainty we tried to estimate here may be getting lost in the uncertainty of how much each SED component contributes at a given wavelength.\\ 
\indent Throughout this exercise, \ncol\,varied by $\lesssim$5\% for Regions 9, 10, and 11, and by $\lesssim$15\% for Region 26.
\section{Background Subtraction---Additional Information}\label{sec:apbg}
\begin{figure}
	\includegraphics[width=\columnwidth,trim={2mm 4mm 2mm 0},clip]{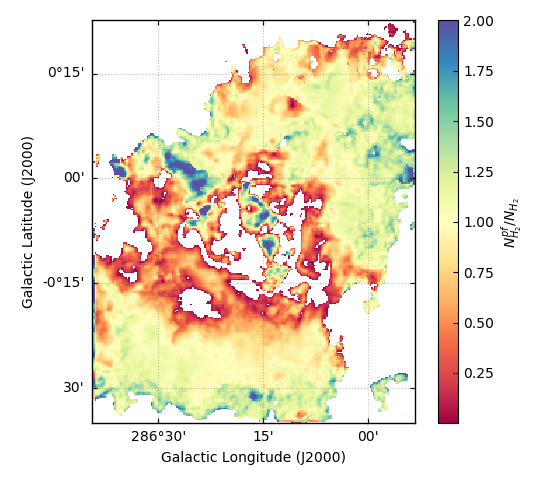} \vspace{-4mm}
    \caption{Ratio of \ncol\,values calculated by subtracting a ``background'' \ncol\,from the \ncol\,maps after SED-fitting (numerator), to \ncol\,values calculated by subtracting a flat background intensity at each wavelength and doing the SED-fitting afterward (denominator) in Region 9.}
    \label{fig:r9prevpost}
\end{figure}
\indent Figure~\ref{fig:r9prevpost} shows that background subtraction in the manner of \citet[][hereafter SOC15]{schneider15}, i.e. subtracting a constant ``background-level'' column density from the \ncol\,maps after SED-fitting, is not equivalent to calculating background-level intensities at each wavelength and performing SED-fitting after subtraction. Both approaches yield similar results where \ncol\,is highest, but diverge in more diffuse areas in a way that looks correlated with temperature. Near the PDR front around NGC 3324 and in the small HII region adjoining BYF 73, \citetalias{schneider15}'s approach tends to yields smaller column densities than we calculated, although there is significant variation inside the PDR and to the northeast. Far from the PDR where the gas is colder, the \citetalias{schneider15} method yields larger \ncol\,values than we calculate. The reason for the apparent temperature correlation is that allowing the background level to vary by wavelength necessarily changes the fitted $T_{\text{d}}$, whereas \citetalias{schneider15}'s method does not. The \citetalias{schneider15} method is certainly easier, and may be suitable for data on optically thin sources that are well-calibrated on an absolute scale (e.g. Planck data). However, our approach does not require absolute calibration of the input data, and is more appropriate where the assumption of a small optical depth can break down.\\
\begin{figure*}
  \includegraphics[width=0.9\textwidth]{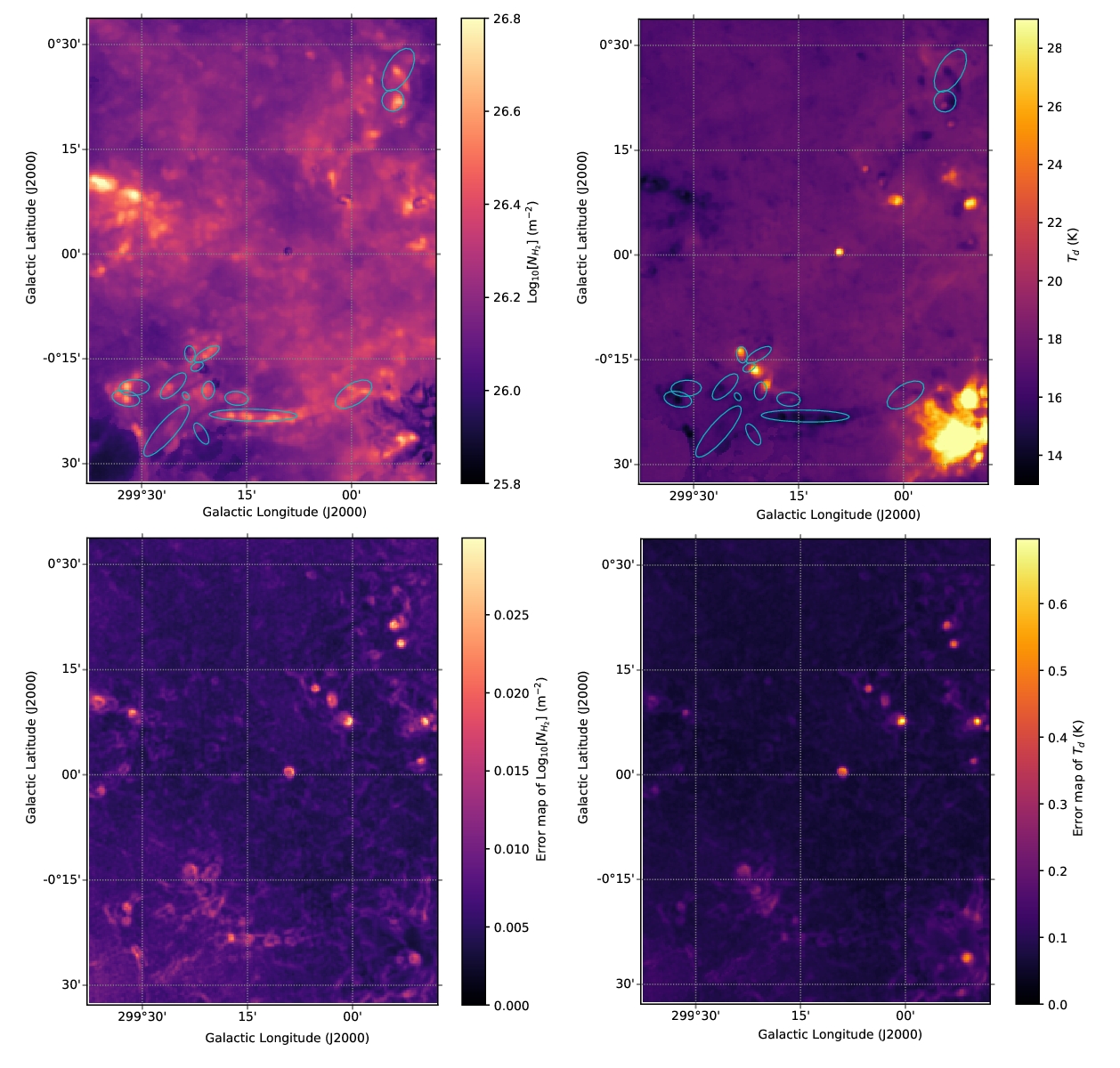} \vspace{-4mm}
  \caption{Parameter maps (top) and their error maps (bottom) for Region 26 before background subtraction. \textbf{Left:} log$_{10}$[\ncol] (log \psqm) derived from dust emission. \textbf{Right:} $T_{\text{d}}$ (K).}
  \label{fig:p26unsub}
\end{figure*}
\indent Figure~\ref{fig:p26unsub} shows the parameter maps and their uncertainties for Region 26 before background subtraction. For all scatter plots using data from the parameter maps of Region 26, we masked pixels where the background-subtracted column densities were $<$15\% of the column densities in the map in Figure~\ref{fig:p26unsub}. Our rationale is covered in \S\ref{ssec:contdat} and \S\ref{ssec:code}.\\
\indent Before modelling the background of Region 26 as a gradient, we tried several other methods. One was to fit a two-component modified blackbody function of the form,
\begin{equation}
    I_{\nu}\approx B_{\nu}(T)[1-e^{-\tau}]+B_{\nu}(T_{\text{BG}})[e^{-\tau}].
	\label{eq:mbb2}
\end{equation}
The fits failed to converge on any real values, probably because the foreground objects of interest are both hotter and colder than the background depending on the location, and both are seen in both emission and absorption. Knowing from the original parameter maps that most of the foreground and background could not be distinguished by temperature, we opted not to try the two-component fit of the form used on Region 9. Our \texttt{Mosaic-Math} program does not currently allow specification of different initial parameters at different spatial locations except by fitting each pixel one at a time, which is prohibitively time-consuming.\\
\begin{figure*}
    \includegraphics[width=\textwidth]{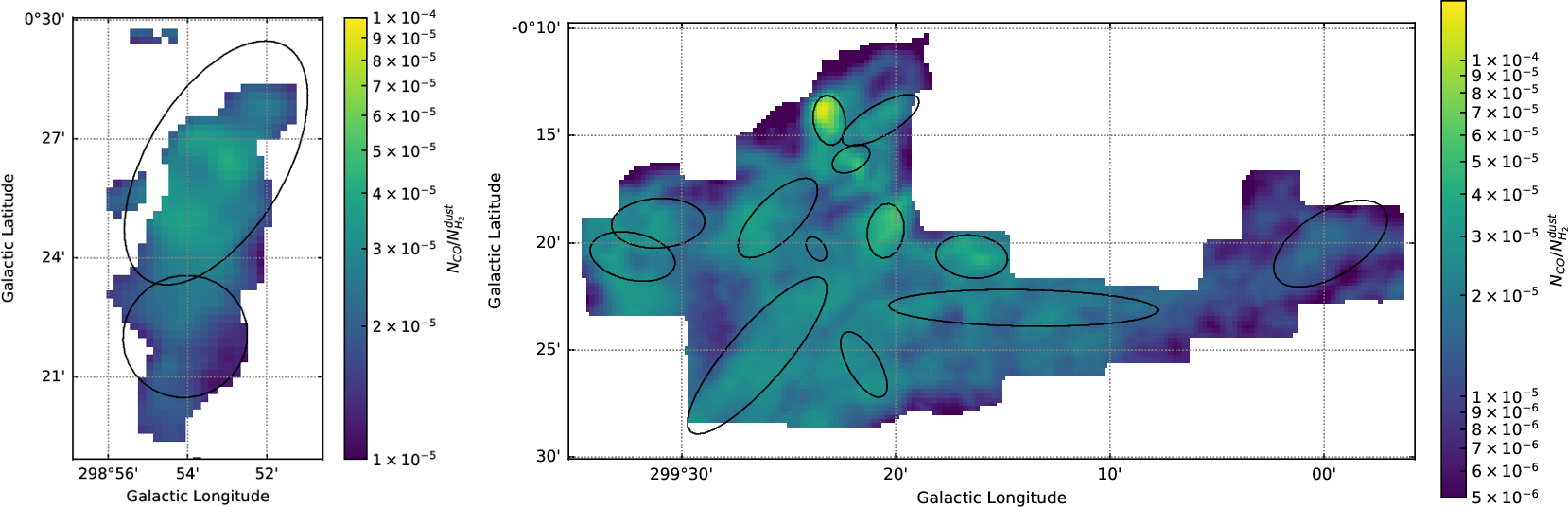}
    \vspace{-4mm}
    \caption{$N_{\text{CO}}$/\ncold\,for Region 26b, but without performing the background subtraction. Note the vastly different result compared to Figure~\ref{fig:r26bco}.}
    \label{fig:nratio26bu}
\end{figure*}
\indent For reference, Figure~\ref{fig:nratio26bu} shows the results we would have gotten had we not subtracted emission from the outskirts of the Dragonfish Nebula from Region 26. The IRDC (BYF 201ab and 202a) is nearly indistinguishable from the gas around the other clumps in BYF 202, and even some parts of the BYF 203 complex would appear to have $N_{\text{CO}}$/\ncold more consistent with what Gerner et al. 2014 predict for IRDCs, but still realistic within the margins of error. Unwary researchers who know the predicted CO abundances in IRDCs but don't know about the Dragonfish Nebula could easily convince themselves that this map is realistic.

\section{Luminosity-to-Mass Ratio Maps}\label{sec:aplm}
\begin{figure}
	\includegraphics[width=\columnwidth]{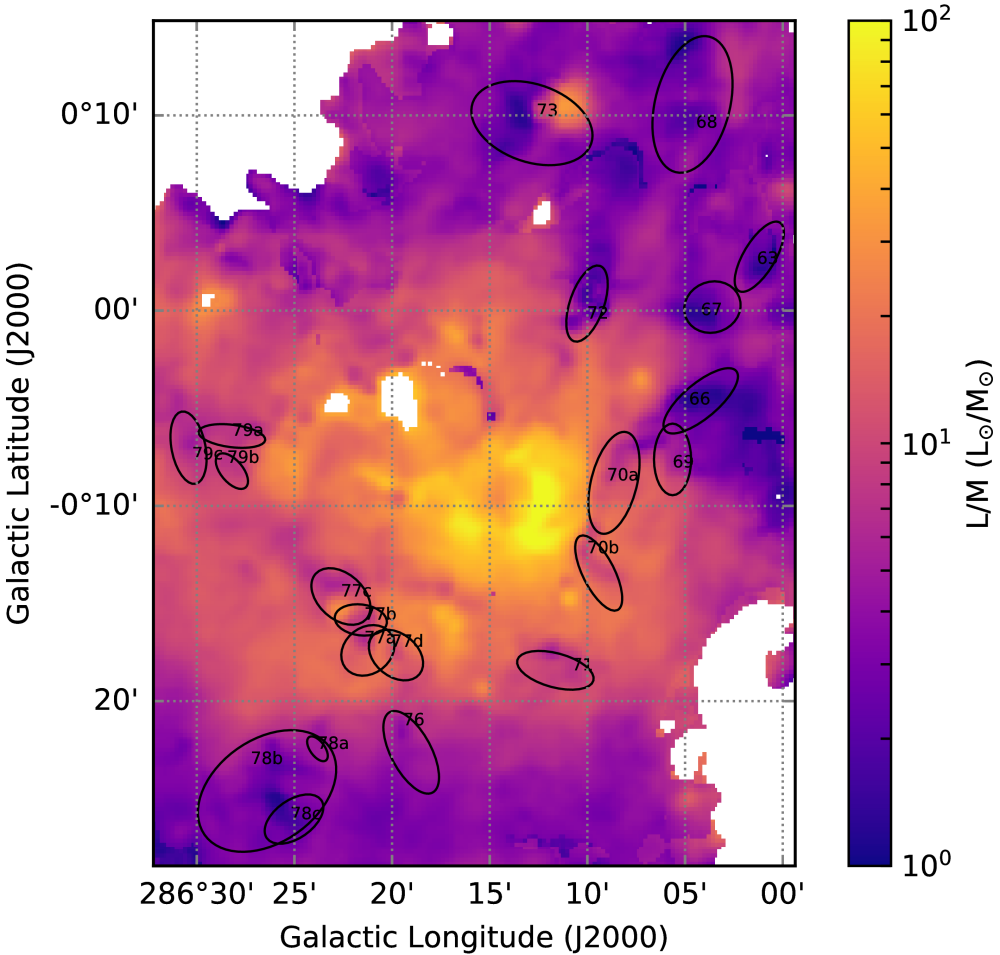} \vspace{-4mm}
    \caption{Map of $L/M$ ($L_{\odot}/M_{\odot}$) for Region 9.}
    \label{fig:r9lm}
\end{figure}
\begin{figure}
	\includegraphics[width=\columnwidth]{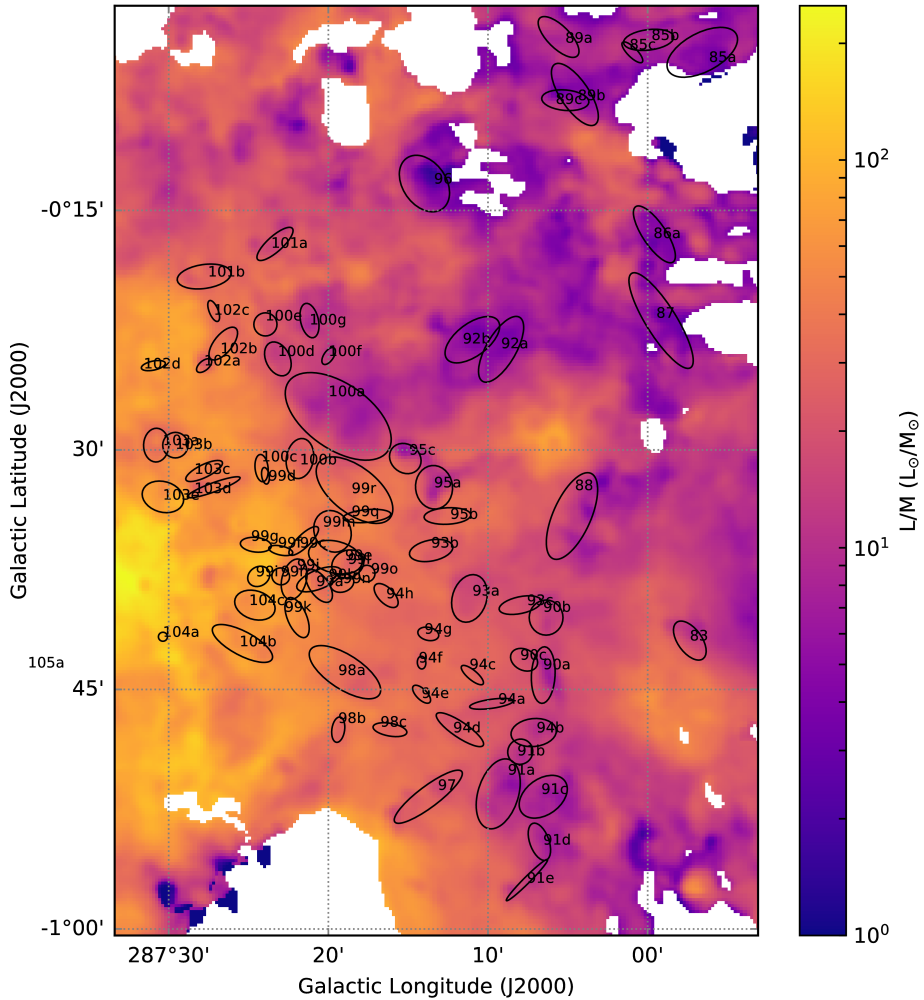} \vspace{-4mm}
    \caption{Map of $L/M$ ($L_{\odot}/M_{\odot}$) for Region 10.}
    \label{fig:r10lm}
\end{figure}
\begin{figure} 
	\includegraphics[width=\columnwidth]{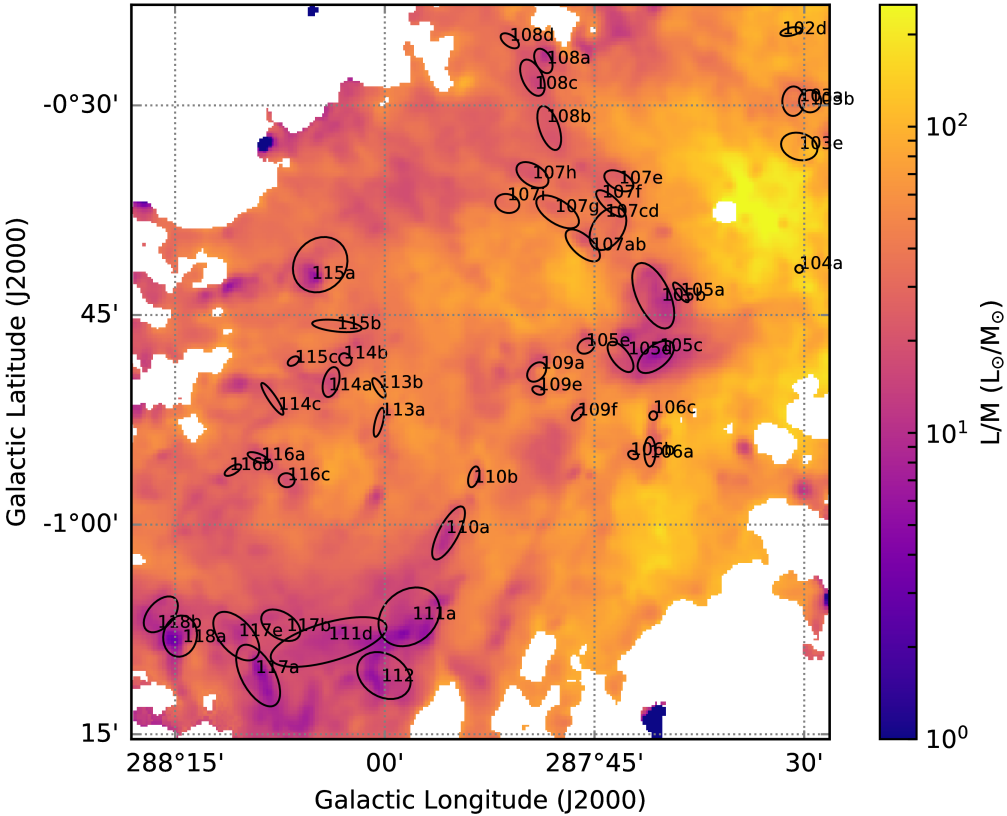} \vspace{-4mm}
    \caption{Map of $L/M$ ($L_{\odot}/M_{\odot}$) for Region 11.}
    \label{fig:r11lm}
\end{figure}
\begin{figure*}
  \begin{tabular}{@{}cc@{}}
	\includegraphics[width=.3\textwidth]{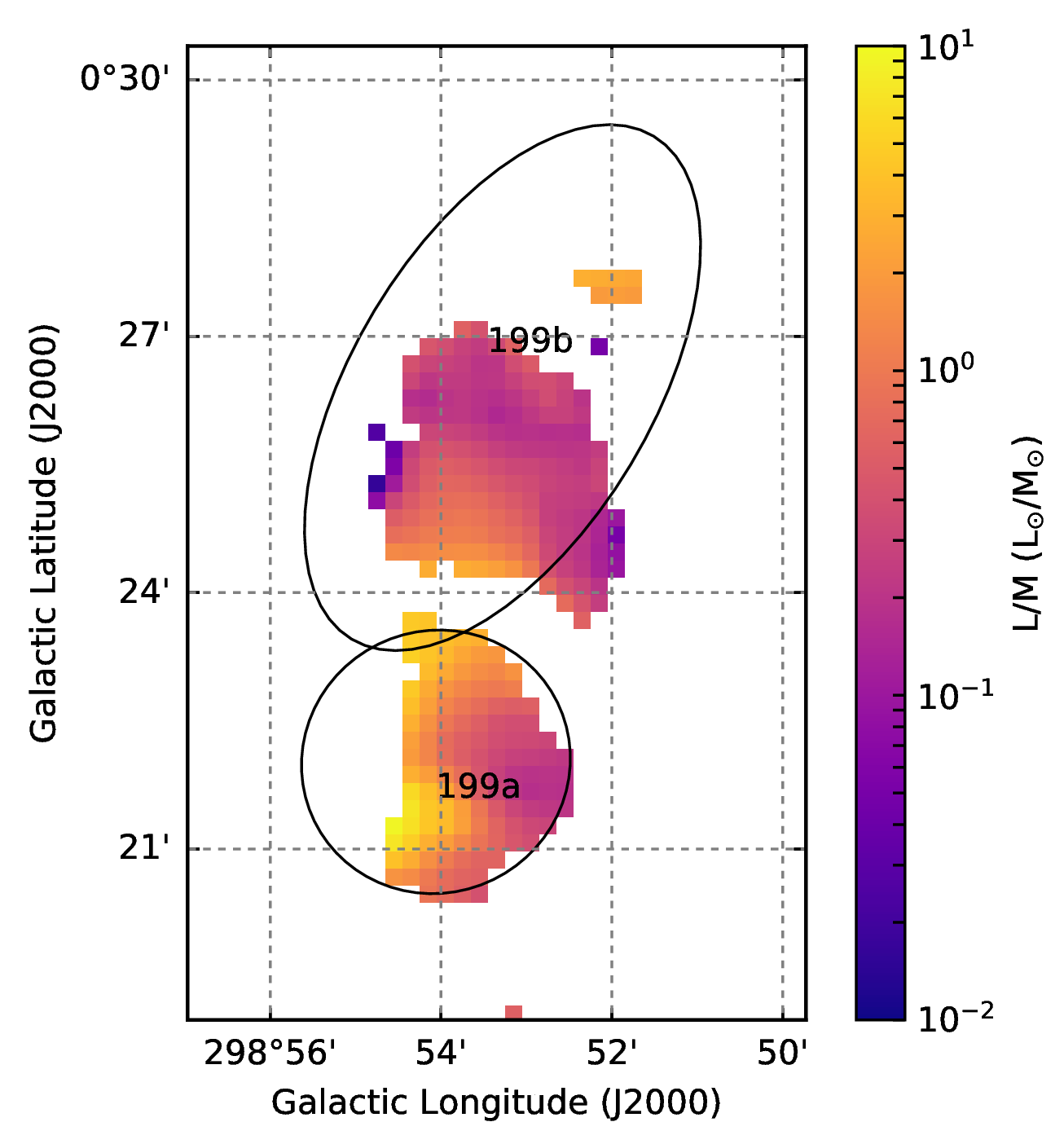} & \hspace{-4mm}
	\includegraphics[width=.72\textwidth]{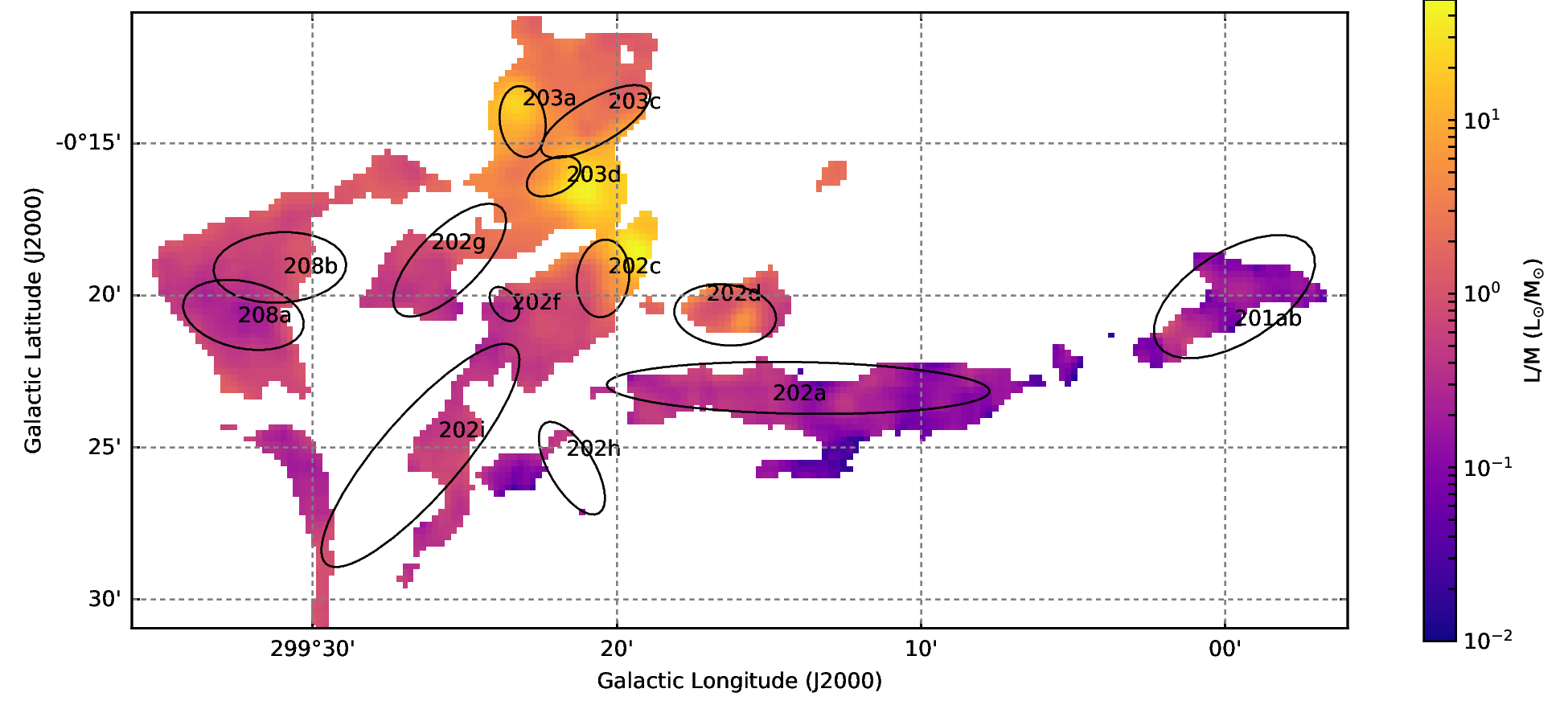}
	\end{tabular} \vspace*{-4mm}
    \caption{Map of $L/M$ ($L_{\odot}/M_{\odot}$) for Region 26A (left) and 26B (right).}
    \label{fig:r26lm}
\end{figure*}
\indent Figures~\ref{fig:r9lm} through \ref{fig:r26lm} show maps of $L/M$ across each Region. Note the resemblance to the temperature maps in Figures~\ref{fig:pmaps9} through \ref{fig:pmaps26}. $L/M$ and its uncertainties were computed assuming a single-temperature component SED for the integral, even where a two-component fit was used. Where a 2-component fit was used, the code simply takes the cool-component parameters and plugs them into the single-component model to pass to the integrating function. Typical uncertainties in these maps are around 20\%.

\section{Region 9 Warm Component}\label{sec:apwc}
\begin{figure*}
  \includegraphics[width=0.9\textwidth]{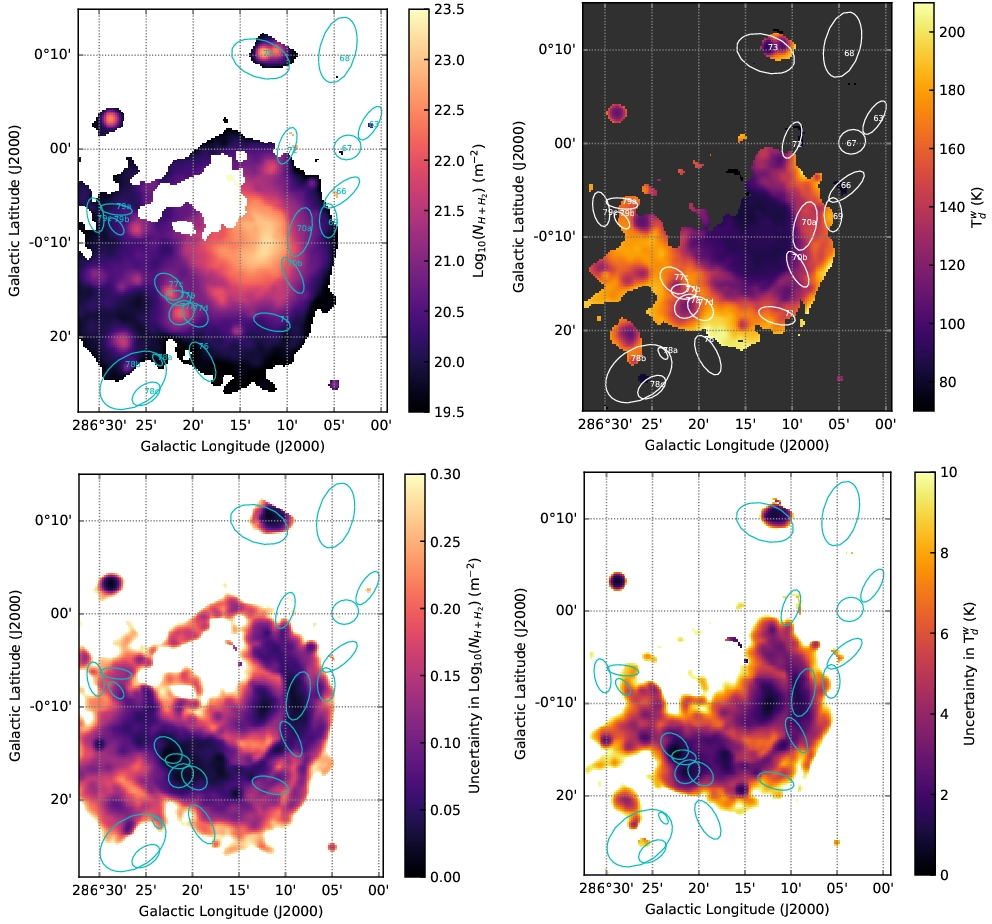}
  \caption{Parameter maps (top) and their error maps (bottom) for the ``warm'' component used to improve the SED fit in Region 9 \textbf{Left:} log$_{10}$[\ncol] (log \psqm) derived from dust emission. \textbf{Right:} $T_{\text{d}}$ (K).}
  \label{fig:p9w}
\end{figure*}
\indent The ``warm'' component used in SED-fitting for Region 9 is poorly constrained in any parameter (see Figure~\ref{fig:p9w}), and the uncertainties are correspondingly high. Reassuringly, its distribution qualitatively reflects the positions of known hot dust emission and embedded sources. There are also few significant discontinuities in the parameter maps of the cool component (Figure~\ref{fig:pmaps9}) where the warm component is too weak to fit and the program reverts to single-temperature fitting. However, in some (usually very low-column-density) pixels, adding a second component can cause fitting to fail for both the initial two-component fit and the back-up single-component fit where fitting a single component from the start succeeded. We ruled out the initial suspect, that the permitted temperature range of the cool component fails to update upon switching to a single-temperature SED fit, so the cause remains undetermined.

\bsp	
\label{lastpage}
\end{document}